\documentclass[aps,prd,showpacs,notitlepage,nofootinbib,superscriptaddress,floatfix,showkeys,twocolumn]{revtex4-2}
\usepackage{blindtext}
\usepackage{hyperref}
\usepackage{amsmath,amssymb}
\usepackage{graphicx}
\usepackage{bm}
\usepackage{latexsym}
\usepackage{epsfig}
\usepackage{psfrag}
\usepackage[normalem]{ulem}
\usepackage{color}
\usepackage[dvipsnames]{xcolor}
\usepackage{subfigure}
\usepackage{soul}

\def\nn{\nonumber} 
\def\f{\frac}

\def\l{\left}
\def\r{\right}
\def\d{{\mathrm{d}}}
\def\pa{\partial}
\def\Mpl{M_{_{\mathrm{Pl}}}}
\def\cA{{\mathcal{A}}}

\def\ps{\mathcal{P}_{_{\mathrm{S}}}}

\def\rem{\rho_{_{\mathrm{EM}}}}
\def\pb{\mathcal{P}_{_{\mathrm{B}}}}
\def\pe{\mathcal{P}_{_{\mathrm{E}}}}

\def\mub{\mu_{_{\mathrm{B}}}}

\def\HI{H_{_\mathrm{I}}}

\def\ee{\eta_{\mathrm{e}}}
\def\e1i{\epsilon_{1\mathrm{i}}}

\def\ke{k_{\mathrm{e}}}

\def\nb{n_{_{\mathrm{B}}}}
\def\nbb{\bar{n}_{_{\mathrm{B}}}}

\def\phii{\phi_{\mathrm{i}}}
\def\phie{\phi_{\mathrm{e}}}
\def\chie{\chi_{\mathrm{e}}}

\allowdisplaybreaks[1]

\begin{document}
\title{Circumventing the challenges in the choice of the non-conformal\\ 
coupling function in inflationary magnetogenesis}
\author{Sagarika Tripathy}
\email{E-mail: sagarika@physics.iitm.ac.in}
\affiliation{Centre for Strings, Gravitation and Cosmology,
Department of Physics, Indian Institute of Technology Madras, 
Chennai~600036, India}
\author{Debika Chowdhury}
\email{E-mail: debika.chowdhury@swansea.ac.uk}
\affiliation{Department of Physics, Swansea University, Swansea, SA2 8PP, U.K.}
\author{H. V. Ragavendra}
\email{E-mail: ragavendra.pdf@iiserkol.ac.in}
\affiliation{Department of Physical Sciences, Indian Institute of 
Science Education and Research Kolkata, Mohanpur, Nadia~741246, India}
\author{Rajeev Kumar Jain}
\email{E-mail: rkjain@iisc.ac.in}
\affiliation{Department of Physics, Indian Institute of Science, 
Bengaluru~560012, India}
\author{L.~Sriramkumar}
\email{E-mail: sriram@physics.iitm.ac.in}
\affiliation{Centre for Strings, Gravitation and Cosmology,
Department of Physics, Indian Institute of Technology Madras, 
Chennai~600036, India}
\begin{abstract}
As is well known, in order to generate magnetic fields of observed amplitudes
during inflation, the conformal invariance of the electromagnetic field has 
to be broken by coupling it either to the inflaton or to the scalar curvature.
Couplings to scalar curvature pose certain challenges even in slow roll 
inflation and it seems desirable to consider couplings to the inflaton.
It can be shown that, in slow roll inflation, to generate nearly scale 
invariant magnetic fields of adequate strengths, the non-conformal
coupling to the inflaton has to be chosen specifically depending on the 
inflationary model at hand.
In a recent work, we had found that, when there arise sharp departures 
from slow roll inflation leading to strong features in the scalar power spectra,
there inevitably arise sharp features in the spectra of the electromagnetic 
fields, unless the non-conformal coupling functions are extremely fine tuned.
In particular, we had found that, if there occurs an epoch of ultra slow roll
inflation (that is often required either to lower power on large scales or to 
enhance power on small scales), then the strength of the magnetic field over
large scales can be severely suppressed. 
In this work, we examine whether these challenges can be circumvented in 
models of inflation involving two fields.
We show that the presence of the additional scalar field allows us to construct 
coupling functions that lead to magnetic fields of required strengths even when 
there arise intermediate epochs of ultra slow roll inflation.
However, we find that the features in the spectra of the magnetic fields that 
are induced due to the departures from slow roll inflation cannot be completely 
ironed out.
We make use of the code MagCAMB to calculate the effects of the magnetic fields 
on the anisotropies in the cosmic microwave background and investigate if the 
spectra with features are broadly consistent with the current constraints.
\end{abstract}
\maketitle



\section{Introduction}

Magnetic fields are ubiquitous in the universe.
They are observed at different strengths over a wide range of scales, ranging 
from planets [$\mathcal{O}(0.5\,{\mathrm{G}})$] and stars [$\mathcal{O}(1\,
\mathrm{G}$)] to galaxies and clusters of galaxies [$\mathcal{O}(10^{-6}\,
\mathrm{G})$] (for reviews on magnetic fields, see Refs.~\cite{Grasso:2000wj,
Giovannini:2003yn,Brandenburg:2004jv,Kulsrud:2007an,Subramanian:2009fu,
Kandus:2010nw,Widrow:2011hs,Durrer:2013pga,Subramanian:2015lua,Vachaspati:2020blt}). 
The Fermi/LAT, HESS and MAGIC observations of TeV blazars over the last decade 
indicate that even the voids in the intergalactic medium may contain
magnetic fields [$\mathcal{O}(10^{-15}\,\mathrm{G}$)]~\cite{Neronov:1900zz,
Tavecchio:2010mk,Dolag:2010ni,Dermer:2010mm,Vovk:2011aa,Taylor:2011bn, 
Takahashi:2011ac,MAGIC:2022piy}.
While astrophysical processes involving the battery mechanism may be sufficient 
to explain the origin of magnetic fields in galaxies and clusters of galaxies 
(in this regard, see, for example, Refs.~\cite{Brandenburg:2004jv,Kulsrud:2007an}),
one may have to turn to a cosmological phenomenon to explain the magnetic 
fields observed in voids (in this context, see the reviews~\cite{Subramanian:2009fu,
Kandus:2010nw,Durrer:2013pga,Subramanian:2015lua,Vachaspati:2020blt}).

Without any doubt, the inflationary scenario is presently the most attractive 
paradigm to explain the origin of perturbations in the early universe.
Hence, it seems natural to turn to inflation for the generation of the 
primordial magnetic fields (PMFs).
However, since the standard electromagnetic action is conformally invariant
and the Friedmann-Lema\^itre-Robertson-Walker~(FLRW) universe is conformally 
flat, the strengths of minimally coupled electromagnetic fields are diluted 
considerably by the end of inflation.
Therefore, it becomes necessary to break the conformal invariance of the 
action governing the electromagnetic field in order to generate magnetic fields 
of observed strengths today.

The conformal invariance of the electromagnetic action is typically broken
by coupling the electromagnetic field to either the scalar curvature or the 
scalar field driving inflation (see, for example, Refs.~\cite{Turner:1987bw,
Ratra:1991bn,Bamba:2003av,Bamba:2006ga,Martin:2007ue,Bamba:2008ja,Watanabe:2009ct,
Kanno:2009ei,Demozzi:2009fu,Byrnes:2011aa,Ferreira:2013sqa,Ferreira:2014hma, 
Markkanen:2017kmy,Bamba:2020qdj,Bamba:2021wyx};
for discussions on effects due to the addition of a parity violating term,
see Refs.~\cite{Anber:2006xt,Durrer:2010mq,Caprini:2014mja,Ng:2014lyb,
Chowdhury:2018mhj,Sharma:2018kgs,Giovannini:2020zjo,Giovannini:2021thf,
Gorbar:2021rlt,Gorbar:2021zlr}).
It can be easily established that, if the non-conformal coupling function, 
say $J$, behaves as $\mathrm{e}^{2\,N}$, where~$N$ denotes the number of
$e$-folds, then one can arrive at a nearly scale invariant spectrum for 
the magnetic field with a strength that is dependent on the fourth power 
of the Hubble scale during inflation.
In a recent work, we had argued that while the coupling to the scalar curvature,
say $R$, works satisfactorily in power law inflation, it poses a problem in slow 
roll inflation~\cite{Tripathy:2021sfb}.
The reason being that, since the scalar curvature hardly varies during slow roll 
inflation, one has to raise~$R$ to a very high power in order to achieve the 
desired variation in the coupling function which leads to magnetic fields with
nearly scale invariant spectra.
In contrast, it is relatively easy to achieve the desired evolution of the 
coupling function (i.e. $J\propto \mathrm{e}^{2\,N}$) when the electromagnetic 
field is coupled to the inflaton.
However, there exists no universal form for the coupling function (in terms
of the dependence on the inflaton) and its form has to be chosen depending on 
the inflationary model being considered.

There has been a constant interest in the literature towards examining 
whether specific features in the inflationary scalar power spectrum 
improve the fit to the cosmic microwave background (CMB) and the large
scale structure data  (in this context, see, for instance, 
Refs.~\cite{Contaldi:2003zv,Sinha:2005mn,Powell:2006yg,Jain:2008dw,Jain:2009pm,
Hazra:2010ve,Benetti:2013cja,Hazra:2014jka,Hazra:2014goa,Chen:2016zuu,Chen:2016vvw,
Ragavendra:2020old,Antony:2021bgp,Braglia:2022ftm}).
Moreover, over the last few years, there has been an interest in 
investigating the non-trivial signatures of strong features at small
scales which can lead to enhanced levels of formation of primordial
black holes~(PBHs) and also generate secondary gravitational waves 
of possibly detectable amplitudes (for a short list of efforts in this regard, 
see Refs.~\cite{Garcia-Bellido:2017mdw,Ballesteros:2017fsr,Germani:2017bcs,
Dalianis:2018frf,Bhaumik:2019tvl,Ragavendra:2020sop,Dalianis:2020cla,
Ragavendra:2021qdu}). 
Such features are often achieved by considering inflationary potentials
that lead to departures from slow roll inflation. 
In our recent work~\cite{Tripathy:2021sfb}, we had shown that, unless the form 
of the non-conformal coupling function is extremely fine tuned, the deviations
from slow roll inflation that lead to features in the scalar power spectrum 
inevitably lead to features in the spectra of the electromagnetic fields as well.
For instance, in the case of single field models of inflation that permit a 
brief phase of ultra slow roll, the spectrum of the magnetic field has a 
strong scale dependence on small scales.
Moreover, the amplitude of the magnetic fields is strongly suppressed on large 
scales depending on the time of onset of the ultra slow roll epoch.

In this work, we shall examine whether these challenges can be circumvented
in two field models of inflation (for some recent discussions on generating
features in two field models at large and small scales, see, for example,
Refs.~\cite{Braglia:2020fms,Palma:2020ejf,Fumagalli:2020adf,Braglia:2020eai,
Pi:2017gih,Braglia:2020taf}).
The presence of the additional field permits a richer dynamics in two field 
models, and one can possibly utilize the second field to overcome the 
challenges faced in single field models.
As we shall see, with suitable choices for the non-conformal coupling function,
we are able to generate magnetic fields of desired strengths even in situations
wherein there arises an intermediate period of ultra slow roll.
However, it seems difficult to avoid the presence of features in the spectra 
of the electromagnetic fields.
In order to understand the viability of such electromagnetic spectra, we
shall consider two specific inflationary models with suitable couplings, 
and roughly compare the smoothed strengths of the generated magnetic 
fields with the constraints from the CMB data~\cite{Planck:2015zrl}.
Moreover, for one of the two models that we consider, we shall also evaluate 
the imprints of the PMFs on the angular power spectra of the CMB using the 
publicly available codes CAMB~\cite{Lewis:1999bs} and MagCAMB~\cite{Zucca:2016iur}.

This paper is organized as follows.
In Sec.~\ref{sec:sfm}, we shall briefly review the challenges that arise in
single field inflationary models and explore possible non-conformal coupling 
functions that can help us overcome the challenges.
In Sec.~\ref{sec:tfm}, we introduce the two field models of inflation that 
we shall consider.
We shall focus on two models that lead either to a suppression in power on 
large scales or to an enhancement in power on small scales.
Thereafter, we shall go on to construct suitable non-conformal coupling functions 
that allow us to arrive at magnetic fields of desired strengths over the CMB scales.
We shall discuss the cases of non-helical as well as helical magnetic fields.
As we shall illustrate, despite the presence of the additional field, it seems 
impossible to completely iron out the features that arise in the spectra of the 
electromagnetic fields.
In Sec.~\ref{sec:i-on-cmb}, we shall first examine if the amplitudes of the magnetic
fields that we obtain in the two inflationary models are broadly consistent with
the constraints on the PMFs from the CMB data.
Then, focusing on the non-helical case, using MagCAMB, we shall compute the angular 
power spectra of the CMB generated by the so-called passive and compensated magnetic
modes~\cite{Zucca:2016iur}.
We shall carry out such an exercise for one of the two models which leads to a 
nearly scale invariant spectrum for the magnetic field over large scales.
We shall also approximately calculate the spectrum of the curvature perturbations
induced by the magnetic field during inflation~\cite{Bonvin:2011dt,Bonvin:2013tba}, 
and compute the corresponding angular power spectra of the CMB using
CAMB~\cite{Lewis:1999bs}.
We shall compare these quantities with the contributions due to the primary scalar 
and tensor power spectra generated from the Bunch-Davies vacuum.
We shall conclude in Sec.~\ref{sec:c} with a summary of the results obtained.
We shall relegate some of the related discussions to three appendices.

At this stage of our discussion, let us clarify a few points regarding our
conventions and notations. 
We shall work with natural units such that $\hbar=c=1$, and set the reduced 
Planck mass to be $\Mpl=\l(8\,\pi\, G\r)^{-1/2}$.
We shall adopt the signature of the metric to be~$(-,+,+,+)$.
Note that Latin indices will represent the spatial coordinates, except for~$k$ 
which will be reserved for denoting the wave number. 
We shall assume the background to be the spatially flat FLRW universe described 
by the following line element: 
\begin{equation}
\d s^2=-\d t^2+a^2(t)\,\d {\bm x}^2
=a^2(\eta)\, \l(-\d \eta^2+\d {\bm x}^2\r),\label{eq:FLRW}
\end{equation}
where $t$ and $\eta$ the denote cosmic time and conformal time coordinates, 
while $a$ represents the scale factor.
Also, an overdot and an overprime will denote differentiation with respect 
to the cosmic and conformal time coordinates, respectively.
Moreover, as mentioned before, $N$ represents the number of $e$-folds.
Lastly, $H=\dot{a}/a$ and $\mathcal{H}=a\,H=a'/a$ shall represent the Hubble 
and the conformal Hubble parameters, respectively.


\section{Challenges in single field models}\label{sec:sfm}

In this section, we shall briefly highlight the challenges one faces in certain 
single field inflationary modes to generate magnetic fields of the desired 
amplitudes and spectral shapes.
Before we go on to describe these challenges, in order for this paper to be 
self-contained, let us quickly recall a few essential points that we will 
require later for our discussion.


\subsection{Electromagnetic modes and power spectra}

We shall consider electromagnetic fields described by the 
action~\cite{Bamba:2003av,Anber:2006xt,Martin:2007ue,Watanabe:2009ct,
Kanno:2009ei,Durrer:2010mq,Caprini:2014mja,Markkanen:2017kmy,Chowdhury:2018mhj,
Sharma:2018kgs,Giovannini:2020zjo,Giovannini:2021thf,Gorbar:2021rlt,Gorbar:2021zlr}
\begin{eqnarray}
S[A^\mu] & = & -\f{1}{16\,\pi}\, \int \d^4x\, \sqrt{-g}\, J^2(\phi)\, 
\biggl[F_{\mu\nu}\,F^{\mu\nu}\nn\\ 
& & -\, \f{\gamma}{2}\, F_{\mu\nu}\,\widetilde{F}^{\mu\nu}\biggr], 
\end{eqnarray}
where $J(\phi)$ denotes the non-conformal coupling function and $\gamma$ is a 
constant.
As usual, the field tensor~$F_{\mu\nu}$ is expressed in terms of the vector 
potential $A_\mu$ as $F_{\mu\nu}=( \pa_{\mu}\,A_{\nu}-\pa_{\nu}\,A_{\mu})$, 
while the dual field tensor $\widetilde{F}^{\mu\nu}$ is defined as 
$\widetilde{F}^{\mu\nu} = (\epsilon^{\mu\nu\alpha\beta}/\sqrt{-g})\, 
F_{\alpha\beta}$, with $\epsilon^{\mu\nu\alpha\beta}$ being the completely 
anti-symmetric Levi-Civita tensor.
The second term in the above action leads to violation of parity and, 
during inflation, this term amplifies the electromagnetic modes associated 
with one of the two states of polarization compared to the 
other~\cite{Anber:2006xt,Durrer:2010mq,Caprini:2014mja,Chowdhury:2018mhj,
Sharma:2018kgs,Giovannini:2020zjo,Giovannini:2021thf,Gorbar:2021rlt,
Gorbar:2021zlr}. 

In the spatially flat FLRW background of our interest, to arrive at the 
solutions describing the electromagnetic field, it proves to be convenient
to work in the Coulomb gauge wherein $A_\eta = 0$ and $\pa_i\,A^i=0$.
We shall denote the Fourier modes of the three-vector potential $A^i$ 
as~$\bar{A}_k$, where the subscript $k$ represents the wave number.
If we write $\bar{A}_k=\cA_k/J$, then, in the Coulomb gauge, the mode 
functions~$\cA_k$  are found to satisfy the differential 
equation~\cite{Anber:2006xt,Durrer:2010mq,
Caprini:2014mja,Chowdhury:2018mhj,Sharma:2018kgs,Giovannini:2020zjo,
Giovannini:2021thf,Gorbar:2021rlt,Gorbar:2021zlr}
\begin{equation}
\cA_k^{\sigma\,\prime\prime}
+ \l(k^2 + \f{2\,\sigma\,\gamma\,k\,J^\prime}{J}
- \f{J^{\prime\prime}}{J}\r)\cA_k^\sigma = 0,\label{eq:cA-h-de1}
\end{equation}
where $\sigma=\pm$ corresponds to the two helicities.
The power spectra of the magnetic and electric fields, viz. $\pb(k)$ 
and~$\pe(k)$, are defined as (see, for example, Refs.~\cite{Martin:2007ue,
Subramanian:2009fu}) 
\begin{equation}
\pb(k) =\f{\d \langle \hat{\rho}_{_{\mathrm{B}}}\rangle}{\d \ln k},\quad
\pe(k) =\f{\d \langle \hat{\rho}_{_{\mathrm{E}}}\rangle}{\d \ln k},
\label{eq:pseb-d}
\end{equation}
where $\rho_{_{\mathrm{B}}}$ and $\rho_{_{\mathrm{E}}}$ are the energy
densities associated with the magnetic and electric fields, respectively,
while the expectation values are to be evaluated in the Bunch-Davies vacuum.
It is also useful to note here that we shall define the spectral index $\nb$
of the magnetic field as $\nb=(\d \ln \pb(k)/\d \ln k)$, and we shall refer
to the case wherein $\nb=0$ as a scale invariant spectrum.
The power spectra $\pb(k)$ and~$\pe(k)$ can be expressed in terms of the 
mode functions $\cA_k$ and their time derivatives $\cA_k'$ as 
follows~\cite{Martin:2007ue,Subramanian:2009fu,Tripathy:2021sfb}:
\begin{subequations}
\begin{eqnarray}\label{eq:psbe-h}
\pb(k) &=& \f{k^{5}}{4\,\pi^2\,a^{4}}\,
\l[\l\vert  \cA_k^{+}\r\vert^2 
+ \l\vert \cA_k^{-}\r\vert^2\r],\label{eq:psb-h1}\\
\pe(k) &=& \f{k^3}{4\, \pi^2\, a^4}\, \l[\l\vert \cA_k^{+\prime}
- \f{J^\prime}{J}\cA_k^+\r\vert^2+\l\vert \cA_k^{-\prime}
- \f{J^\prime}{J}\cA_k^-\r\vert^2\r].\nn\\
\label{eq:pse-h1}
\end{eqnarray}
\end{subequations}

In a de Sitter universe, one often chooses the non-conformal coupling function 
to be of the form $J(\eta) = \l[a(\eta)/a(\ee)\r]^2$, where $\ee$ denotes the 
conformal time coordinate towards the end of inflation. 
Such a choice for the coupling function leads to a scale invariant spectrum for 
the magnetic field (in this context, see, for example, Refs~\cite{Martin:2007ue,
Subramanian:2009fu,Subramanian:2015lua}).
In models allowing slow roll inflation, there exists no universal or model 
independent form of $J(\phi)$ that leads to the above-mentioned behavior in 
terms of the scale factor. 
However, given a model of inflation that permits slow roll, based on the evolution 
of the scalar field, it is easy to construct a function~$J(\phi)$ that approximates
the desired behavior of~$J \propto a^2$ fairly well.  
For such a choice of the non-conformal coupling function, the power spectra of the
electromagnetic fields, evaluated at late times, i.e. as $(k\, \ee) \to 0$, can be 
expressed as (see, for instance, Ref.~\cite{Tripathy:2021sfb})
\begin{subequations}\label{eq:ne2h1}
\begin{eqnarray}
\f{\pb(k)}{\Mpl^4}
&=&\f{9\,\HI^4}{4\,\pi^2}\,f(\gamma)
=\f{9\,\pi^2}{16}\,\l(r\,A_\mathrm{s}\r)^2\,
f(\gamma),\label{eq:pb-ne2h}\\
\f{\pe(k)}{\Mpl^4}
&=&\f{\pb(k)}{\Mpl^4}\,
\biggl[\gamma^2 -\f{\mathrm{sinh}^2(2\,\pi\,\gamma)}{3\,\pi\,
\l(1+\gamma^2\r)\,f(\gamma)}\,(-k\,\ee)\nn\\
& &+\,\f{1}{9}\,\l(1+23\,\gamma^2+40\,\gamma^4\r)\,\l(-k\,\ee\r)^{2}\biggl],
\label{eq:ne2h2}
\end{eqnarray}
\end{subequations}
where $\HI$ represents the Hubble scale during inflation, $A_\mathrm{s}
=2.1\times 10^{-9}$ denotes the observed amplitude of the scalar power 
spectrum at the pivot scale, and $r$~represents the tensor-to-scalar
ratio~\cite{Planck:2018jri,BICEPKeck:2021gln}.
Also, the function $f(\gamma)$ is given by~\cite{Tripathy:2021sfb}
\begin{equation}
f(\gamma)=\f{\mathrm{sinh}\,(4\,\pi\,\gamma)}{4\,\pi\,\gamma\,
\l(1+5\,\gamma^2+4\,\gamma^4\r)},\label{eq:fg1}
\end{equation}
and we should be point out that $f(\gamma)$ reduces to unity in the limit of
vanishing~$\gamma$.

We shall now make a few clarifying remarks regarding the results we have quoted 
above.
Let us first discuss the shape of the electromagnetic spectra in slow roll 
inflation before we turn to comment on their amplitudes.
In the case of helical fields (i.e. when the parameter~$\gamma$ is non-zero), it 
is the first term within the square brackets in the expression~\eqref{eq:ne2h2} 
for~$\pe(k)$ that dominates, and hence one finds that the power spectra of both 
the magnetic and electric fields are scale invariant, with their amplitudes being
determined by the tensor-to-scalar ratio~$r$ (or, equivalently, $\HI$) and the
function~$f(\gamma)$.
When $\gamma$ vanishes (i.e. in the case of non-helical fields), the function 
$f(\gamma)$ reduces to unity and one finds that the spectrum of the magnetic 
field $\pb(k)$ continues to remain scale invariant.
However, in such a limit, it is the last term within the square brackets
in the expression for~$\pe(k)$ that survives, indicating that the power spectrum of 
the electric field behaves as~$k^2$.

Let us now understand the amplitudes of the spectra in slow roll inflation.
Clearly, in the helical case, for $\gamma \simeq \mathcal{O}(1)$, the strengths 
of the scale invariant spectra of the magnetic and electric fields are comparable
and are primarily determined by the tensor-to-scalar ratio~$r$.
But, in the non-helical case, due to the $k^2$ dependence, the spectrum of the 
electric field is considerably suppressed over large scales when compared to 
the scale invariant amplitude of the magnetic field. 
We find that, for $10^{-12}\lesssim r \lesssim 10^{-2}$, upon assuming instantaneous 
reheating, inflationary magnetogenesis leads to non-helical magnetic fields of 
strength in the range of $10^{-17} \lesssim B_0 \lesssim 10^{-11}\, \mathrm{G}$ 
{\it today}.\/
It should be clear that the function $f(\gamma)$ grows exponentially with $\gamma$
[see Eq.~\eqref{eq:fg1}].
As a result, the amplitude of the helical fields can be considerably enhanced
at late times when compared to the non-helical case.
It can be shown that, for inflationary models wherein $r \simeq 10^{-2}$, if the 
backreaction due to the helical electromagnetic fields has to be negligible, 
then one has to work with $\gamma \lesssim 2.5$~\cite{Tripathy:2021sfb}.
When considering helical fields, we shall work with $\gamma=0.25$.
For~$\gamma =0.25$, we find that $f(0.25) \simeq 3$, which implies that the 
strengths of the helical magnetic fields will be higher by such a factor 
when compared to the non-helical case.


\subsection{Difficulty in ultra slow roll inflation}\label{subsec:d-usr}

\begin{figure*}
\centering
\includegraphics[width=8.5cm]{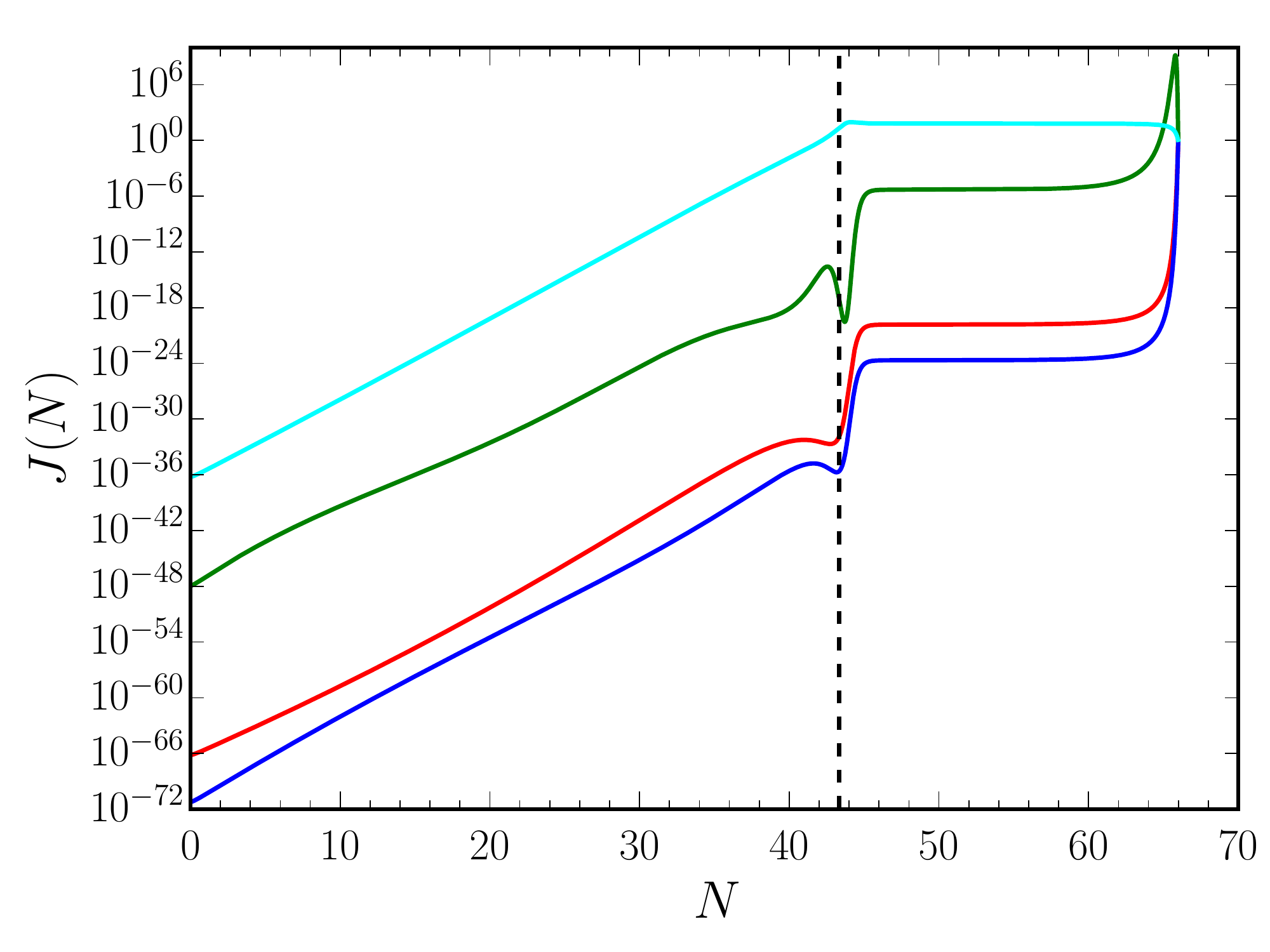}
\includegraphics[width=8.5cm]{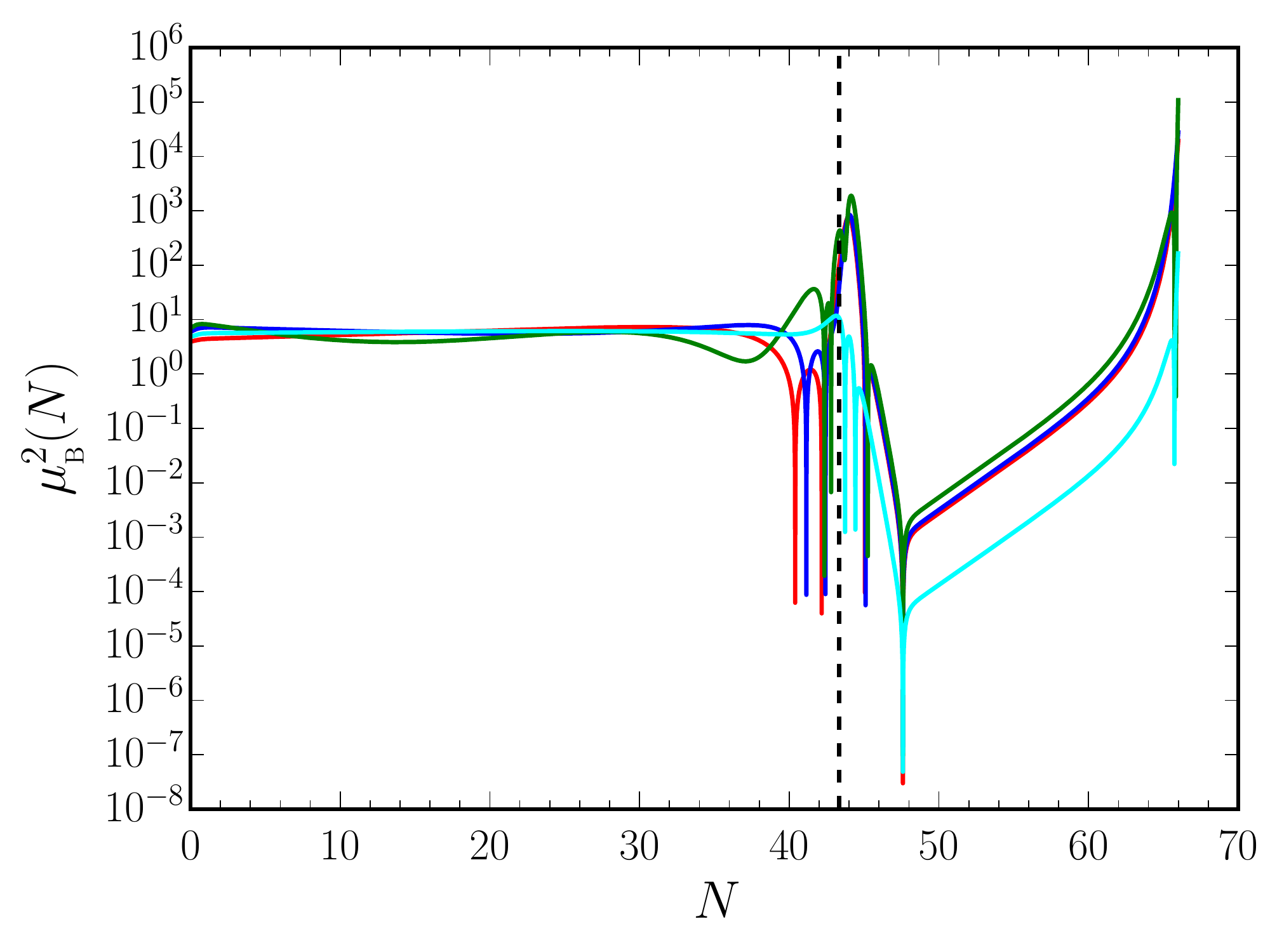}
\caption{The evolution of the non-conformal coupling function $J$ (on the left) 
and the quantity~$\mub^2=J^{\prime\prime}/(J\,a^2\,H^2)$ (on the right) in a model
involving a single, canonical scalar field that leads to an epoch of ultra 
slow roll inflation---the potential~(58) in our earlier 
paper~\cite{Tripathy:2021sfb}---have been plotted as functions of $e$-folds~$N$.
These plots illustrate the challenge faced in such scenarios.
In our previous paper, we had worked with a coupling function $J(\phi)$ that 
was arrived at by fitting the numerical solution for the scalar field with 
a fourth order polynomial {\it until}\/ the onset of the ultra slow roll regime 
(plotted here in cyan). 
Apart from such a choice for the coupling function, we have plotted the coupling 
function $J(\phi)$ as well as the quantity $\mub^2$ wherein the entire evolution of 
the field (i.e. from the initial time until the end of inflation) has been fit to 
fourth, sixth and eighth order polynomials (in red, blue and green, respectively).
Note that $J\propto \mathrm{e}^{2\,N}$ and $\mub^2\simeq 6$ until the onset 
of the ultra slow roll regime (indicated by the vertical dashed black lines in 
the two figures), which are required to lead to a scale invariant spectrum for 
the magnetic field.
However, it seems impossible to achieve such a behavior for $J$ and $\mub^2$
after the onset of ultra slow roll.
This can be primarily attributed to the fact that the field hardly evolves
during this period.}\label{fig:J-usr}
\end{figure*}
We had mentioned above that, in models permitting slow roll inflation, based 
on the evolution of the field arrived at in the slow roll approximation, it 
is possible to construct a function $J(\phi)$ so that the desired behavior of
$J\propto a^2$ is achieved. 
Now, consider situations wherein there arise deviations from slow roll.
In single field models of inflation involving the canonical scalar field, 
typically, departures from slow roll occur because of features in the
inflationary potential, such as a step, a bump, a dip, a burst of 
oscillation, or a point of inflection.
If the deviations from slow roll are small, then one can work with the form of 
$J(\phi)$ that is constructed using the slow roll approximation in the absence 
of the feature in the potential.
Under such conditions, in our earlier work~\cite{Tripathy:2021sfb}, we had shown
that the departures from slow roll inflation generate features in the spectra of 
the electromagnetic fields in much the same manner as they produce features in the 
scalar power spectrum.
The small deviations from slow roll induce brief departures from scale invariance
in the spectrum of the magnetic field.
However, we had found that, for a given choice of the coupling function $J(\phi)$,
say, chosen based on the slow roll evolution at early or late times, strong 
departures from slow roll inflation generically lead to prominent features in the 
spectra of the electromagnetic fields.

Strong departures from slow roll inflation are usually considered in two contexts.
They are invoked either to suppress the scalar power over large scales in order 
to explain the lack of power observed at the low multipoles~\cite{Contaldi:2003zv,
Jain:2008dw,Jain:2009pm,Hazra:2014jka,Hazra:2014goa,Ragavendra:2020old}
or to boost the power over small scales leading to enhanced formation of 
PBHs~\cite{Garcia-Bellido:2017mdw,Ballesteros:2017fsr,Germani:2017bcs,
Dalianis:2018frf,Bhaumik:2019tvl,Ragavendra:2020sop,Dalianis:2020cla,
Ragavendra:2021qdu}.
These features are often achieved with the aid of an epoch of ultra slow 
roll inflation during which the first slow roll parameter decreases
exponentially~\cite{Tsamis:2003px,Kinney:2005vj}.
While the first slow roll parameter remains small during this period, the 
second and higher order slow roll parameters prove to be large resulting
in a violation of the slow roll conditions.
In single field models of inflation driven by the canonical scalar field, 
a period of ultra slow roll, in turn, seems guaranteed, if there is a point
of inflection in the potential.
In our earlier work~\cite{Tripathy:2021sfb}, we had found that, in models 
which permit a period of ultra slow roll inflation, the non-conformal 
coupling function hardly evolves during the phase of ultra slow roll.
Due to this reason, the spectra of both the magnetic and electric fields 
behave as $k^4$ for wave numbers that leave the Hubble radius after the 
onset of ultra slow roll inflation.
Moreover, the amplitude of the spectra on large scales are suppressed by 
the factor of $\mathrm{e}^{-4\,(N_\mathrm{e}-N_1)}$, where $N_1$ and 
$N_\mathrm{e}$ represent the $e$-folds at the onset of the epoch of ultra 
slow roll and the end of inflation, respectively.
In arriving at these spectra, we had considered coupling functions that 
are based on the behavior of the scalar field during the initial slow roll
regime.
One may wonder if it is possible to arrive at the desired non-minimal 
coupling function (i.e. one wherein $J(\phi)\propto a^2$) by fitting for
the entire evolution of the scalar field.
As we have illustrated in Fig.~\ref{fig:J-usr}, we find that this is indeed 
difficult to achieve.
This primarily occurs due to the fact that, generically, the scalar field 
virtually ceases to evolve once the epoch of ultra slow roll begins, until
the very end of inflation.
As we shall discuss in this work, due to the additional degree of freedom 
available, it is possible to circumvent such a challenge in the case of
two field models.


\section{Circumventing the challenges in two field models}\label{sec:tfm}

In this section, we shall illustrate the manner in which the challenges with 
the epochs of ultra slow roll inflation can be circumvented in two field models.
We shall begin by introducing the inflationary models of our interest before we 
go on to discuss the choice of the non-conformal coupling functions and the 
resulting spectra of electromagnetic fields.


\subsection{Models of interest}\label{subsec:bg}

We shall consider a system of two scalar fields, say, $\phi$ and $\chi$, that
are described by the action~\cite{Lalak:2007vi}
\begin{eqnarray}
S[\phi,\chi] 
&=& \int \d^4x\, \sqrt{-g}\,
\biggl[-\f{1}{2}\,\pa_\mu{\phi}\, \pa^\mu {\phi}\nn\\
& & -\, \f{f(\phi)}{2}\,\pa_\mu\chi\, \pa^\mu {\chi}-V(\phi,\chi)\biggr].
\end{eqnarray}
Clearly, while $\phi$ is a canonical scalar field, $\chi$ is a non-canonical 
scalar field due to the presence of the function $f(\phi)$ in the term describing 
its kinetic energy. 
We shall work with potentials $V(\phi, \chi)$ that are separable.
As a result, the two fields essentially interact through the function $f(\phi)$,
which we shall assume to be of the form $f(\phi)= \mathrm{e}^{2\,b(\phi)}$.

The equations of motion describing the evolution of the scalar fields are 
given by~\cite{Lalak:2007vi}
\begin{subequations}\label{eq:bg-phi-chi}
\begin{eqnarray}
\ddot{\phi}+3\,H\,\dot{\phi}
+ V_{\phi}&=&b_{\phi}\,\mathrm{e}^{2\,b}\,\dot{\chi}^2,\label{eq:bg-phi}\\
\ddot{\chi}+(3\,H+2\,b_{\phi}\,\dot{\phi})\,\dot{\chi}
+\mathrm{e}^{-2\,b}\,V_{\chi}&=&0,\label{eq:bg-chi}
\end{eqnarray}
\end{subequations}
where the subscripts $\phi$ and $\chi$ denote differentiation of the potential 
$V(\phi,\chi)$ and the function $b(\phi)$ with respect to the corresponding fields. 
Also, it is useful to note that the Hubble parameter and its time derivative 
are governed by the following equations:
\begin{eqnarray}
H^2 &=& \f{1}{3\,\Mpl^2}\, \l(\f{\dot{\phi}^2}{2}+\mathrm{e}^{2\,b}\,
\f{\dot{\chi}^2}{2}+V\r),\\
\dot{H}&=&-\f{1}{2\,\Mpl^2}\,\l(\dot{\phi}^2+\mathrm{e}^{2\,b}\,\dot{\chi}^2\r).
\end{eqnarray}
Let us now discuss the specific models that we shall consider.


\subsubsection{Suppression of power on large scales} \label{subsec:ls_sup}

The first of the two models that we shall consider leads to a suppression 
of power on large scales.
In our earlier work, we had discussed the so-called punctuated inflationary 
models which result in a suppression of power over large scales that are 
comparable to the Hubble radius today.
We had also mentioned that such models can mildly improve the fit to the CMB
data (for early discussions in this context, see Refs.~\cite{Contaldi:2003zv,
Sinha:2005mn,Powell:2006yg,Jain:2008dw,Jain:2009pm,Hazra:2014jka,Hazra:2014goa}; 
for a recent discussion, see Ref.~\cite{Ragavendra:2020old}).
We had shown that the punctuated inflationary models leave strong imprints on 
the spectra of the electromagnetic fields.
In particular, we had found that the strengths of the magnetic fields on large
scales are considerably suppressed and their spectra behave as~$k^4$ on small
scales.
Our aim in this section is to investigate whether such challenges can be overcome 
in inflationary models involving two fields.

To achieve a suppression in the spectrum of curvature perturbations on the 
largest scales, we shall consider a simple quadratic potential for the 
field~$\phi$ and a KKLTI-like potential for the field~$\chi$~\cite{Kallosh:2018zsi},
so that the complete potential is given by~\cite{Braglia:2020fms}
\begin{equation}
V(\phi,\chi)=\f{m_\phi^2}{2}\,\phi^2 + V_0\,\f{\chi^2}{\chi_0^2+\chi^2}.\label{eq:sup-ls}
\end{equation}
Moreover, we shall assume that $b(\phi)=\bar{b}\,\phi$, where $\bar{b}$ is a 
constant.
We shall work with the following two sets of values of the parameters 
involved:~$(m_\phi/\Mpl,V_0/\Mpl^4,\chi_0/\Mpl,\bar{b}\,\Mpl)=(1.672 \times10^{-5},
2.6\times10^{-10}, \sqrt{3},1.0)$ and $(1.688\times10^{-5},2.65\times10^{-10}, 
\sqrt{3},2.0)$.
We shall choose the initial values of the fields to be $\phi_\mathrm{i}= 8.8\,\Mpl$, 
$\chi_\mathrm{i} =5.76 \, \Mpl$, and set $\epsilon_{1\mathrm{i}}=2.47 \times 10^{-2}$,
for both these sets of parameters.
In fact, we shall choose a very small value of $\dot{\chi}$ so that $\chi$ does 
not evolve at all during the initial phase.
For these choices of the parameters and initial conditions, there arise two stages 
of inflation with distinct values of the first slow roll parameter $\epsilon_1$.
\begin{figure*}
\includegraphics[width=0.475\linewidth]{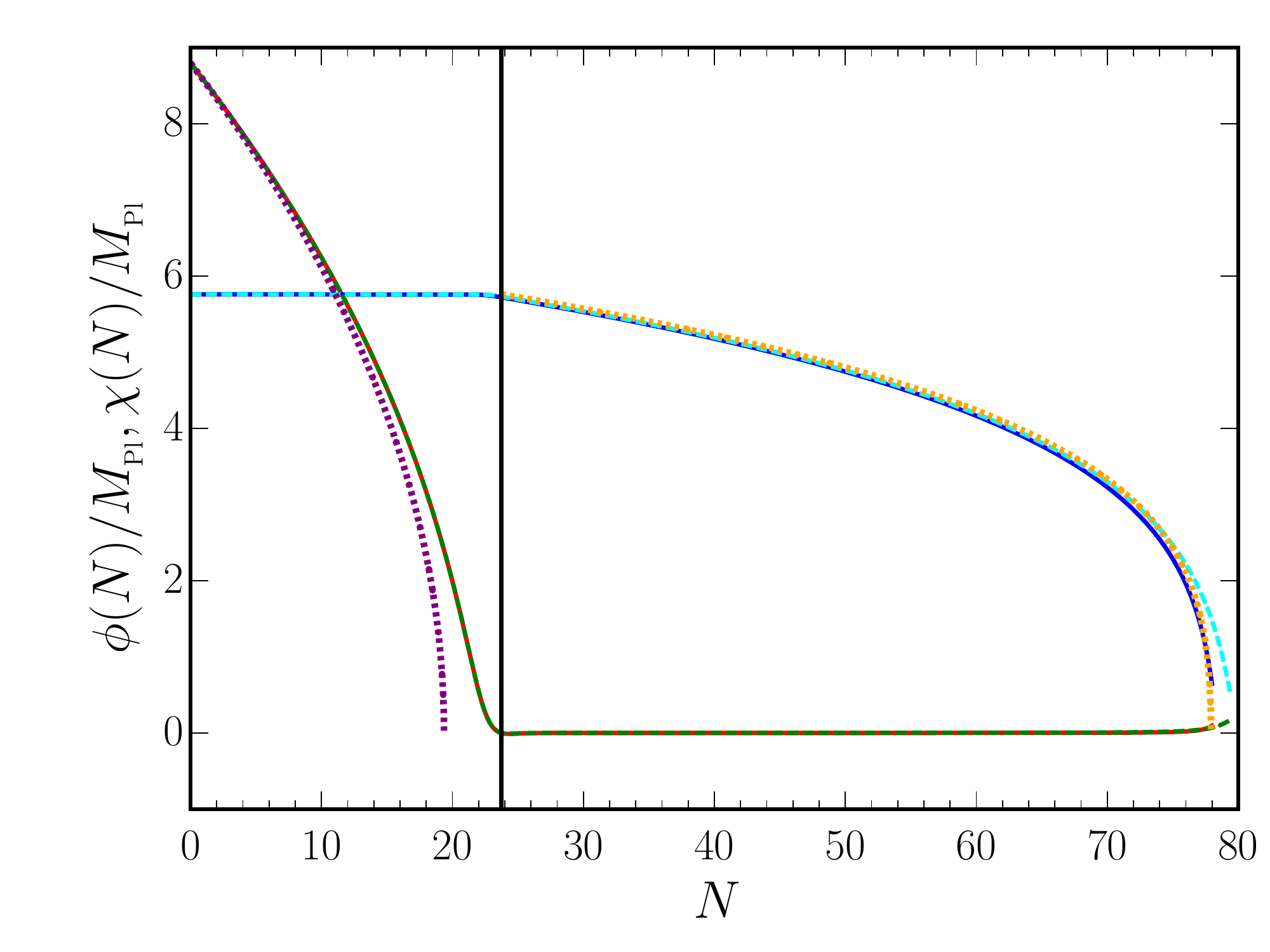}
\includegraphics[width=0.475\linewidth]{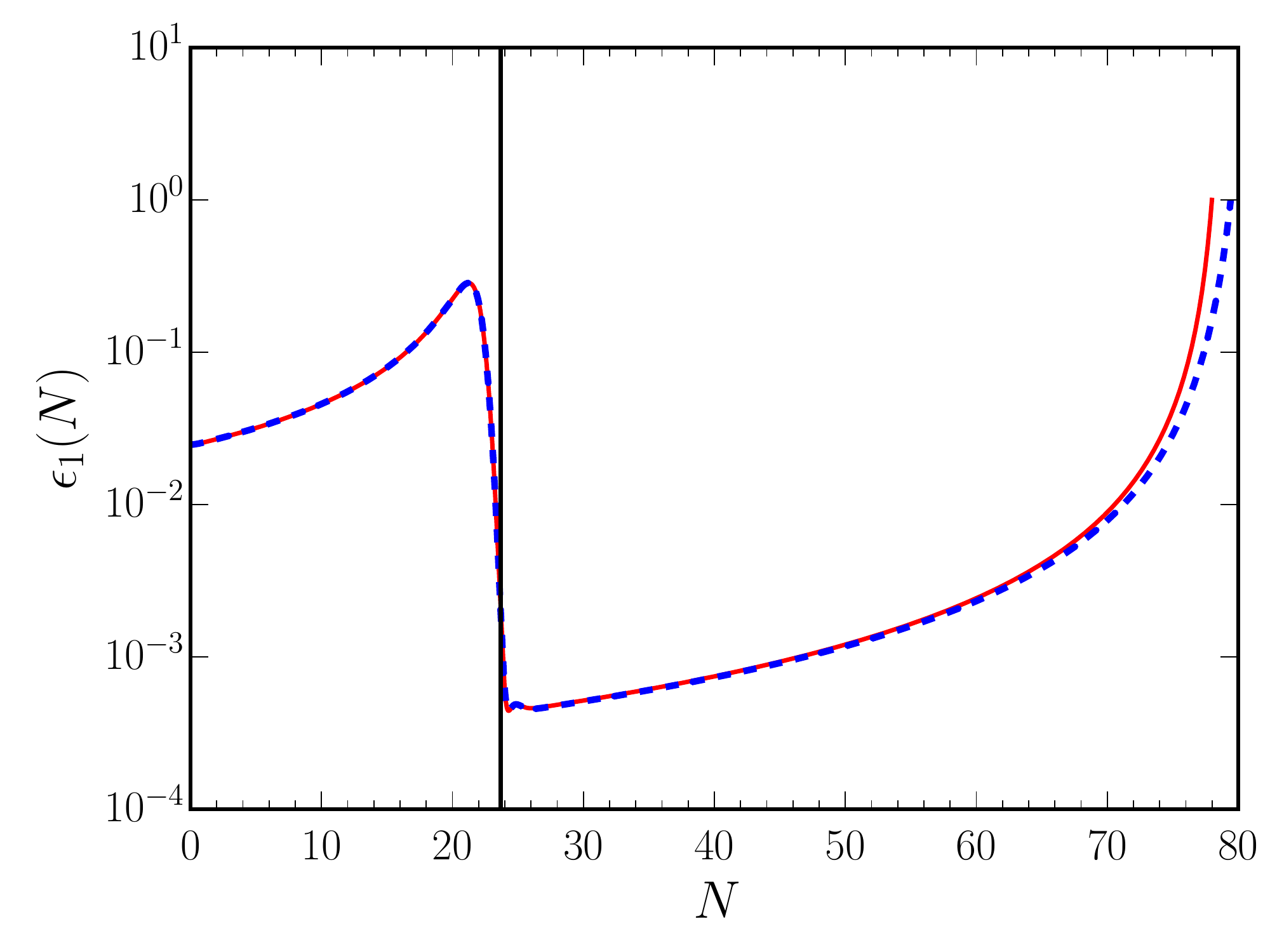}
\includegraphics[width=0.475\linewidth]{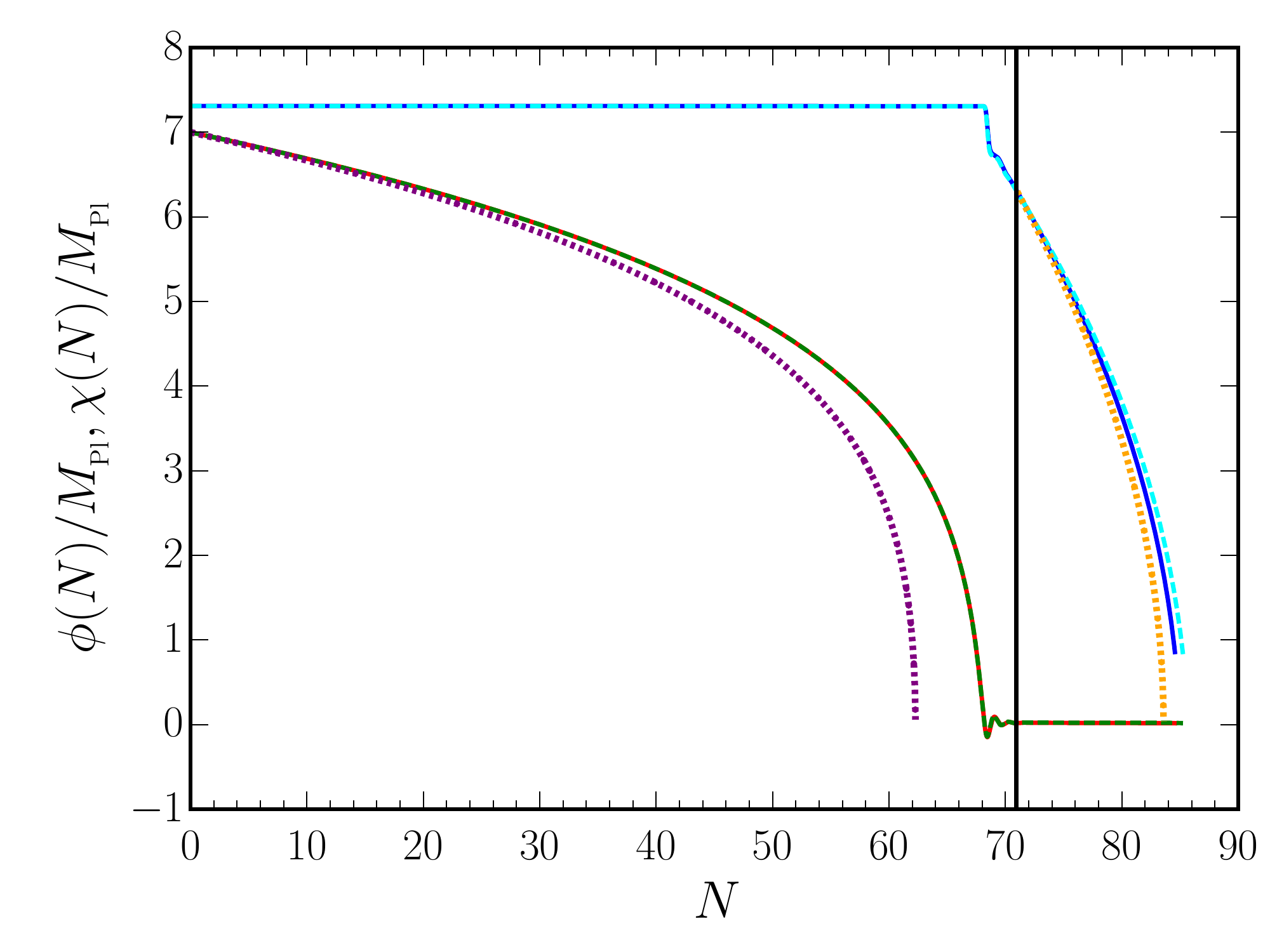}
\includegraphics[width=0.475\linewidth]{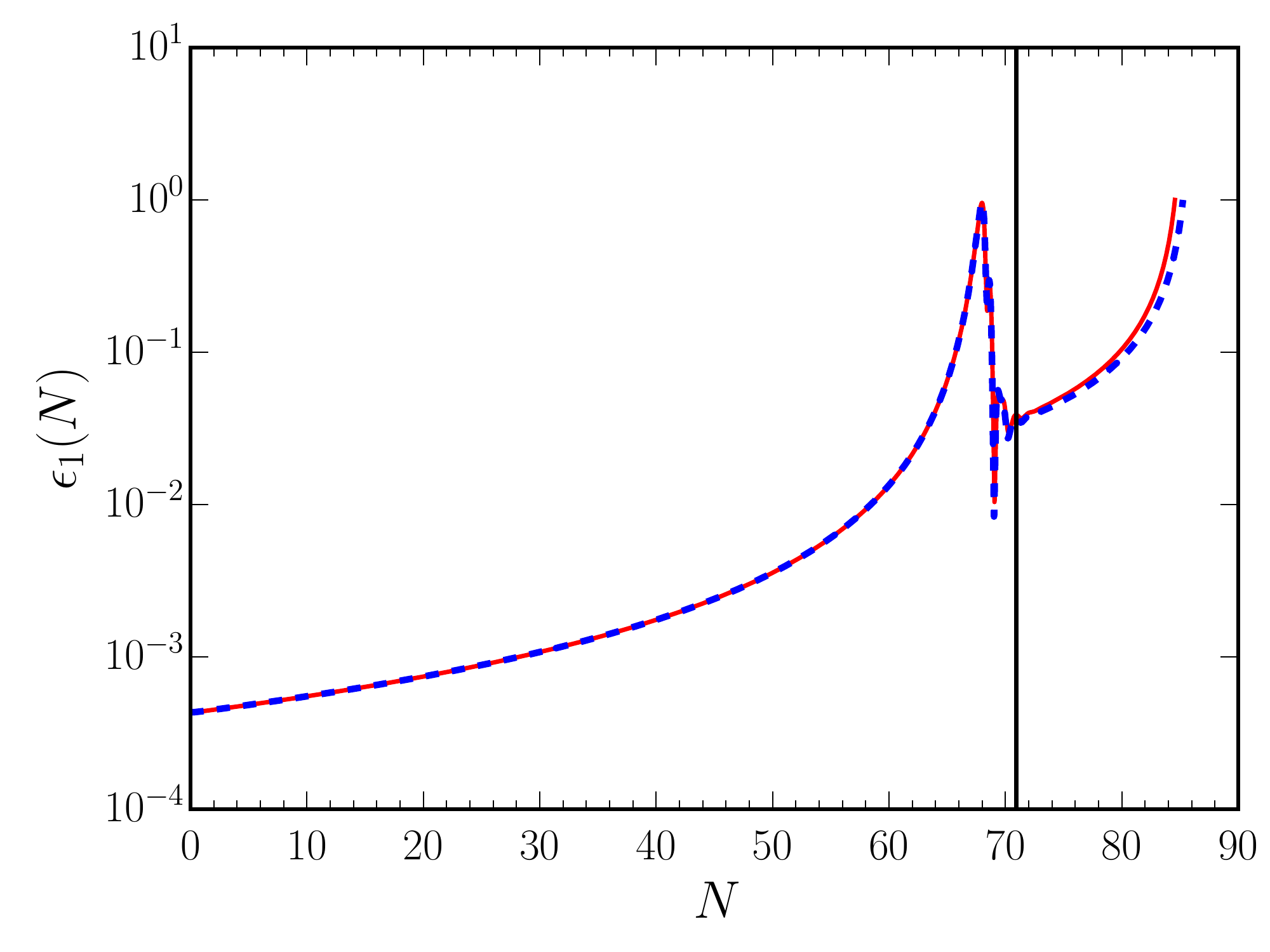}
\caption{The evolution of the two scalar fields, viz. $\phi$ (in solid red and 
dashed green) and $\chi$ (in solid blue and dashed cyan), in the models described 
by the potentials~\eqref{eq:sup-ls} and~\eqref{eq:ss-peak} have been plotted (on
the left, in the top and bottom panels, respectively) as functions of $e$-folds.
We have plotted the results for two sets of values of the parameters involved 
(with the first set in solid lines and the second in dashed lines).
We have also plotted the corresponding evolution of the first slow parameter (in 
solid red and dashed blue, on the right).
Moreover, we have indicated the $e$-folds (as vertical black lines) when the 
transition from the first to the second stage of inflation occurs in the two models 
of our interest, viz. around $N\simeq 23.7$ in the first model (on top) and $N\simeq 71$ 
in the second model (at the bottom), respectively.
Note that, for a given potential, the primary difference between the values of the 
two sets of parameters is the value of~$\bar{b}$.
However, it should be clear from the above plots that the difference in $\bar{b}$ 
does not lead to a significant difference in the evolution of the fields.
In the figure, we have also indicated (as dotted curves) the analytical solutions 
for the fields $\phi$ (in purple) and $\chi$ (in orange) that can be arrived at 
in the slow roll approximation (for details, see App.~\ref{app:as}).
It should be clear that the analytical solutions are a reasonably good approximation 
to the exact numerical results during the two slow roll regimes. 
As one would expect, the analytical solutions fail to capture the dynamics 
around the point of transition from the first to the second stage of
inflation.}\label{fig:ebg}
\end{figure*} 
In Fig.~\ref{fig:ebg}, we have plotted the evolution of the two scalar fields and
the first slow roll parameter in the model for the above sets of parameters and
initial conditions.
It should be clear from the figure that the first stage is driven by the field 
$\phi$ with $\epsilon_1\simeq 10^{-2}$.
The second stage begins when the field $\phi$ has reached the bottom of the 
quadratic potential and the field $\chi$ begins to drive the accelerated 
expansion.
In other words, there arises a turning in field space.  
During the transition, the first slow roll parameter falls exponentially in 
a manner somewhat similar to the single field models that admit an epoch of 
ultra slow roll inflation.
The first slow roll parameter is very small (with $\epsilon_1\simeq 10^{-3}$) 
during the early phase of the second stage and it slowly begins to rise 
leading to the end of inflation.
We find that for the parameters and initial conditions that we have worked with, 
inflation lasts for $78$--$79$~$e$-folds.


\subsubsection{Enhancement of power on small scales}\label{subsec:ss_enh}

The second model we shall consider leads to enhanced power on small scales.
As in the first model, this is achieved through a turning in field space,
which briefly increases the strength of the coupling between the curvature 
and the isocurvature perturbations as well as induces a tachyonic instability.
If the turning occurs at a sufficiently late stage of inflation, these two 
effects combine to lead to an enhancement in the spectrum of curvature
perturbations on smaller scales~\cite{Braglia:2020fms,Palma:2020ejf,
Fumagalli:2020adf,Braglia:2020eai}.

To obtain a peak in the power spectrum at smaller scales, we interchange the 
potentials for the two fields we had considered earlier [see Eq.~\eqref{eq:sup-ls}].
In other words, we consider a model of inflation driven by a KKLTI-like potential 
for~$\phi$ and a simple quadratic potential for $\chi$, so that the complete
potential is given by~\cite{Braglia:2020eai}
\begin{equation}
V(\phi,\chi)=V_0\,\f{\phi^2}{\phi_0^2+\phi^2}
+\f{m_\chi^2}{2}\,\chi^2.\label{eq:ss-peak}
\end{equation}
We shall again assume that $b(\phi)=\bar{b}\,\phi$. 
We shall work with the following two sets of values of the 
parameters:~$(V_0/\Mpl^4,\phi_0/\Mpl,m_\chi/\Mpl,\bar{b}\,\Mpl)=(7.1 \times10^{-10}, 
\sqrt{6}, 1.19164\times10^{-6},7.0)$ and $(7.31 \times10^{-10}, \sqrt{6},
1.209\times10^{-6},7.8)$.
We assume that $\phi_\mathrm{i}=7.0\,\Mpl$, $\chi_\mathrm{i}=7.31\,\Mpl$ and
$\e1i=4.32 \times 10^{-4}$.
Also, as in the earlier model, we shall choose a small value of $\dot{\chi}$
so that $\chi$ hardly evolves during the first phase.
With these choices of the parameters, we obtain about $84$--$85$ $e$-folds 
of inflation.
In Fig.~\ref{fig:ebg}, we have plotted the evolution of the two fields as
well as the behavior of the first slow roll parameter.
Clearly, as in the previous case, there arise two stages of inflation, with
the first stage again driven by the field~$\phi$ and the second stage driven 
by the field~$\chi$.
Moreover, at the transition, the first slow roll parameter $\epsilon_1$ decreases
briefly before increasing to unity leading to the termination of inflation.
Further, we find that, in contrast to the single field case, the first slow roll
parameter does not decrease to considerably low values [say, to 
$\mathcal{O}(10^{-9}$--$10^{-7})$] in order to lead to a significant enhancement
in power.


\subsection{Scalar and tensor power spectra}

Let us now briefly discuss the spectra of curvature and isocurvature perturbations 
that arise in the two models we discussed above.
Let us begin by recalling a few essential points regarding the scalar perturbations
in two field models.
As is well known, in two field models of inflation, the scalar perturbations can 
be decomposed into the so-called adiabatic (say, $\delta\sigma$) and entropy (say, 
$\delta s$) components~\cite{Gordon:2000hv,Bassett:2005xm,Lalak:2007vi}. 
In field space, while the adiabatic perturbations are parallel to the background 
trajectory, the entropy perturbations are orthogonal to it.

If $\delta \phi$ and $\delta \chi$ denote the perturbations in the two scalar 
fields, the adiabatic and entropic perturbations are defined 
as~\cite{Lalak:2007vi,Raveendran:2018yyh}
\begin{subequations}
\begin{eqnarray}
\delta\sigma &=& \cos{\theta}\,\delta\phi + \mathrm{e}^b\,\sin{\theta}\,\delta\chi,\\
\delta s &=& -\sin{\theta}\,\delta\phi + \mathrm{e}^b\,\cos{\theta}\,\delta\chi,
\end{eqnarray}
\end{subequations}
where $\cos\,{\theta}=\dot{\phi}/\dot{\sigma}$, $\sin{\theta}=\dot{\chi}/\dot{\sigma}$, 
and $\dot{\sigma}^2=\dot{\phi}^2+\mathrm{e}^{2\,b}\,\dot{\chi}^2$.
Upon using the background equations~\eqref{eq:bg-phi-chi}, one can arrive at the following
equations that govern the adiabatic field~$\sigma$ and the angle~$\theta$:
\begin{subequations}
\begin{eqnarray}
\ddot{\sigma}+3\,H\,\dot{\phi}+V_{\sigma} &=&0, \\
\dot{\theta}&=& -\f{V_s}{\dot{\sigma}}-b_\phi\, \dot{\sigma}\,\sin{\theta},
\end{eqnarray}
\end{subequations}
where the quantities $V_\sigma$ and $V_s$ are given by
\begin{subequations}
\begin{eqnarray}
V_\sigma &=&  \cos{\theta}\,V_\phi+\mathrm{e}^{-b}\,\sin{\theta}\,V_\chi,\\
V_s &=& -\sin{\theta}\,V_\phi+\mathrm{e}^{-b}\,\cos{\theta}\,V_\chi.
\end{eqnarray}
\end{subequations}

In the spatially flat gauge, the Mukhanov-Sasaki variables associated with 
the curvature and isocurvature perturbations are given by $v^{\sigma}=a\,
\delta \sigma$ and $v^{s}=a\,\delta s$. 
The equations of motion describing the evolution of the Mukhanov-Sasaki variables 
can be obtained to be~\cite{Lalak:2007vi,Raveendran:2018yyh,Braglia:2020eai,
Braglia:2020fms}
\begin{subequations}
\begin{eqnarray}
{v_k^{\sigma}}{''} +\l(k^2-\f{z''}{z}\r)\,v_k^{\sigma} 
&=&\f{1}{z}\,(z\, \xi\, v_k^s)',\\
{v_k^{s}}{''}+\l(k^2-\f{a''}{a}+\mu_s^2\,a^2\r)\,v_k^{s} 
&=&-z\,\xi\, \l(\f{v_k^{\sigma}}{z}\r)',\qquad
\end{eqnarray}
\end{subequations}
where $z=a\,\dot{\sigma}/H$, $\xi=-2\,a\,V_s/\dot{\sigma}$, while $\mu_s^2$ is 
given by
\begin{eqnarray}
\mu_s^2 &=& V_{ss}-\l(\f{V_s}{\dot{\sigma}}\r)^2
+b_\phi\,\l(1+\sin^2{\theta}\r)\,\cos\,{\theta}\,V_\sigma\nn\\
& &+\,b_\phi\,\cos^2{\theta}\,\sin{\theta}\,V_s
-\l(b_\phi^2+b_{\phi\phi}\r)\,\dot\sigma^2,\label{eq:mus2}
\end{eqnarray}
with $V_{ss}$ being defined as 
\begin{equation}
V_{ss} =\sin^2{\theta}\, V_{\phi\phi}
-\mathrm{e}^{-b}\,\sin{2\,\theta}\, V_{\phi\chi}
+ \mathrm{e}^{-2\,b}\,\cos^2{\theta}\, V_{\chi\chi}.\quad
\end{equation}

As is done in the case of single field models, the Bunch-Davies initial conditions 
are imposed on the Fourier modes when they are sufficiently inside the Hubble radius,
and the scalar and tensor power spectra are evaluated when the modes are well outside 
the Hubble radius.
While computing the scalar power spectra numerically, we impose the initial conditions
when $k \simeq 10^2\, \sqrt{z''/z}$ and evaluate the spectra  at the end of inflation.
To ensure that there are no correlations between the curvature and the isocurvature 
perturbations at early times, when the modes are inside the Hubble radius, 
the scalar perturbations are evolved from two sets of initial 
conditions~\cite{Gordon:2000hv,Lalak:2007vi}. 
In the first set, the standard Bunch-Davies initial conditions are imposed on the 
Mukhanov-Sasaki variable~$v_k^\sigma$, while the variable $v_k^s$ is set to be zero. 
In the second set, the initial conditions on $v_k^\sigma$ and $v_k^s$ are interchanged. 
The curvature perturbation $\mathcal{R}_k$ and the isocurvature perturbation 
$\mathcal{S}_k$ are related to the Mukhanov-Sasaki variables as
follows:~$\mathcal{R}_k=v_k^\sigma/z$ and
$\mathcal{S}_k=v_k^s/z$~\cite{Braglia:2020eai,Braglia:2020fms}.
Let $(\mathcal{R}_{k1},\mathcal{S}_{k1})$ and $(\mathcal{R}_{k2},\mathcal{S}_{k2})$ 
denote the curvature and the isocurvature perturbations evolved from the two sets 
of initial conditions mentioned above.
The spectra of curvature and isocurvature perturbations are evaluated from both 
these sets of solutions and are given by~\cite{Lalak:2007vi,Raveendran:2018yyh,
Braglia:2020eai,Braglia:2020fms}
\begin{subequations}
\begin{eqnarray}
\mathcal{P}_{\mathcal{R}}(k) 
&=& \f{k^3}{2\, \pi^2}\,\l(\vert \mathcal{R}_{k1}\vert^2 
+\vert \mathcal{R}_{k2}\vert^2\r),\\
\mathcal{P}_{\mathcal{S}}(k)  
&=& \f{k^3}{2\, \pi^2}\,\l(\vert \mathcal{S}_{k1}\vert^2 
+\vert \mathcal{S}_{k2}\vert^2\r).\label{eq:ps-d}
\end{eqnarray}
\end{subequations}
In our discussion below, we shall focus on the spectrum of curvature 
perturbations~$\mathcal{P}_{\mathcal{R}}(k)$. 
Also, we should mention that the tensor power spectrum is evaluated in
the same manner as in single field inflation.

\begin{figure*}
\centering
\includegraphics[width=0.475\linewidth]{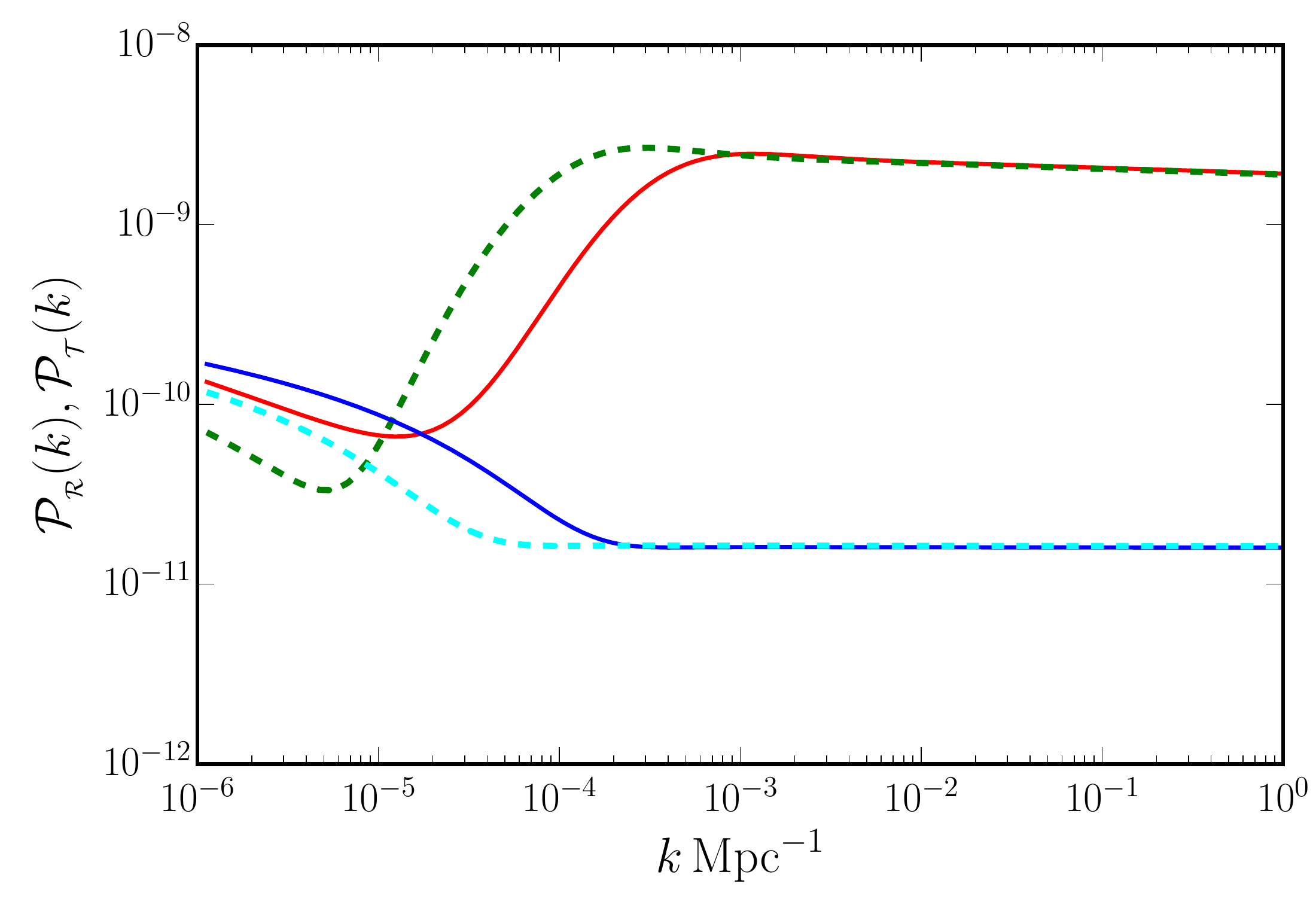}
\includegraphics[width=0.475\linewidth]{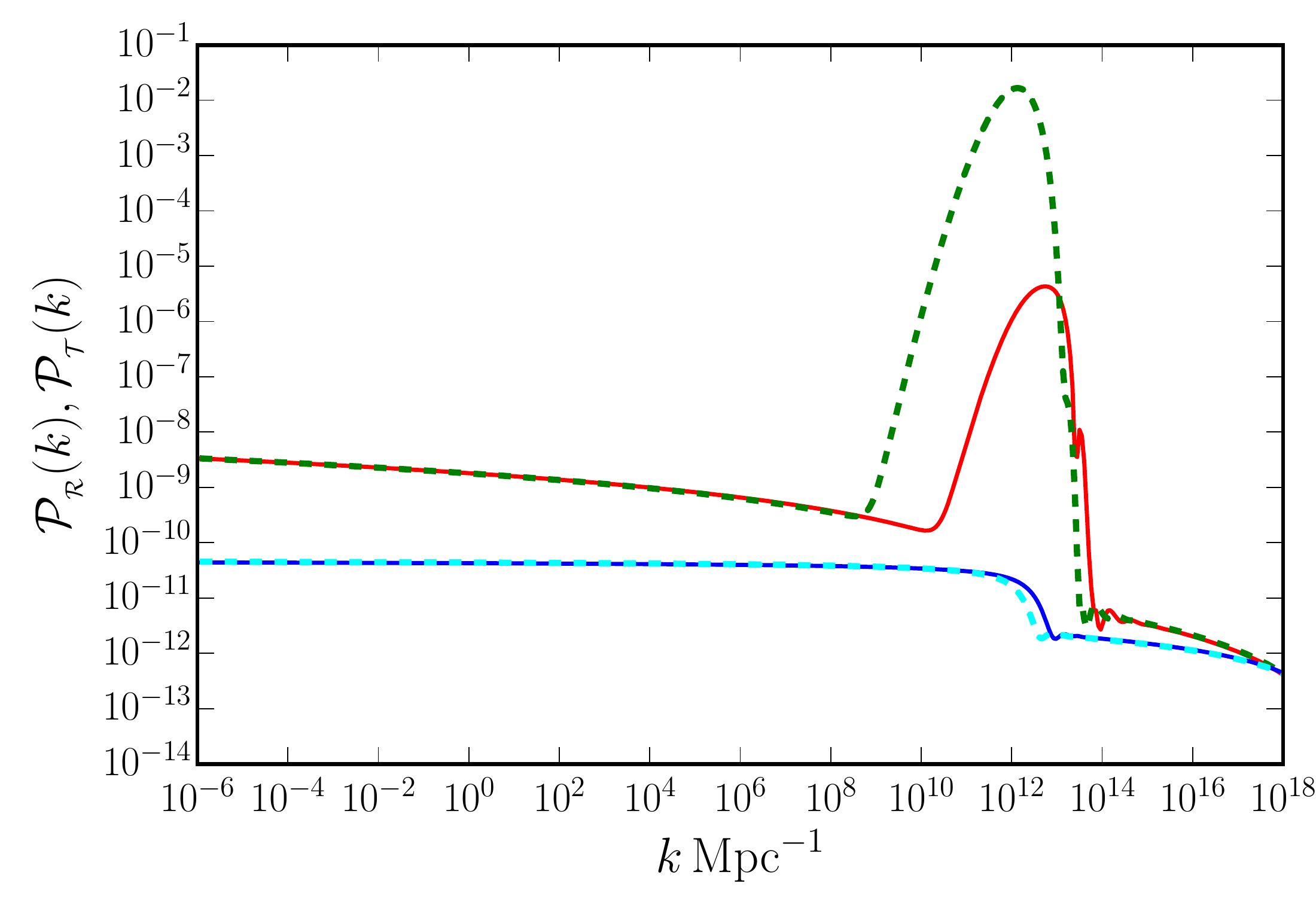}
\caption{The spectra of curvature (in solid red and dashed green) and tensor 
(in solid blue and dashed cyan) perturbations, viz.~$\mathcal{P}_{\mathcal{R}}(k)$ 
and~$\mathcal{P}_{\mathcal{T}}(k)$, have been plotted for the two 
field inflationary models that we have considered.
We have plotted the spectra with features over the CMB scales (on the left) 
arising in the potential~\eqref{eq:sup-ls} and with a peak in the scalar power 
at small scales (on the right) occurring in the potential~\eqref{eq:ss-peak},
for the two sets of parameters (as solid and dashed lines) we have mentioned 
earlier.
In arriving at these spectra, we have assumed that the pivot scale $k_\ast
=0.05\,{\rm Mpc}^{-1}$ leaves the Hubble radius $50$ $e$-folds before the end 
of inflation.
It is the scalar spectra with a sharp rise in power on small scales that are
often considered to produce significant number of PBHs.
Recall that, for a given potential, the two sets of parameters primarily 
differed in the value of $\bar{b}$.
As we had seen in the previous figure, there was hardly any difference in the 
evolution of the background for the two sets of parameters.
However, note that the spectra of curvature perturbations differ significantly
for these two sets.
This can be attributed to the tachyonic instability that arises for non-zero
values of~$\bar{b}$.
It is found that, even small differences in $\bar{b}$ can significantly alter 
the evolution of the curvature perturbations, leading to very different scalar 
power spectra.}
\label{fig:stps}
\end{figure*}
The evolution of the scalar fields in the two models of our interest can be 
obtained by solving the background equations~\eqref{eq:bg-phi-chi} numerically 
as discussed in the previous subsection.
Recall that, we had illustrated the behavior of the scalar fields and the 
first slow roll parameter as functions of $e$-folds in Fig.~\ref{fig:ebg}.
In the case of the potential~\eqref{eq:sup-ls}, the field $\phi$ slowly rolls 
down the potential until it reaches the bottom of the potential when $N\simeq 
23.7$, while the field $\chi$ remains frozen during this period. 
At this point of transition, for the set of parameters we have worked with, the
values of the fields $\phi$ and $\chi$ are $\phi_1=6.55\times10^{-4} \Mpl$ and
$\chi_1=5.722\,\Mpl$, respectively.
After the transition, while $\phi$ oscillates about the minimum of the potential, 
the field~$\chi$ drives inflation until the end.
Also, the first slow roll parameter $\epsilon_1$ decreases exponentially soon 
after the transition, giving rise to a brief period of ultra slow roll, before 
it eventually rises to unity leading to the end of inflation. 
A similar behavior of the fields and the slow roll parameter are observed in 
the case of the potential~\eqref{eq:ss-peak} as well, with the
transition point occurring at a much later time, viz. at the $e$-fold $N\simeq 71$. 
In this case, we choose the point of transition to be when the oscillations of 
the field $\phi$ have substantially died down.
At the transition point, the values of the fields $\phi$ and $\chi$ are found
to be $\phi_1=1.694 \times 10^{-2}\, \Mpl$ and $\chi_1=6.3\, \Mpl$, respectively.
In Fig.~\ref{fig:stps}, we have presented the spectra of the curvature and tensor 
perturbations arising in the two models for the two sets of parameters we have
considered.
In the case of the potential~\eqref{eq:sup-ls}, we obtain a suppression in power
in the spectrum of curvature perturbations on the largest observable scales, while 
over the CMB and smaller scales, the scalar power spectrum is nearly scale invariant. 
The imprints of these scalar and tensor power spectra on the anisotropies in the CMB 
have been discussed earlier (in this context, see Ref.~\cite{Braglia:2020fms}).
For the potential~\eqref{eq:ss-peak}, we obtain nearly scale invariant scalar 
and tensor power spectra over the CMB scales, whereas there is a significant 
enhancement in scalar power on small scales. 
We find that, at the pivot scale of $k_\ast=0.05\, \mathrm{Mpc}^{-1}$, the scalar 
spectral index and the tensor-to-scalar ratio turn out to be $n_{_{\mathrm{S}}}=0.96$ 
and $r=0.02$, which are consistent with the constraints from the CMB data~\cite{Planck:2018jri}. 
As has been illustrated earlier in the literature, the turning in field space briefly
increases the strength of the coupling between the isocurvature and the curvature 
perturbations.
It also induces a tachyonic instability.
These two effects combine to lead to the increased scalar power on smaller scales over 
modes which leave the Hubble radius just prior to or during the turning in field
space~\cite{Braglia:2020fms,Palma:2020ejf,Fumagalli:2020adf,Braglia:2020eai}.
Earlier, we had seen that for a given potential, despite the difference in the 
values of the parameter $\bar{b}$, the evolution of background scalar fields were 
very similar (see Fig.~\ref{fig:ebg}).
However, as should be clear from Fig.~\ref{fig:stps}, the resulting inflationary 
scalar power spectra are considerably different.
This can be attributed to the difference in $\mu_s^2$ [cf.~Eq.~\eqref{eq:mus2}]
that arises due to the difference in the values of $\bar{b}$ and the resulting
amplitude of the tachyonic instability that occurs due to the turning in field
space (for a detailed discussion in this context, see Ref.~\cite{Braglia:2020fms}).


\subsection{Construction of the non-conformal coupling function}

Recall that, our main reason for considering two field models of inflation in 
this paper is to circumvent the challenges that we face in single field models,
especially those that permit an epoch of ultra slow roll inflation.
Our goal is to overcome these hurdles and arrive at electromagnetic spectra of 
desired shapes and strengths.
In our earlier work~\cite{Tripathy:2021sfb}, we had shown that, in slow roll inflation
driven by a single field, say, $\phi$, it is possible to construct analytical forms 
for the non-conformal coupling function $J(\phi)$ that lead to the required time 
dependence (viz. $J \propto  a^2$) and therefore generate nearly scale invariant 
spectra for the magnetic field.
However, when strong departures from slow roll occur, we had to turn to a numerical 
approach to construct the coupling function.
Since the dynamics in the two field models of our interest is fairly non-trivial, we 
shall adopt the numerical approach here as well.
We shall now outline the procedure to construct the required~$J(\phi,\chi)$.
We should mention that, in App.~\ref{app:as}, we have discussed the construction
of an analytical form for the coupling function $J(\phi,\chi)$ and the resulting 
power spectra of the electromagnetic fields that we obtain in such a case.

In fact, the procedure that needs to be adopted to construct the desired 
non-conformal coupling function $J(\phi,\chi)$ is fairly straightforward.
In the two models of our interest, we have seen that, at any given time during 
inflation, one of the two slowly rolling fields largely determines the dynamics 
of the background. 
Essentially, we need to make use of the dominant field to construct the coupling
function in a given domain.
Thereafter, we can utilize the step function to stitch together the coupling 
functions in the two domains to arrive at the complete function.
Let us first consider the model described by the potential~\eqref{eq:sup-ls}.
In the model, during the first domain, the field~$\phi$ rolls down the potential 
largely determining the background dynamics, while the field~$\chi$ remains frozen. 
After having solved the equations~\eqref{eq:bg-phi-chi} numerically to arrive 
at~$\phi(N)$, we assume an ansatz for the functional form of $N(\phi)$.
We choose the functional form of $N(\phi)$ in the first regime to be given by
the following fourth order polynomial:
\begin{equation}
N(\phi) = a_1\, \f{\phi^4}{\Mpl^4} + b_1\,\f{\phi^3}{\Mpl^3}
+c_1\,\f{\phi^2}{\Mpl^2} + d_1\,\f{\phi}{\Mpl}+e_1.\quad\label{eq:N-phi}
\end{equation}
We then determine the values of the constants $(a_1,b_1,c_1,d_1,e_1)$ by fitting the
polynomial to the numerical solution for $\phi(N)$ during the initial regime, thereby
arriving at $N(\phi)$.
Similarly, after the transition, as the field $\chi$ starts to dominate the background
dynamics, we choose $N(\chi)$ to be a fourth order polynomial of the following form:
\begin{equation}
N(\chi) =a_2\,\f{\chi^4}{\Mpl^4} +b_2\,\f{\chi^3}{\Mpl^3}
+c_2\,\f{\chi^2}{\Mpl^2}+ d_2\,\f{\chi}{\Mpl}+e_2.\quad\label{eq:N-chi}
\end{equation}
We fit the polynomial to the solution $\chi(N)$ in the second regime to determine the 
constants $(a_2,b_2,c_2,d_2,e_2)$ and arrive at $N(\chi)$.
We should clarify here that we have chosen to work with fourth order polynomials 
for $N(\phi)$ and $N(\chi)$ above since they seem sufficient to lead to the desired 
behavior of~$J$ in the two slow roll regimes on either side of the transition. 
With the forms of $N(\phi)$ and $N(\chi)$ at hand, we can combine them to construct the 
complete non-conformal coupling function to be 
\begin{eqnarray}
J(\phi,\chi) 
&=& \f{J_0}{2}\, \biggl\{\l[1+\mathrm{tanh}\l(\f{\chi -\chi_1}{\Delta\chi}\r)\r]
\mathrm{exp}\l[n\,N(\phi)\r]\nn\\
& &+\l[1-\mathrm{tanh}\l(\f{\chi -\chi_1}{\Delta\chi}\r)\r]\,
\mathrm{exp}\l[n\,N(\chi)\r]\biggr\},\qquad\label{eq:J-phi-chi}
\end{eqnarray}
where $\chi_1$ is the value of the field $\chi$ at the transition (that we had 
mentioned earlier).
Note that the quantity within the square brackets involving the hyperbolic 
tangent function in the above form for $J(\phi,\chi)$ essentially acts as a 
step function for a suitably small value of $\Delta \chi$.
Evidently, it is the first and second terms in the above expression that 
contribute prior to and after the field crossing $\chi_1$. 
Lastly, we should mention that we need to choose $J_0$ suitably so that 
$J(\phi,\chi)$ reduces to unity at the end of inflation.
As we shall soon illustrate, for $n=2$, the above coupling function largely 
behaves as $J\propto a^2$. 

\begin{figure*}
\centering
\includegraphics[width=0.475\linewidth]{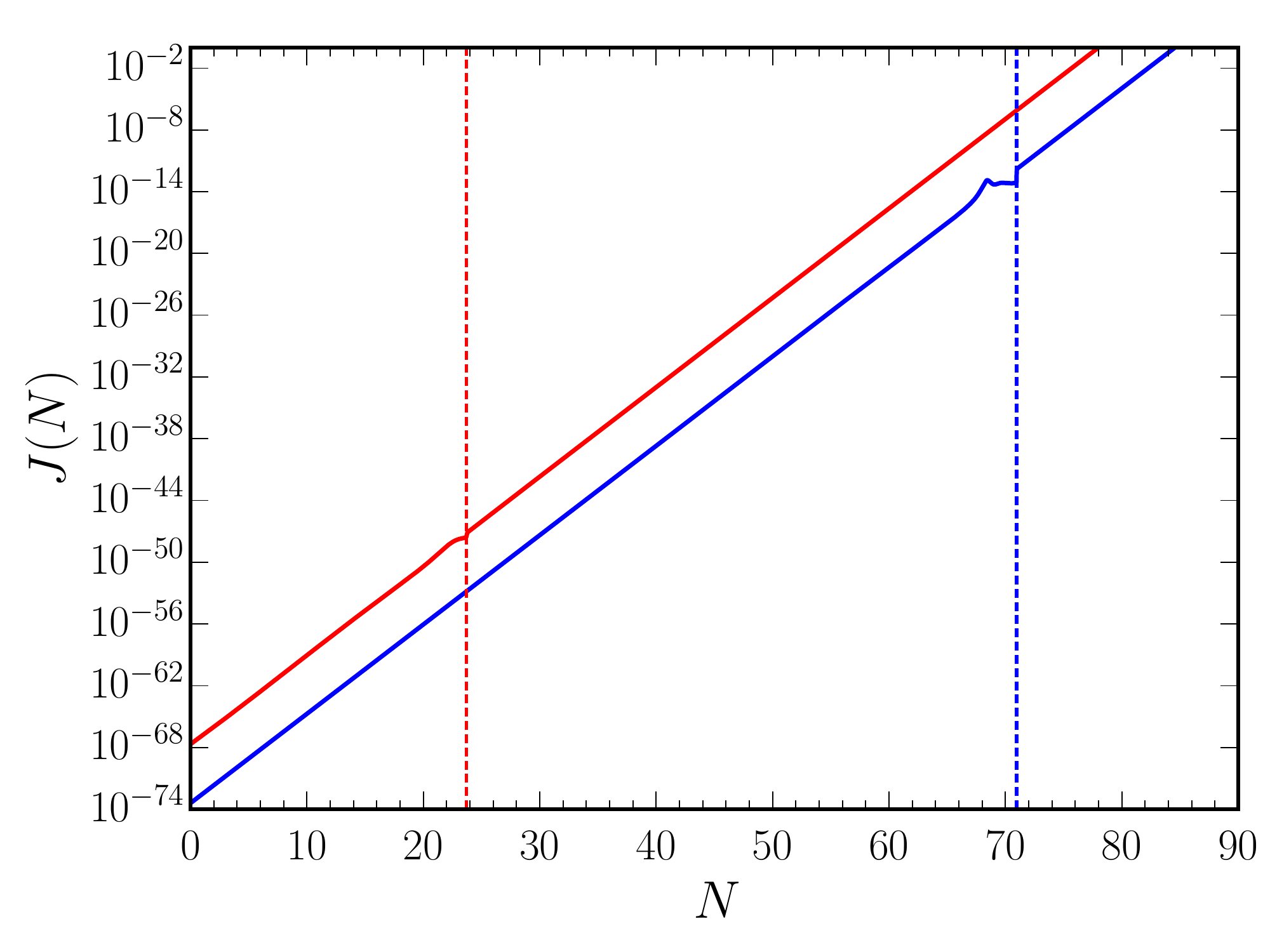}
\includegraphics[width=0.475\linewidth]{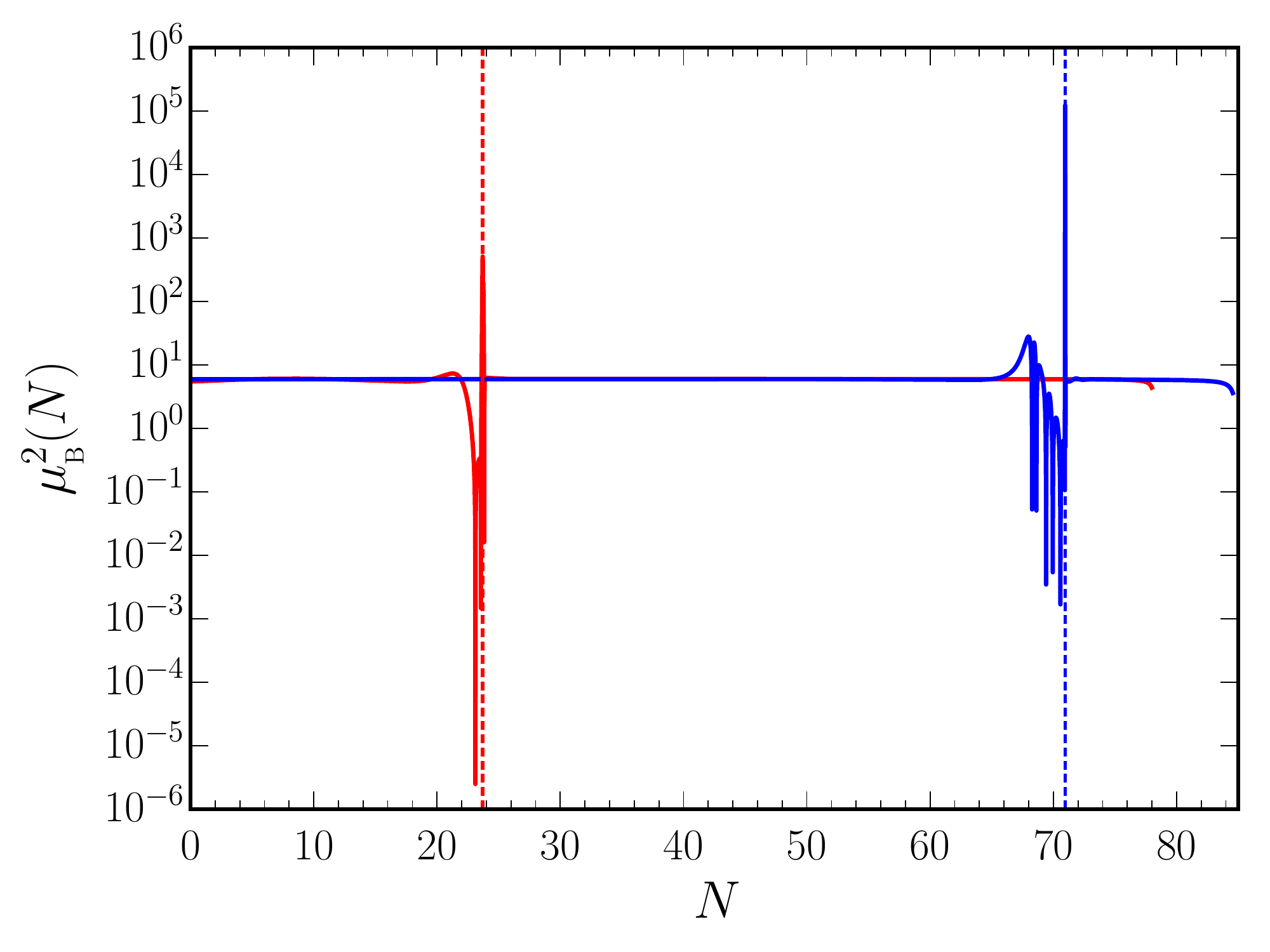}
\caption{The evolution of the non-conformal coupling function $J(\phi,\chi)$ given
by Eq.~\eqref{eq:J-phi-chi} (on the left) and the quantity $\mub^2$ (on the right) 
have been plotted as a function of $e$-folds for the two models described by the 
potentials~\eqref{eq:sup-ls} (in red) and~\eqref{eq:ss-peak} (in blue). 
Since the background evolution is very similar for the two sets of parameters we
have worked with, the corresponding $J(N)$ prove to be essentially the same in
both the models.
The vertical lines (in corresponding colors) represent the points of transition at 
$N\simeq 23.7$ and $N\simeq 71$, respectively. 
It is clear from the figures that $J\propto a^2$ and $\mub^2\simeq 6$ for most of 
the evolution except for the domain near the transition.
Recall that, in the single field case, it was impossible to achieve such a behavior 
for $J$ and $\mub^2$ after the onset of ultra slow roll.
Clearly, the presence of the additional field in the two field models allows us 
to circumvent this difficulty.}\label{fig:J-mub2}
\end{figure*}
Recall that, in each of the two models described by the potentials~\eqref{eq:sup-ls} 
and~\eqref{eq:ss-peak}, we had worked with two sets of the parameters involved.
Since the dynamics of the background fields for these two sets of parameters are not
significantly different (in this context, see Fig.~\ref{fig:ebg}), we find that the
coefficients characterizing the polynomial fitting functions $N(\phi)$ and $N(\chi)$ 
[cf. Eqs.~\eqref{eq:N-phi} and~\eqref{eq:N-chi}] largely prove to be the same.
For the model described by the potential~\eqref{eq:sup-ls}, we obtain the values 
of the fitting parameters to be $(a_1,b_1,c_1,d_1,e_1)=(2.7558\times 10^{-3}, -0.06, 
0.227, -1.8556, 23.1717)$ and $(a_2,b_2,c_2,d_2,e_2)=(-0.0421, 5.47\times 10^{-3}, 
-0.2443, -0.4516, 78.3998)$.
Similarly, in the case of the potential~\eqref{eq:ss-peak}, we find that the 
fitting parameters are given by $(a_1,b_1,c_1,d_1,e_1) =(-0.01619, -0.1027, 
0.3857, -2.0208,69.4198)$ and $(a_2,b_2,c_2,d_2,e_2) =(2.7745 \times 10^{-4},
-6.0702 \times 10^{-3}, -0.2810, -0.2805, 85.0065)$.
Moreover, we shall assume that the width of the step described by the hyperbolic
tangent function is given by $\Delta\chi=10^{-3}\,\Mpl$.
In App.~\ref{app:e-of-p}, we have discussed the effects of modifying the 
transition point~$\chi_1$ and the width~$\Delta\chi$ on the spectra of
the electromagnetic fields.
We find that the values of~$\chi_1$ and~$\Delta\chi$ we shall work with 
are optimal, as they do not introduce spurious features in the spectra 
of the magnetic field.

\begin{figure*}
\includegraphics[width=0.475\linewidth]{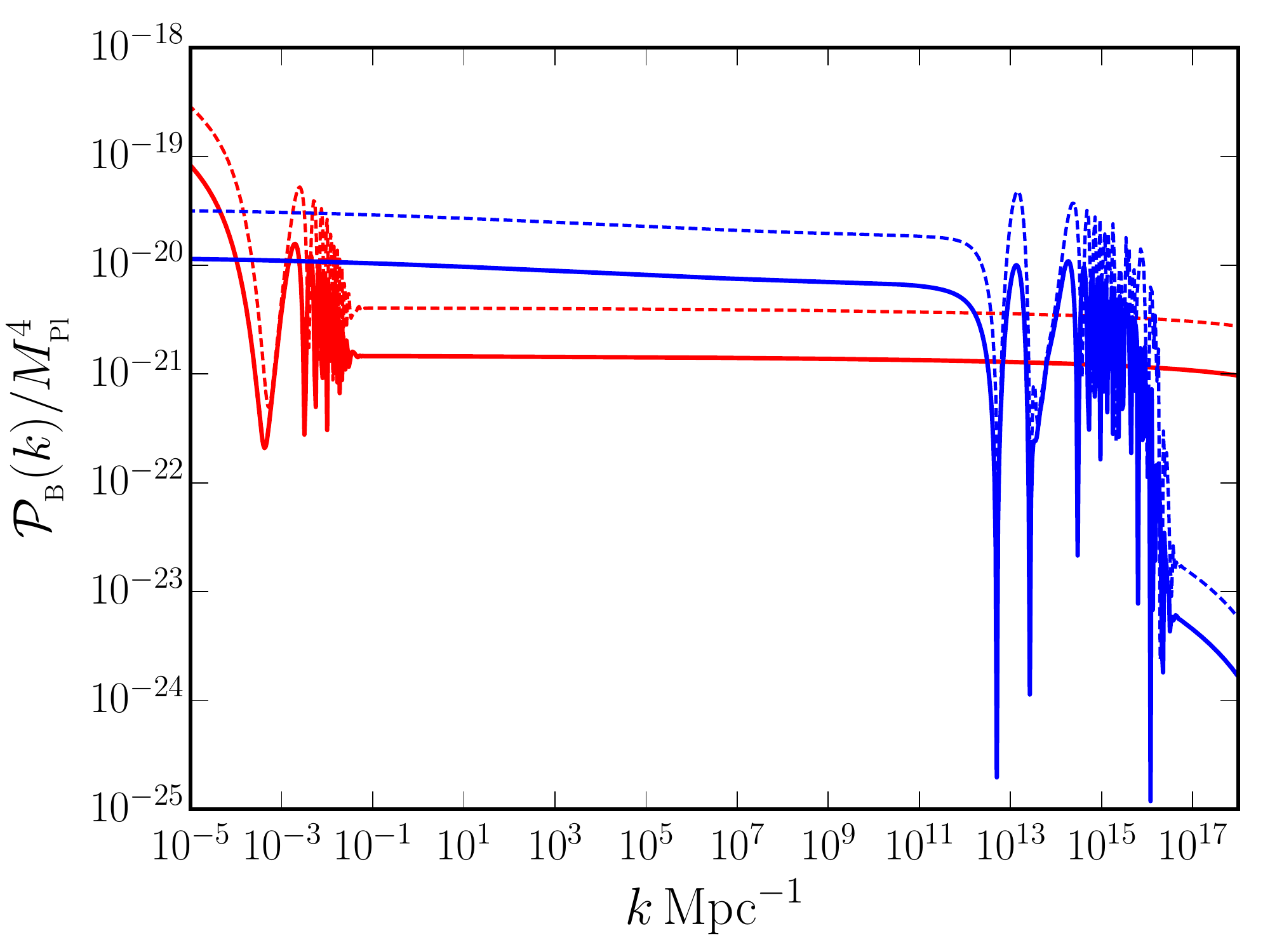}
\includegraphics[width=0.475\linewidth]{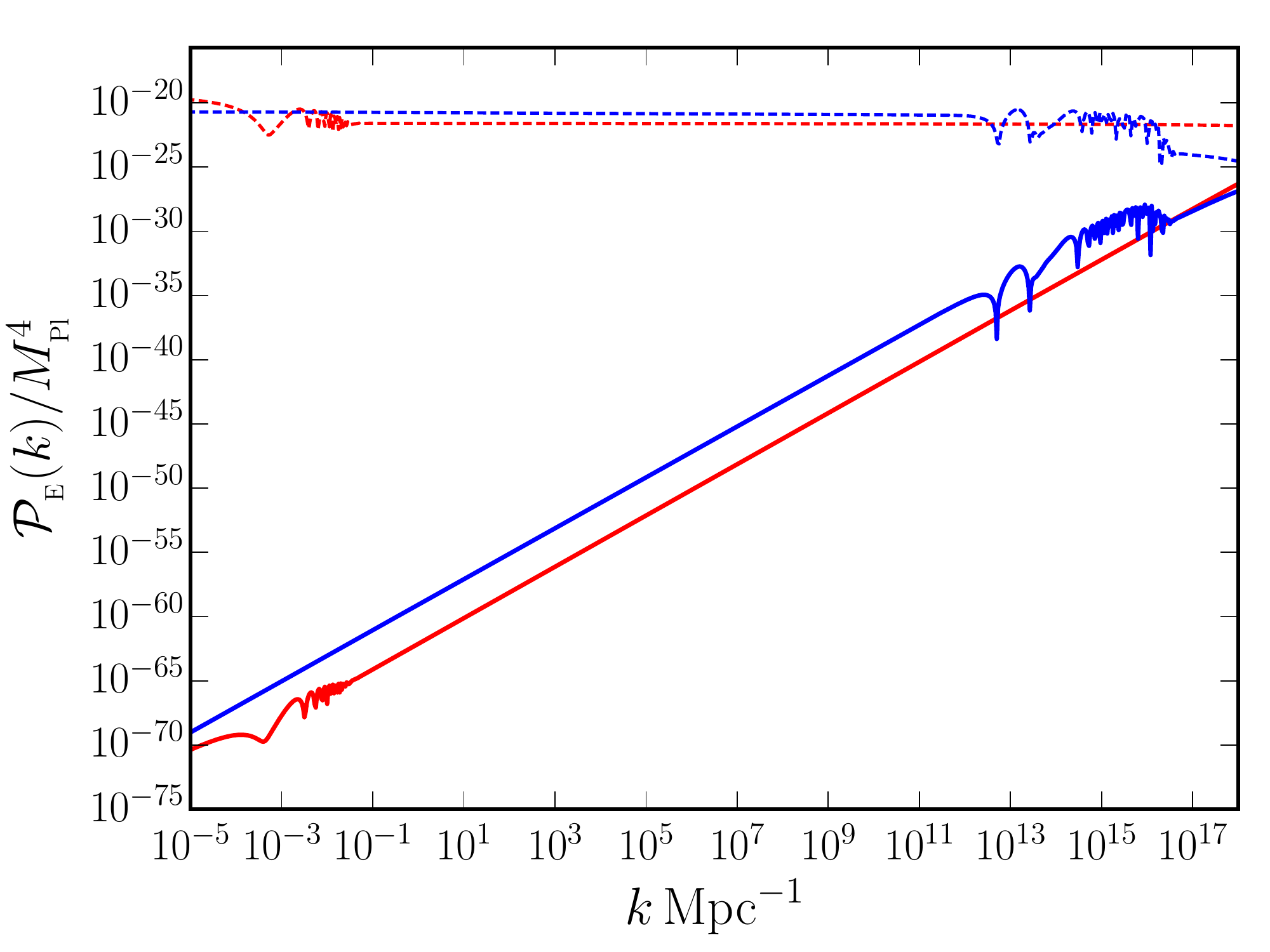} 
\caption{The power spectra of the magnetic (on the left) and the electric (on the 
right) fields have been plotted in the cases of the models described by the 
potentials~\eqref{eq:sup-ls} (in red) and~\eqref{eq:ss-peak} (in blue) for
the coupling function $J(\phi,\chi)$ given by Eq.~\eqref{eq:J-phi-chi}.
Apart from the non-helical case (plotted as solid curves), we have also plotted
the results for the helical electromagnetic fields (as dashed curves) with
$\gamma=0.25$.
The spectra of both the magnetic and electric fields are nearly scale invariant
in the helical case.
Also, in the non-helical case, while the spectra of the magnetic field are nearly
scale invariant, the spectra of the electric field behave as~$k^2$.
Moreover, as expected, all the spectra exhibit bursts of oscillations over wave 
numbers which leave the Hubble radius around the time of the turning in the field 
space.
This is because of the fact that the coupling function $J(\phi,\chi)$ contains 
deviations from the behavior $J\propto a^2$ during the time of the transition.
We should mention that we have worked with $\gamma=0.25$ so that the amplitudes
of the present day magnetic field generated in the two models of our interest
are approximately consistent with the current constraints (see our discussion
in Sec.~\ref{sec:i-on-cmb}).
Note that, since $f(0.25)\simeq 3$, the scale invariant non-helical and helical
amplitudes differ by a factor of three.}\label{fig:pb-pe}
\end{figure*}

In Fig.~\ref{fig:J-mub2}, we have plotted the evolution of the coupling function
$J(\phi,\chi)$ as well as the quantity $\mub^2=J''/(J\,a^2\,H^2)$ for the two 
potentials.
In contrast to the single field case, where the coupling function almost ceased 
to evolve after the onset of ultra slow roll (see Fig.~\ref{fig:J-usr}), we find 
that the $J$'s we have constructed in the two field models grow as $a^2$ even 
after the transition. 
Moreover, clearly, $\mub^2\simeq 6$ for most of the evolution apart from the 
domain around the transition.
This behavior suggests that the spectrum of the magnetic field will remain scale
invariant apart from the effects arising due to the transition.
In Fig.~\ref{fig:pb-pe}, we have plotted the resulting spectra of the electromagnetic
fields arising in the two models for the above choices of the coupling functions.
In addition to the non-helical case, in Fig.~\ref{fig:pb-pe}, we have plotted 
the spectra in the helical case.
It is evident that, in the helical case, the spectra of the magnetic and electric 
fields are nearly scale invariant and are of the same amplitude apart from the domain
over wave numbers which leave the Hubble radius around the time of the turning in 
the field space.
Around these wave numbers, the spectra exhibit a burst of oscillations. 
These oscillations occur over large scales in the first model described by the
potential~\eqref{eq:sup-ls}, whereas they occur over small scales in the second 
model governed by the potential~\eqref{eq:ss-peak}.
While we have been able to largely iron out very strong features in the 
power spectra of the electromagnetic fields, the oscillations are unavoidable
unless we further fine tune the form of the non-minimal coupling function
$J(\phi,\chi)$.
We shall make some additional comments on this point in the concluding section.


\section{Imprints on the CMB}\label{sec:i-on-cmb}

In this section, we shall  examine the observational imprints of the PMFs on 
the anisotropies in the CMB, which have been extensively discussed in the 
literature~\cite{Seshadri:2000ky,Mack:2001gc,Lewis:2004ef,Finelli:2008xh, 
Paoletti:2008ck,Shaw:2009nf,Adamek:2011pr,Bonvin:2011dt, Bonvin:2013tba,
Durrer:2013pga,Zucca:2016iur}. 
In what follows, we shall adopt the approach discussed earlier~\cite{Bonvin:2011dt,
Bonvin:2013tba} and make use of the publicly available packages CAMB~\cite{Lewis:1999bs}
and MagCAMB~\cite{Zucca:2016iur} to calculate the angular power spectra of 
the anisotropies in the CMB generated by the PMFs.

Cosmological magnetic fields can be constrained via the measurement of the
anisotropies in the temperature~(T) and polarization~(E and B modes) of the
CMB (see Ref.~\cite{Shaw:2010ea}; for bounds from Planck, see 
Ref.~\cite{Planck:2015zrl}). 
We are specifically interested in the angular spectra of the CMB sourced by
the PMFs in the epochs before and after neutrino decoupling.  
Recall that, neutrino decoupling takes place at an energy scale of about~$1$ MeV, 
after which they start streaming freely.
However, before that epoch, neutrinos are strongly bound to the photons and
baryons. 
During this regime, the anisotropic stresses in the magnetic fields source the 
scalar and tensor perturbations, and these contributions are referred to as the 
passive magnetic modes~\cite{Shaw:2009nf,Bonvin:2011dt}. 
After the neutrinos decouple from the photons, they begin to stream freely and, 
in the process, they can develop a non-zero anisotropic stress that compensates 
the anisotropic stress of the PMFs~\cite{Adamek:2011pr}. 
During this period, the PMFs generate the so-called compensated modes which are
somewhat similar to the isocurvature perturbations~\cite{Shaw:2009nf,Bonvin:2011dt}.
Apart from the passive and compensated modes, it has been shown that there arises a 
contribution to the angular power spectra of the CMB due to the curvature perturbations 
induced by the magnetic fields generated during inflation (in this context, see
Refs.~\cite{Bonvin:2013tba,Markkanen:2017kmy}).
The spectrum of these secondary curvature perturbations depend on the model 
being considered for the generation of the magnetic field.  
Also, we should clarify that the secondary curvature perturbations are induced
in addition to the primary adiabatic perturbations generated during inflation. 
In our analysis below, we shall take into account the effects arising from all 
these contributions in the calculation of the angular power spectra of the
anisotropies in the CMB.
We should stress that, in this section, we shall confine our discussion to 
non-helical magnetic fields.


\subsection{Contributions due to the passive and compensated modes}

In order to evaluate the contributions due to the passive and compensated 
modes of the PMFs to the angular power spectra of the CMB, we shall make 
use of the publicly available package MagCAMB~\cite{Zucca:2016iur}, which is 
a modification of CAMB~\cite{Lewis:1999bs}. 
Similar to CAMB, the package computes the multipole moments of the CMB, 
viz. the $C_\ell$'s, arising due to various contributions of the PMFs, for 
a given cosmological model.
However, to reduce the computational complexity in estimating the integrals 
involved in arriving at the $C_\ell$'s, MagCAMB assumes a power law spectrum 
for the PMFs.
In fact, we find that the power law form for the spectrum is hardcoded in 
the package.
Note that, among the two inflationary models we have considered---viz. the 
models described by the potentials~\eqref{eq:sup-ls} and~\eqref{eq:ss-peak}---it 
is only the second potential which leads to a nearly scale invariant power 
law spectrum for the magnetic field over the CMB scales (see Fig.~\ref{fig:pb-pe}).
Therefore, using MagCAMB, we shall explicitly compute the angular power spectra
of the anisotropies in the CMB generated by the passive and the compensated modes 
for the case of the second potential~\eqref{eq:ss-peak}.
Moreover, since the potential~\eqref{eq:ss-peak} permits slow roll inflation 
during the early stages, as we shall discuss in the next subsection, it is also 
possible to approximately evaluate the angular power spectra of the CMB due 
to the curvature perturbations induced by the magnetic field. 
For the first model described by the potential~\eqref{eq:sup-ls}, we find
that it is challenging to carry out such an analysis, due to the complicated
nature of the power spectrum for the magnetic field over large scales. 
Hence, in what follows, we shall only check if the strength of the magnetic 
field generated in this model is roughly compatible with the observational 
constraints.

Before going on to compute the angular power spectra of the CMB generated by
the passive and compensated modes of the PMFs, let us examine if the magnetic 
fields generated in the two inflationary models of interest are broadly
consistent with the current constraints.
To do so, we need to evaluate the amplitude of the magnetic field, say 
$B_\lambda$, that has been smoothed over a coherence 
scale~$\lambda$~\cite{Durrer:2013pga,Zucca:2016iur}.
In practice, the quantity $B_\lambda^2$ is obtained by integrating the 
spectral energy density of the PMFs [i.e. the spectrum of the magnetic 
field, see Eq.~\eqref{eq:pseb-d}] with a Gaussian window function of 
width $\lambda=1\, \mathrm{Mpc}$, and it is defined as 
\begin{eqnarray}
B_\lambda^2 = \int \f{{\d}^3 \bm{k}}{4\,\pi\,k^3}\, 
\mathrm{e}^{-k^2 \lambda^2}\, \pb(k).\label{eq:B-lambda} 
\end{eqnarray}
For instantaneous reheating, the smoothed amplitude {\it today}\/, say 
$B_\lambda^0$, is given by
\begin{eqnarray}
B_\lambda^{0}=B_\lambda\,\l(\f{a_\mathrm{e}}{a_0}\r)^2, \label{eq:B-lambda-today}
\end{eqnarray}
where $B_\lambda$ is the smoothed strength of the magnetic field generated during
inflation, while $a_\mathrm{e}$ and $a_0$ denote the scale factors at the end of 
inflation and today, respectively.
The ratio of these scale factors is given by~\cite{Tripathy:2021sfb}
\begin{equation}
\f{a_0}{a_\mathrm{e}} \simeq  2.8\times 10^{28}\, 
\l(\f{\HI}{10^{-5}\,\Mpl}\r)^{1/2},
\end{equation}
where $\HI$ is the Hubble parameter during inflation.

Let us now evaluate the quantity $B_\lambda^{0}$ in the two inflationary 
models of our interest.
In the case of the first potential~\eqref{eq:sup-ls}, upon using the resulting power 
spectrum for the magnetic field (as illustrated in Fig.~\ref{fig:pb-pe}) and carrying 
out the integral~\eqref{eq:B-lambda} numerically over all scales (viz. $10^{-5}\lesssim 
k \lesssim 10^{19}\,\mathrm{Mpc}^{-1}$), we obtain that $B_\lambda^2 = 1.079\times 
10^{-20}\, \Mpl^4$.
Thereafter, upon using the relation~\eqref{eq:B-lambda-today}, we obtain an 
estimate of the smoothed strength of the magnetic field today to be 
$B_{\lambda}^0\simeq 2.77\times 10^{-2}\,\mathrm{nG}$, corresponding to $\HI 
\simeq 4.07 \times 10^{-6}\,\Mpl$. 
We should mention that, to arrive at this result, we have used the conversion factors
$1\, \Mpl=2.43\times 10^{18}\, \mathrm{GeV}$ and $1\,\mathrm{G}= 6.91\times10^{-20}\,
\mathrm{GeV}^2$.
Similarly, in the case of the second potential~\eqref{eq:ss-peak}, for the spectrum
of the magnetic field illustrated in Fig.~\ref{fig:pb-pe}, we obtain that $B_\lambda^2
= 9.69 \times 10^{-21}\,\Mpl^4$, which leads to the present day strength of 
$B_\lambda^0 \simeq 2.05 \times 10^{-1}\,\mathrm{nG}$, corresponding to $\HI \simeq 
5.26 \times 10^{-7}\,\Mpl$. 
These estimates suggest that the spectra of the magnetic fields from the two 
inflationary models are broadly in agreement with the observational bound of 
$B_\lambda^0 \lesssim 1\, \mathrm{nG}$ on the strength of the magnetic field today
(in this context, see, for instance, Ref.~\cite{Markkanen:2017kmy}).

Let us now turn to the explicit evaluation of the imprints of the passive and 
compensated modes induced by the PMFs on the CMB using MagCAMB.
As we have already mentioned, in MagCAMB, the primordial power spectrum of the 
magnetic field is assumed to be of the power law form, say, $\pb(k) \propto k^{\nbb}$, 
where $\nbb$ is the spectral index. 
We find that the spectral index $\nb$ we have defined [see our comments 
following Eq.~\eqref{eq:pseb-d}] is related to the spectral index $\nbb$ of 
MagCAMB as $\nbb=-3+\nb$. 
The quantities required to compute $C_\ell$'s due to the PMFs using MagCAMB are 
the smoothed amplitude $B_\lambda^0$ that we discussed above and the spectral 
index~$\nbb$~\cite{Zucca:2016iur}. 
As mentioned earlier, in the scenario described by the potential~\eqref{eq:sup-ls}, 
since the magnetic power spectrum contains strong features over large scales,  
we are unable to use MagCAMB.
For the model described by the potential~\eqref{eq:ss-peak}, as the magnetic power
spectrum is nearly scale invariant over large scales, we have provided MagCAMB
with the smoothed amplitude $B_\lambda^0$ and the spectral index $\nbb$ to arrive
at the angular spectra of the CMB corresponding to the passive and compensated modes.
We find that the spectral index over the CMB scales for the spectrum of the magnetic
field illustrated in Fig.~\ref{fig:pb-pe} is $\nb=-0.0112$. 
So, we have supplied the following values of the parameters to 
MagCAMB:~$B_\lambda^0= 2.05 \times 10^{-1}\,\mathrm{nG}$, $\nbb= -3.0112$, and 
set the pivot scale to be $k_\ast=0.05\,\mathrm{Mpc}^{-1}$. 
Using these parameters, we have computed the contributions of the PMFs to the 
angular power spectra of the CMB through the passive and compensated modes.
We shall present and discuss the results in subsection~\ref{subsec:angps-cmb}.


\subsection{Contributions due to the induced curvature perturbations}

Next, we investigate the contributions to the angular power spectra of the CMB
due to the curvature perturbations induced by the magnetic fields generated 
during inflation, which are often referred to as the inflationary magnetic
modes~\cite{Bonvin:2011dt,Bonvin:2013tba}. 
These modes are unique to inflationary magnetogenesis and are absent if the PMFs 
are generated after inflation. 
They remain unaffected by the behavior of magnetic fields after the termination 
of inflation. 
Once again, to examine the imprints on the angular power spectrum of the CMB due 
to these modes, we restrict ourselves to the model described by the
potential~\eqref{eq:ss-peak}, as the sharp features in the spectrum arise only 
over small scales and we can work with the de Sitter approximation to compute 
the observables over the CMB scales.

In the slow roll approximation, the strength of the curvature perturbation,
say $\mathcal{R}_k^{\mathrm{mag}}$, induced by the magnetic fields during 
inflation (for the case wherein $J \propto a^2$) can be written 
as~\cite{Bonvin:2013tba,Bonvin:2011dt}
\begin{eqnarray}
k^{3/2}\,\mathcal{R}_k^{\mathrm{mag}}(\ee)
=\f{2\,\HI^2}{3\,\Mpl^2\,\epsilon_1}\,C_{_{\mathrm{EM}}}(k)\,
\mathrm{ln}\l({\f{k}{\ke}}\r),\label{eq:icp}
\end{eqnarray}
where $\epsilon_1$ is the first slow roll parameter and $\ke$ represents the 
wave number that leaves the Hubble radius at the end of inflation (i.e. at $\ee$), 
when the strength of the perturbations is evaluated. 
The quantity $C_{_{\mathrm{EM}}}(k)$ is determined by the expression
\begin{equation}
\sqrt{\f{k^3\,P_{_\mathrm{EM}}(k)}{\rho_\phi^2}}
=\f{\HI^2}{3\,\Mpl^2}\,C_{_{\rm EM}}(k),\label{eq:cem}
\end{equation}
where $P_{_\mathrm{EM}}(k)$ is the power spectrum of the fluctuations in the
energy density of a given mode of the electromagnetic field which is defined 
through the relation~\eqref{eq:P-EM}, and $\rho_\phi$ denotes the energy 
density of the scalar field(s) driving the inflationary background.
In App.~\ref{app:pem}, we have initially arrived at a generic expression for 
the power spectrum~$P_{_\mathrm{EM}}(k)$ and have then gone on to evaluate 
the quantity for the case wherein $J\propto a^2$, which leads to a scale
invariant spectrum for the magnetic field $\pb(k)$.
But, evidently, in the model described by the potential~\eqref{eq:ss-peak},
there arise deviations from slow roll.
Also, the resulting spectrum of the magnetic field is not scale invariant, 
as should be clear from Fig.~\ref{fig:pb-pe}. 
However, note that the departures from slow roll occur at late times and, 
due to this reason, the deviations from the nearly scale invariant behavior 
arise only over very small scales.
Moreover, the deviations from scale invariance are mostly in the form of 
oscillations.
Therefore, {\it over the CMB scales},\/ we believe that the slow roll approximation 
leads to a reasonable estimate of the power spectrum of the curvature perturbations 
induced by the magnetic field.
We can arrive at the strength of the induced curvature perturbation at late
times, i.e. $\mathcal{R}^{\mathrm{mag}}_k(\ee)$, by using the result~\eqref{eq:pem-fv} 
for $P_{_\mathrm{EM}}(k)$ in Eqs.~\eqref{eq:icp} and~\eqref{eq:cem} and the
fact that $\rho_\phi=3\,H^2\,\Mpl^2$.
The scalar power spectrum associated with the inflationary magnetic mode can 
be obtained to be
\begin{equation}
\mathcal{P}_{\mathcal{R}}^{\mathrm{mag}}(k)  
\simeq \f{24}{\pi^3}\,\l(\ps^0\r)^2\,\ln\l(\f{k}{k_{\mathrm{min}}}\r)\,
\l[\ln\l(\f{k}{\ke}\r)\r]^2,\label{eq:ps-inf-mag}
\end{equation}
where $\ps^0={\HI^2}/{(8\,\pi^2\,\epsilon_1)}$ is the standard scalar power 
spectrum evaluated in the slow roll approximation.

Having obtained the spectrum of curvature perturbations induced by the magnetic 
field, we proceed to compute the corresponding contributions to the angular power 
spectrum of the CMB.
We should stress that the contributions due to the inflationary magnetic mode  
arise {\it in addition}\/ to the contributions due to the primary curvature
perturbations generated from the quantum vacuum during inflation.
This enables us to treat it in the same manner as the primary curvature perturbations 
and use the standard apparatus of CAMB to compute the corresponding angular power 
spectra.
To evaluate the $C_\ell$'s, we make use of the scalar power spectrum obtained in
Eq.~\eqref{eq:ps-inf-mag}. 
Since we are working with the de Sitter approximation over the CMB scales, 
the parameters that we require to compute the amplitude $\ps^0$ are $\HI$ and 
$\epsilon_1$, evaluated at the $e$-fold when the pivot scale exits the Hubble 
radius.
Moreover, to obtain $\mathcal{P}_\mathcal{R}^{\mathrm{mag}}(k)$, we require 
$k_{\mathrm{min}}$ and $\ke$.
We assume $k_{\mathrm{min}}$ to be $10^{-7}\,\mathrm{Mpc}^{-1}$. 
In the model of our interest [viz. the potential~\eqref{eq:ss-peak}], for the 
values of parameters we have worked with, we find that $\ke \simeq 
10^{19}\,\mathrm{Mpc}^{-1}$. 
We should further note that, since the electromagnetic field possesses anisotropic
stress, apart from inducing secondary scalar perturbations, they will also generate
secondary tensor perturbations (in this context, see, for example,
Refs.~\cite{Caprini:2003vc,Caprini:2014mja,Ballardini:2014jta,Sharma:2019jtb,
Okano:2020uyr,Brandenburg:2021bfx,Brandenburg:2021pdv}).  
Such a tensor mode will also contribute to the B-mode polarization of the CMB, 
apart from the contributions to the temperature and E-mode polarizations. 
In this work, we have {\it not}\/ calculated these contributions due to the 
induced tensor perturbations.


\subsection{Angular power spectra of the CMB}\label{subsec:angps-cmb}

\begin{figure*}
\centering
\includegraphics[width=0.475\linewidth]{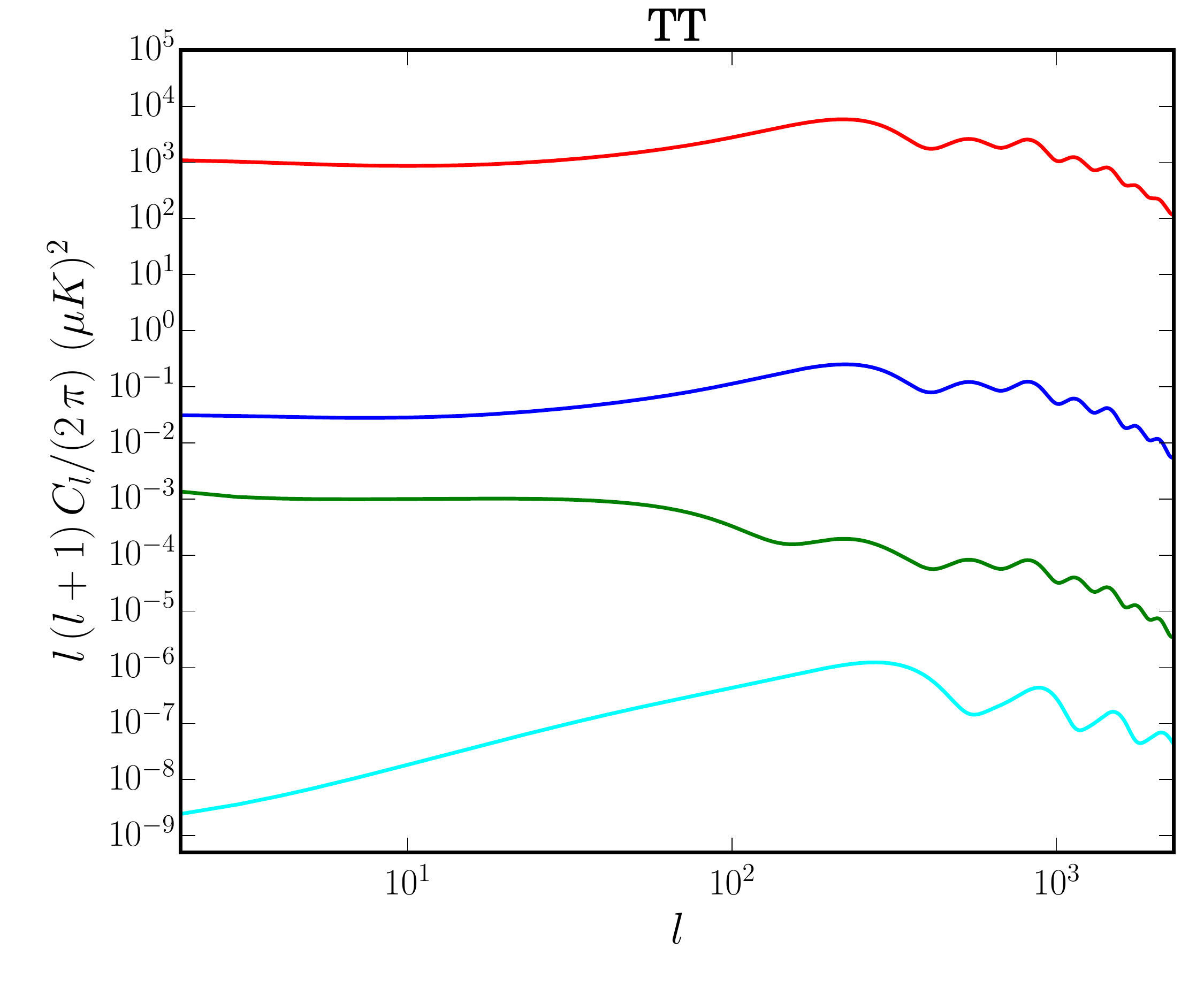}
\includegraphics[width=0.475\linewidth]{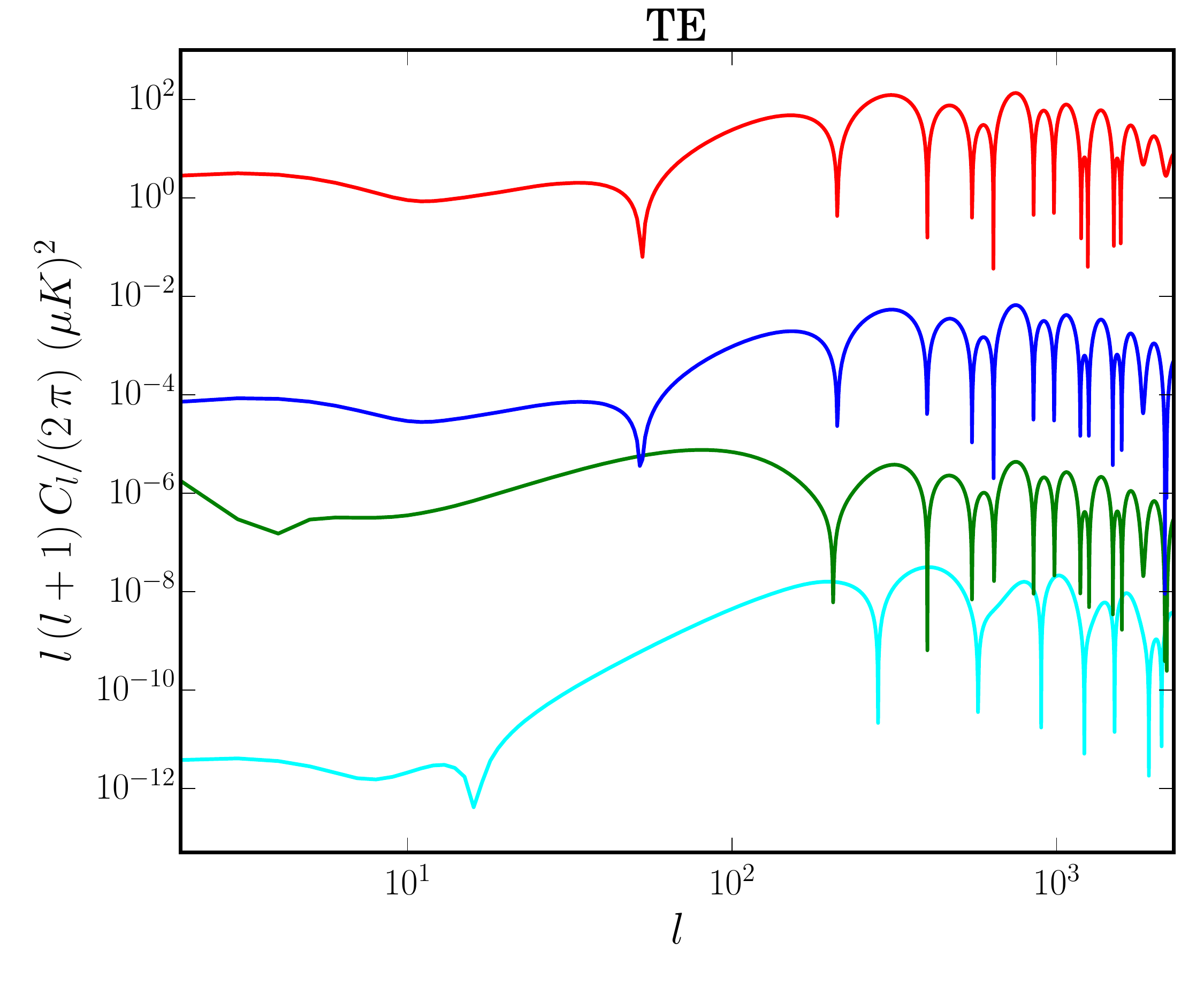}
\includegraphics[width=0.475\linewidth]{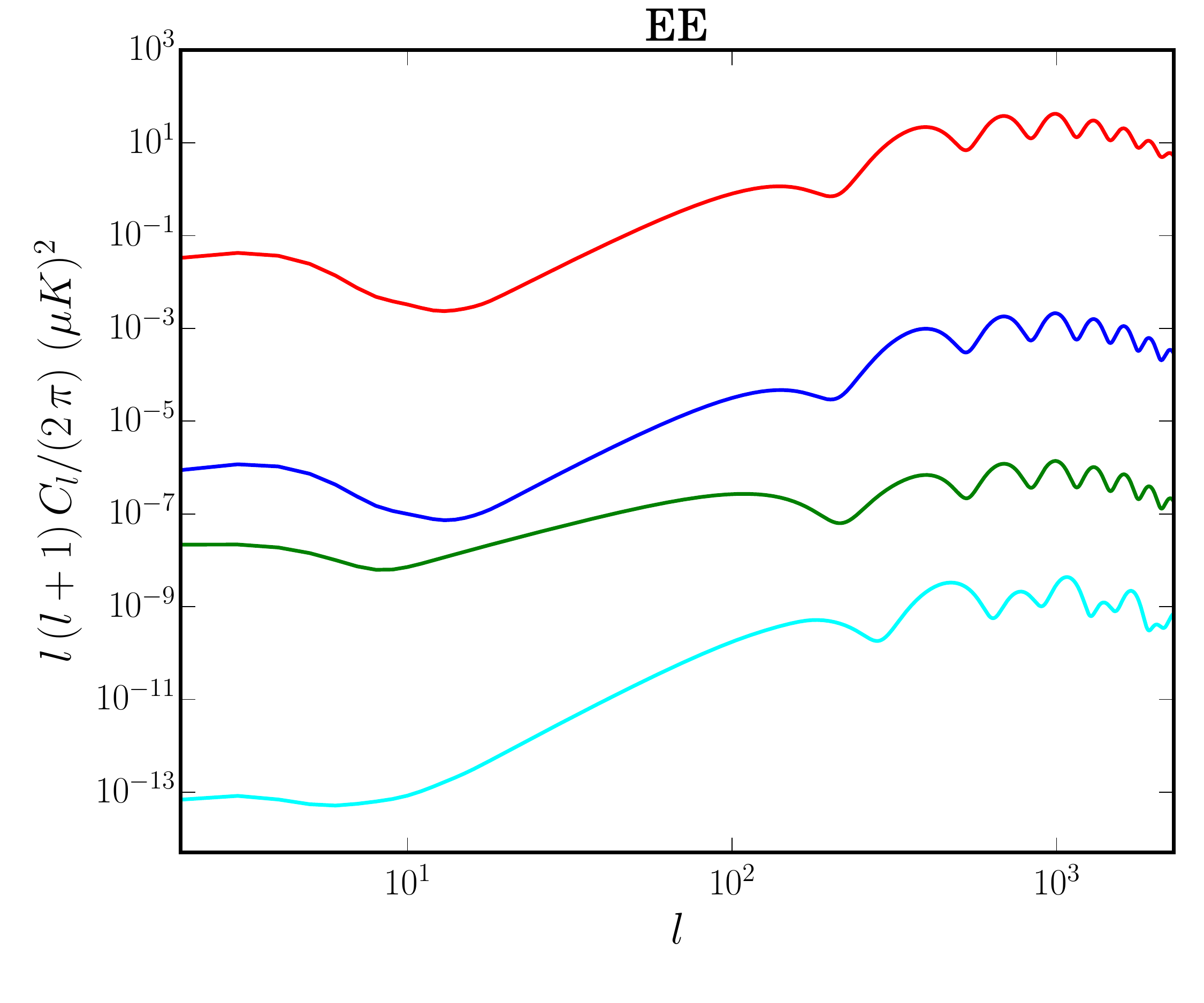}
\includegraphics[width=0.475\linewidth]{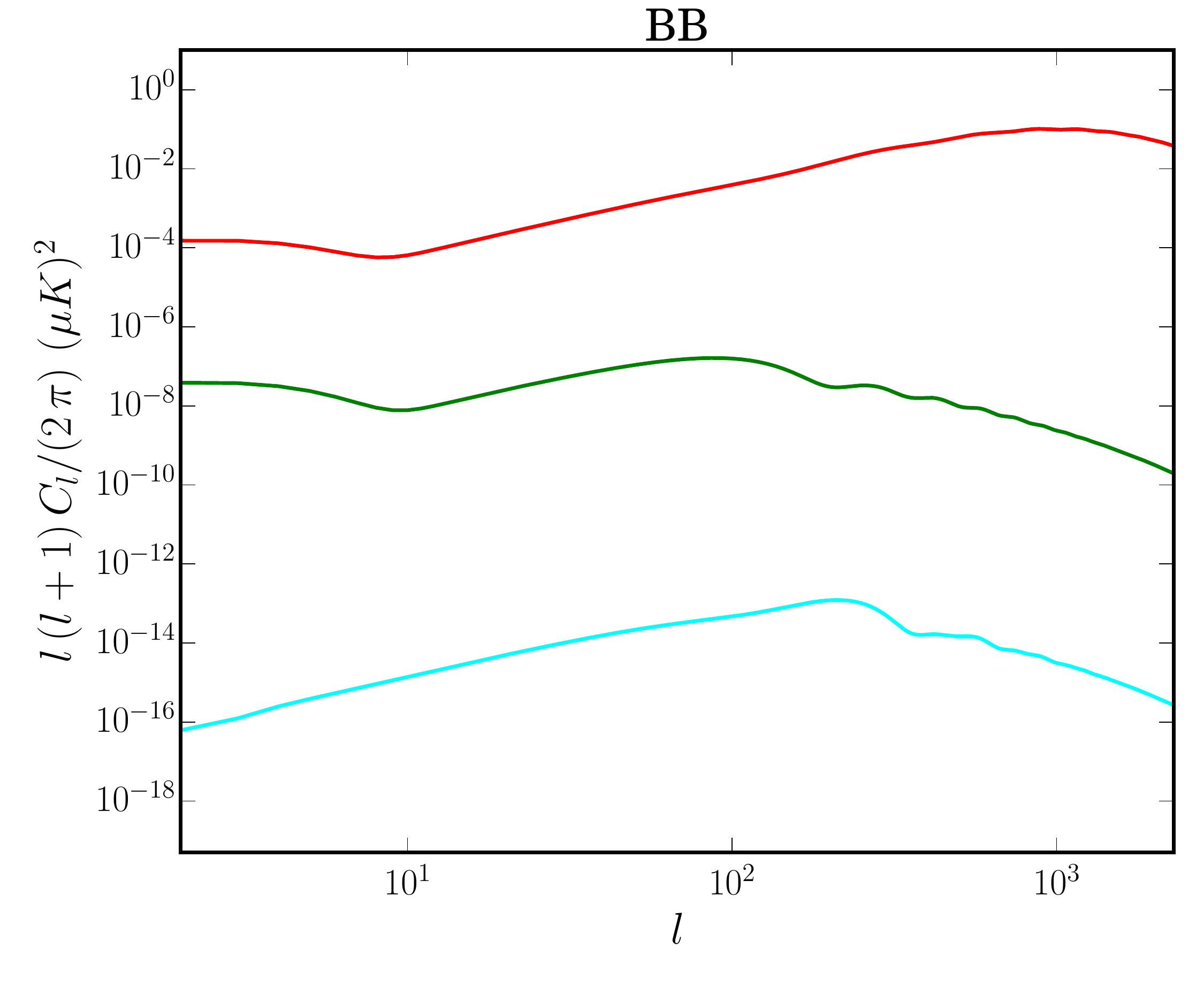}
\caption{We have illustrated the contributions of the magnetic modes to the 
temperature and polarization angular power spectra of the CMB due to the total 
(i.e. scalar plus tensor) passive (in green) and the total compensated (in cyan) 
modes.
We have arrived at these quantities using MagCAMB corresponding to a magnetic 
field with smoothed strength of $B_{1\,\mathrm{Mpc}}^0 = 2.05 \times 10^{-1}\,
\mathrm{nG}$ today and a spectral index of $\nbb=-3.0112$. 
In addition, using CAMB, we have plotted the standard angular power spectra
of the CMB (in red) induced by the primary scalar and tensor perturbations. 
Moreover, we have also presented the contribution due to the curvature 
perturbation induced by the magnetic field during inflation (in blue), 
which we have computed using CAMB.
Note that, apart from the contributions due to the primary tensor perturbations
to the angular power spectrum of the CMB (in particular, to the B-mode polarization,
which we have illustrated in red in the plot on the bottom right corner), there 
will also arise a contribution due to the tensor perturbations induced by the 
magnetic field during inflation. 
We should mention that we have not calculated this additional contribution in 
this work.}\label{fig:Cl}
\end{figure*}

In Fig.~\ref{fig:Cl}, we have illustrated all the contributions to the CMB 
angular spectra arising due to the PMFs for the model described by the 
potential~\eqref{eq:ss-peak} which generates enhanced scalar power on small
scales. 
For reference, we have also plotted the standard CMB spectra obtained from CAMB, 
where there are no contributions from the PMFs.
We should mention that these standard spectra of TT, TE, EE and BB are obtained 
by supplying the numerically computed scalar and tensor power spectra from our 
inflationary model (illustrated in Fig.~\ref{fig:stps}) to CAMB.
In arriving at these spectra, we have included the effects due to non-linear 
lensing.
We have then presented the contributions due to the scalar inflationary magnetic 
mode, obtained using our modified setup of CAMB as described above. 
This mode contributes only to the CMB temperature and E-mode polarization spectra. 
We should clarify that, in computing these spectra, we have ignored
the effects due to non-linear lensing.
As is evident from the plots, this contribution is lower in amplitude when compared
to the standard CMB spectra by $\mathcal{O}(10^4)$. 
We have further illustrated the contributions due to the passive and compensated 
modes to the angular power spectra of the CMB, which have been computed using MagCAMB.
We have computed these contributions using the parameters obtained from the power
spectrum of magnetic field arising in the model, viz. $B_\lambda^0 = 7.25\times 10^{-1}
\mathrm{nG}$ and $\nbb=-3.0112$.
As can be seen clearly, although the spectra due to the passive and compensated 
modes have different amplitudes, their shape is roughly similar to the standard 
CMB spectra.
It is evident from the figure that the contributions to the CMB angular spectra due 
to the PMFs are substantially smaller in magnitude for the model we have considered. 
The largest contribution arises from the inflationary magnetic mode and it is still
at least ${\mathcal{O}}(10^4)$ lesser in magnitude than the standard spectra.

Though we could not carry out a similar exercise for the model described by the
potential~\eqref{eq:sup-ls}, which leads to a suppression in the scalar power 
spectrum over large scales, the estimate of $B_\lambda^0$ in that case suggests 
that the corresponding contributions to $C_\ell$'s would be of amplitudes lesser 
than the case that we have discussed. 
Recall that, for these additional contributions, $C_\ell \propto (B_\lambda^0)^4$ 
and hence the overall amplitude of the passive and compensated contributions can 
be expected to be of lesser magnitude for the model~\eqref{eq:sup-ls} than in the 
model~\eqref{eq:ss-peak} wherein we have explicitly have computed these contributions. 
However, there can arise a difference in the shape of $C_\ell$'s at the lower 
multipoles due to the sharp features in $\pb(k)$ over large scales. 
In particular, the scalar power spectrum associated with the inflationary magnetic 
mode in this case can have interesting features, and being the largest of the
contributions, it may leave discernible imprints on the total angular power 
spectrum of the CMB.
But it is challenging to compute these contributions induced by the magnetic 
field using analytical methods for spectra with strong features.
We would have to employ numerical procedures to compute the induced scalar spectra 
over large scales. 
This is a non-trivial exercise and we believe that it is beyond the scope of 
the  current work.


\section{Conclusions}\label{sec:c}

In our earlier work~\cite{Tripathy:2021sfb}, we had shown that, in the case of 
single field models involving strong deviations from slow roll, there arise
certain challenges in obtaining nearly scale invariant power spectra for the 
magnetic field. 
We had shown that even finely tuned non-conformal coupling functions may not 
help us avoid strong features generated in scenarios involving an epoch of 
ultra slow roll inflation. 
To overcome such challenges, in this work, we have examined two field models of 
inflation where a turning in the background trajectory in the field space gives
rise to departures from slow roll. 
We have considered two field models where such deviations from slow roll lead to
either a suppression in the scalar power over large scales or to an enhancement
over small scales.

We have constructed model dependent coupling functions numerically using the 
background dynamics of the two fields in these models. 
Using these coupling functions, we have been able to obtain the desired amplitude and 
a nearly scale invariant form for the power spectra of magnetic fields in the models 
of interest.
While we have been able to mostly circumvent the challenges faced in single field 
models, we find that it is not entirely possible to remove the strong features over 
the range of scales that leave the Hubble radius during deviations from slow roll.
For the potential that gives rise to an enhancement in scalar power over small scales, 
we obtain a power spectrum for the magnetic field which is nearly scale invariant 
over large scales, but contains a rapid burst of oscillations over small scales. 
Similarly, for the potential that generates a suppression in the scalar power over 
large scales, the power spectrum of the magnetic field exhibits strong oscillations 
over very large scales and turns scale invariant over smaller scales. 
In the first model, the oscillations have higher amplitude than the scale invariant 
part, whereas they have same amplitude in the second model. 
In both these models, we obtain amplitudes of the smoothed magnetic fields, which 
lie in the current range of observations, i.e. $10^{-16}$--$10^{-9}\, \mathrm{G}$.

Further, for the model that generates enhancement in scalar power over small scales, 
we have also computed the contributions of the PMFs to the anisotropies in the CMB
using MagCAMB.
Apart from calculating the contributions due to the passive and compensated modes,
using CAMB, we have also evaluated the contribution due to the curvature perturbation
induced by the magnetic field during inflation.
These contributions to the angular power spectra of the CMB are of roughly similar
shape as the standard spectrum, but are of lower amplitudes. 
Moreover, we find that, the corresponding value of the smoothed amplitude $B_\lambda^0$ 
is well within the upper bound on the parameter obtained earlier (in this context, 
see Ref.~\cite{Zucca:2016iur}).

To summarize, using two field models, along with suitable choices of coupling
functions, we have been able to largely overcome the challenges faced in the generation 
of PMFs in single field models of inflation permitting an epoch of ultra slow roll.
Also, we have been able to approximately evaluate the imprints of the PMFs on the
CMB in the second model that leads to a scale invariant spectrum for the magnetic
field over large scales. 
But, clearly, there are some limitations to the approach we have adopted. 
For instance, it seems fair to assume that the small scale features in the power
spectrum of the magnetic field are unlikely to affect the angular power spectra 
of the CMB.
However, since there arise departures from slow roll at late times in the second 
model, the de Sitter approximation we have worked with to estimate the induced 
spectrum of curvature perturbations is likely to be inadequate. 
Moreover, as we mentioned, the first model which leads to features in the spectrum
of the magnetic field over the CMB scales needs to be analyzed numerically to evaluate
the induced spectrum of curvature perturbations and the corresponding imprints on
the CMB.
In addition, to compute the signatures of the passive and compensated modes in 
such models, MagCAMB needs to be suitably modified to take into account features 
in the power spectra of the electromagnetic fields.
We are presently investigating such issues.


\section*{Acknowledgments}

The authors wish to thank Matteo Braglia, Dhiraj Kumar Hazra and Shiv Kumar 
Sethi for discussions and detailed comments on the manuscript.
ST would like to thank the Indian Institute of Technology Madras, Chennai, 
India, for support through the Half-Time Research Assistantship.
DC's work is supported by the STFC grant ST/T000813/1.
HVR thanks the Indian Institute of Science Education and Research Kolkata, for
support through a postdoctoral fellowship.
LS and RKJ wish to acknowledge support from the Science and Engineering 
Research Board, Department of Science and Technology, Government of India, 
through the Core Research Grant~CRG/2018/002200.
RKJ also acknowledges financial support from the new faculty seed start-up 
grant of the Indian Institute of Science, Bengaluru, India.
RKJ further wishes to thank the Infosys Foundation, Bengaluru, India, for 
support through the Infosys Young Investigator Award. 
For the purpose of open access, the authors have applied a Creative Commons 
Attribution (CC BY) license to any Author Accepted Manuscript version arising.


\appendix

\section{Analytical construction of the non-conformal coupling 
function}\label{app:as}

In this appendix, using the solutions for the fields~$\phi$ and~$\chi$ that 
can be arrived at in the slow roll approximation, we shall construct analytical
forms for the non-conformal coupling function $J(\phi,\chi)$ in the two 
inflationary models we have considered.
We shall then make use of the analytical forms for~$J(\phi,\chi)$ to numerically
compute the resulting spectra of the magnetic field and compare them with the 
spectra we have obtained earlier.

Let us first discuss the model described by the potential~\eqref{eq:sup-ls}.
As we have seen, in the two field models of our interest, there arise two stages 
of inflation, with each regime being driven by one of the two fields.
In the case of inflation driven by the potential~\eqref{eq:sup-ls}, during the 
first stage, the field $\phi$ rolls down the potential, while the field~$\chi$ 
remains frozen. 
During this phase, the evolution of the field $\phi$ in the slow roll approximation
can be expressed as~\cite{Braglia:2020eai}
\begin{equation}
\phi^2(N)=\phii^2-4\,\Mpl^2\,N,
\end{equation}
where we have assumed that $\phi=\phii$ at $N=0$.
To achieve the desired behavior of $J\propto a^2$, in the first stage, we can 
assume that
\begin{equation}
J(\phi) \propto \mathrm{exp}\l[-\f{1}{2}\l(\f{\phi^2}{\Mpl^2}
-\f{\phi_\mathrm{i}^2}{\Mpl^2}\r)\r].\label{eq:J-phi-1}
\end{equation}
The first stage  dominated by the field $\phi$ eventually ends and, after a few 
damped oscillations, the field settles down at the value
\begin{equation}
\phi_{\mathrm{min}} \simeq \f{1}{2\,\bar{b}}\
W\l(\f{8\,V_0\,\bar{b}^2\,\Mpl^2\,\chi_0^4}{3\,m_\phi^2\,\chi_\mathrm{i}^6}\r),
\end{equation}
where $W(z)$ is the so-called Lambert or the product logarithmic
function~\cite{corless1996lambertw}. 
The field $\chi$ drives the second stage of inflation and, during this period, 
the solution for the field in the slow roll approximation can be written as
\begin{eqnarray}
\chi^2(N)&=&\bigl[{\l(\chi_0^2+\chi_\mathrm{i}^2\r)^2
-8\,\Mpl^2\,\chi_0^2\,\mathrm{e}^{-2\,\bar{b}\,\phi_{\mathrm{min}}}\,
(N-N_1)}\bigr]^{1/2}\nn\\
& &-\,\chi_0^2,
\end{eqnarray}
where $N_1$ is the $e$-fold when $\phi=\phi_\mathrm{min}$.
In a fashion similar to the first phase, to achieve $J \propto a^2$, we can 
choose the coupling function during the second stage of slow roll inflation 
to be
\begin{eqnarray}
J(\chi) &=& \mathrm{exp}\,\biggl\{2\,N_1
-\f{\mathrm{e}^{2\,\bar{b}\,\phi_\mathrm{min}}}{4\,\Mpl^2\,\chi_0^2}\,
\biggl[\l(\chi^2+\chi_0^2\r)^2\nn\\
& &-\,\l(\chi_\mathrm{i}^2+\chi_0^2\r)^2\biggr]\biggr\}.\label{eq:J-chi-1}
\end{eqnarray}

Let us now turn to the second model described by the potential~\eqref{eq:ss-peak}.
During the first stage driven by the field~$\phi$, in the slow roll approximation,
the evolution of the field can be expressed as
\begin{eqnarray}
\phi^2(N)=\l[\l(\phi_{\mathrm{i}}^2+\phi_0^2\r)^2
-8\,\Mpl^2\,\phi_0^2\,N\r]^{1/2}-\phi_0^2,\qquad
\end{eqnarray}
where we have assumed that the field is at $\phii$ when $N=0$.
To achieve $J\propto a^2$, the coupling function can be chosen to be
\begin{eqnarray}
J(\phi) &\propto& 
\mathrm{exp}\l\{-\f{1}{4\,\Mpl^2\,\phi_0^2}\,
\l[\l(\phi^2+\phi_0^2\r)^2-\l(\phi_{\mathrm{i}}^2-\phi_0\r)^2\r]\r\}.\nn\\
\label{eq:J-phi-2}
\end{eqnarray}
In between the two stages of inflation, $\phi$ behaves like a massive scalar 
field and undergoes damped oscillations around the minimum~\cite{Braglia:2020eai}.
It seems difficult to obtain an analytical solution during this period since 
the Hubble parameter~$H$ and the field $\chi$ experience a jump.
We find that $\phi$ eventually approaches a constant value $\phi_\mathrm{min}$, 
given by the minimum of its effective potential. 
The value of~$\chi$ at the onset of this period can be written as~$\chi_{1\mathrm{i}} 
= \chi_{\mathrm{i}} - \Delta\chi$, where $\Delta\chi$ is the jump in $\chi$. 
During the second stage of slow roll inflation, the solution for~$\chi$ 
can be written as
\begin{eqnarray}
\chi^2(N)= \chi_{{\mathrm i}}^2-4\,\mathrm{e}^{-2\,\bar{b}\,
\l[\phi_{\mathrm{min}}+\Delta\phi(N)\r]}\,\Mpl^2\, (N-N_1),\qquad
\end{eqnarray}
where $\phi_{\mathrm{min}}=\bar{b}\,\Mpl^2\,m_{\chi}^2\,\phi_0^2/(3\,V_0)$ and 
the quantity $\Delta\phi$ is governed by the equation
\begin{eqnarray}
\f{\d^2 \Delta\phi}{\d N^2}+(3-\epsilon_1)\,\f{\d\Delta\phi}{\d N}
+\f{m_{\Delta\phi}^2}{H^2}\,\Delta\phi=0
\end{eqnarray}
with $m_{\Delta\phi}^2$ being given by
\begin{equation}
m_{\Delta\phi}^2=\f{2\,V_0}{\phi_0^2}+\f{4}{3}\,\bar{b}^2\,m_\chi^2\,\phi_0^2.
\end{equation}
Therefore, the coupling function during the second stage can be chosen to be
\begin{equation}
J(\chi) = \mathrm{exp}\l\{2\,N_1-\f{1}{2}\,
\mathrm{e}^{2\,\bar{b}\,\l[\phi_{\mathrm{min}}+\Delta\phi(N)\r]}\,
\l(\f{\chi^2}{\Mpl^2}-\f{\chi_{\mathrm{i}}^2}{\Mpl^2}\r)\r\}.\label{eq:J-chi-2}
\end{equation}

With the solutions of the coupling functions in the two stages at hand, we can 
combine them [in a manner similar to Eq.~\eqref{eq:J-phi-chi}] to arrive at the
following coupling function:
\begin{eqnarray}
J(\phi,\chi) 
&=& J_0\,\biggl\{\f{1}{2}\,\l[1+\mathrm{tanh}\l(\f{\chi-\chi_1}{\Delta\chi}\r)\r]\,
J(\phi)\nn\\
& &+\,\f{1}{2}\,\l[1-\mathrm{tanh}\l(\f{\chi -\chi_1}{\Delta\chi}\r)\r]\,
J(\chi)\biggr\},\qquad\quad\label{eq:aJ}
\end{eqnarray}
where $\chi_1$ is the value of $\chi$ around the $e$-fold when the transition 
from the the first stage of slow roll region to the second stage occurs. 
Since we require $J$ to reduce to unity at the end of inflation, we have 
\begin{eqnarray}
J_0 &=& \biggl\{\f{1}{2}\,\l[1+\mathrm{tanh}\l(\f{\chie -\chi_1}{\Delta\chi}\r)\r]\, 
J(\phie)\nn\\ 
& &+\,\f{1}{2}\,\l[1-\mathrm{tanh}\l(\f{\chie-\chi_1}{\Delta\chi}\r)\r]\,
J(\chie)\biggr\}^{-1},\qquad\quad
\end{eqnarray}
where $\phie$ and $\chie$ denote the values of the fields at the end of inflation.

\begin{figure*}
\includegraphics[width=0.32\linewidth]{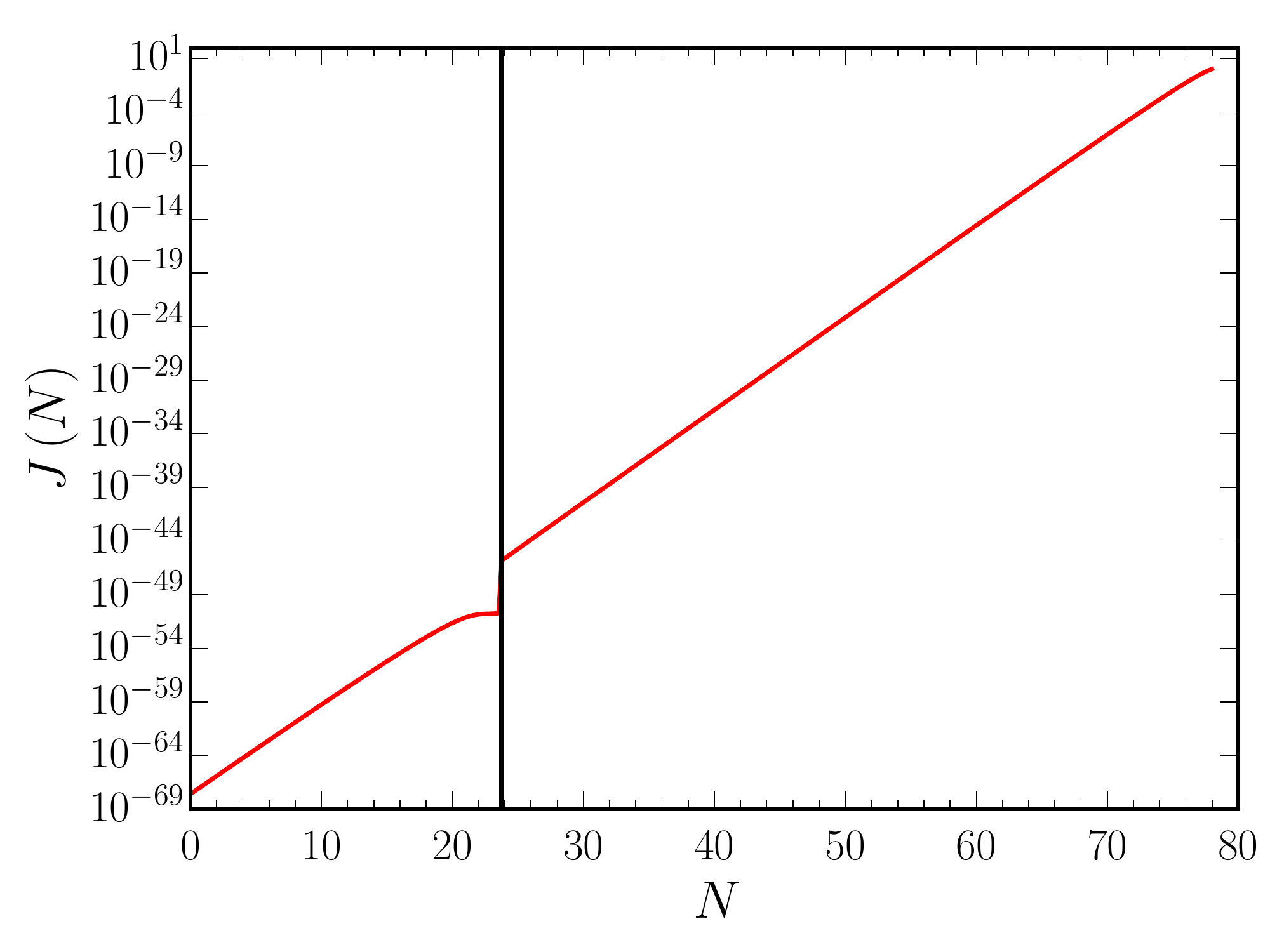}
\includegraphics[width=0.32\linewidth]{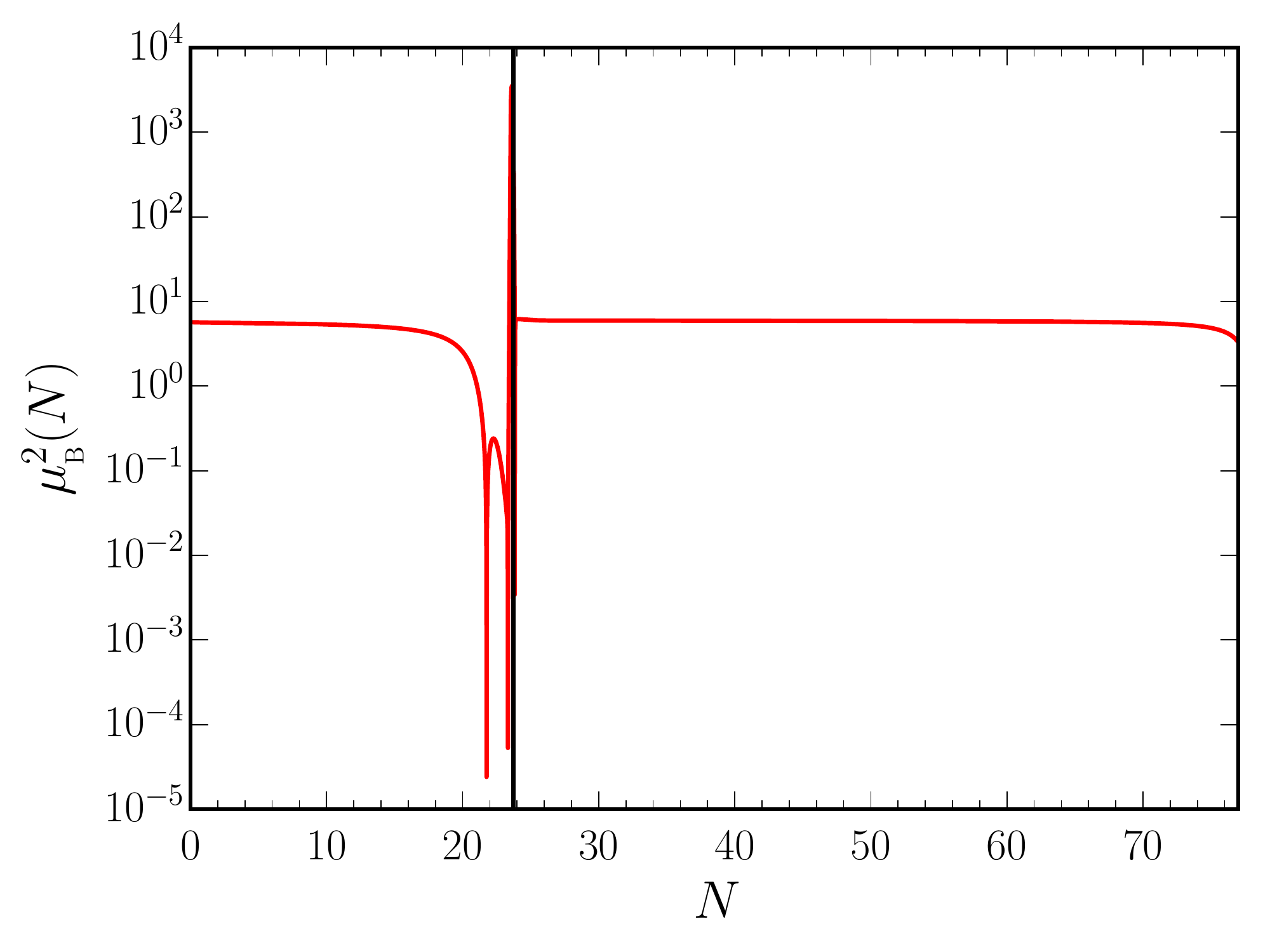}
\includegraphics[width=0.32\linewidth]{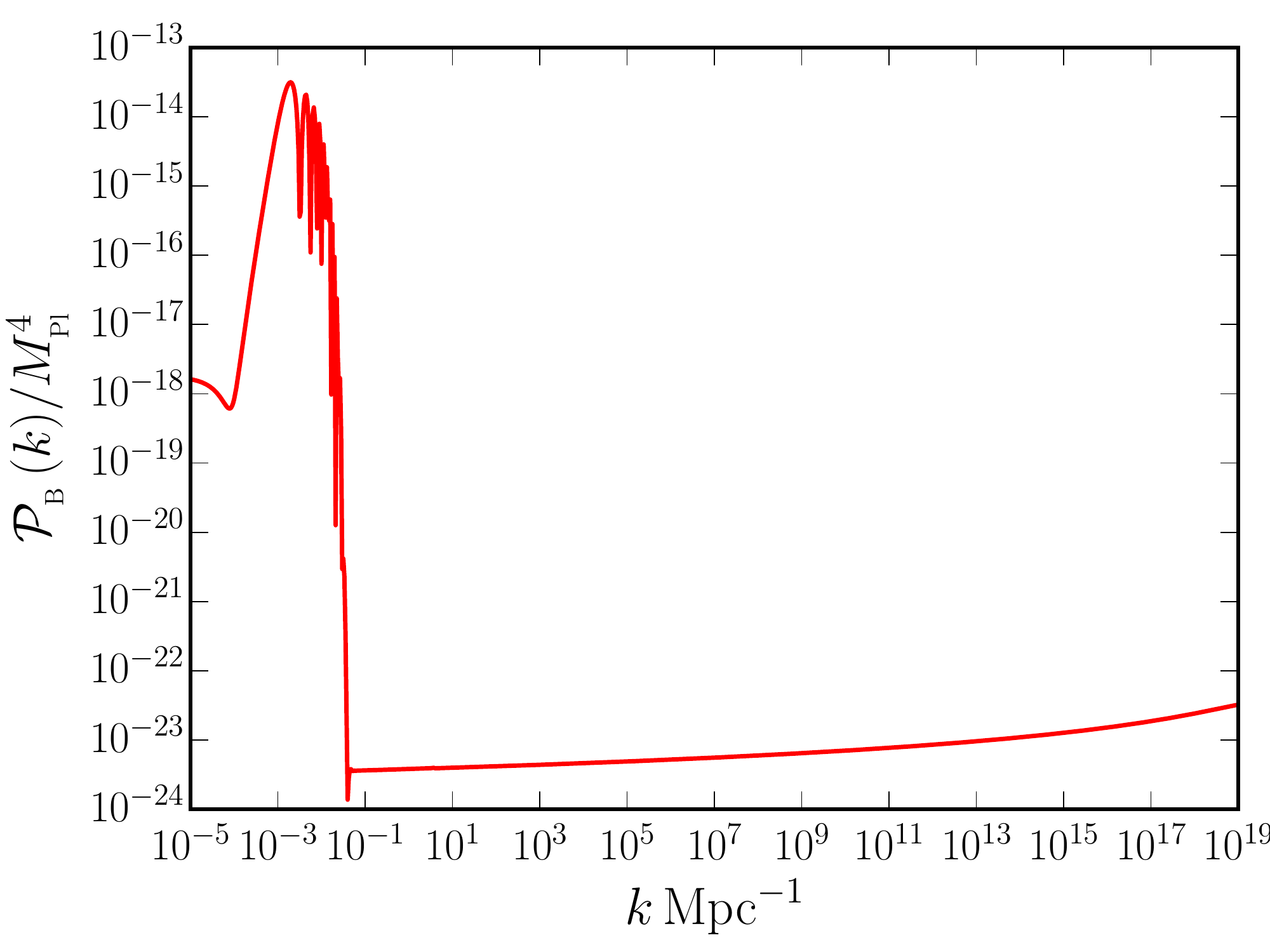}
\includegraphics[width=0.32\linewidth]{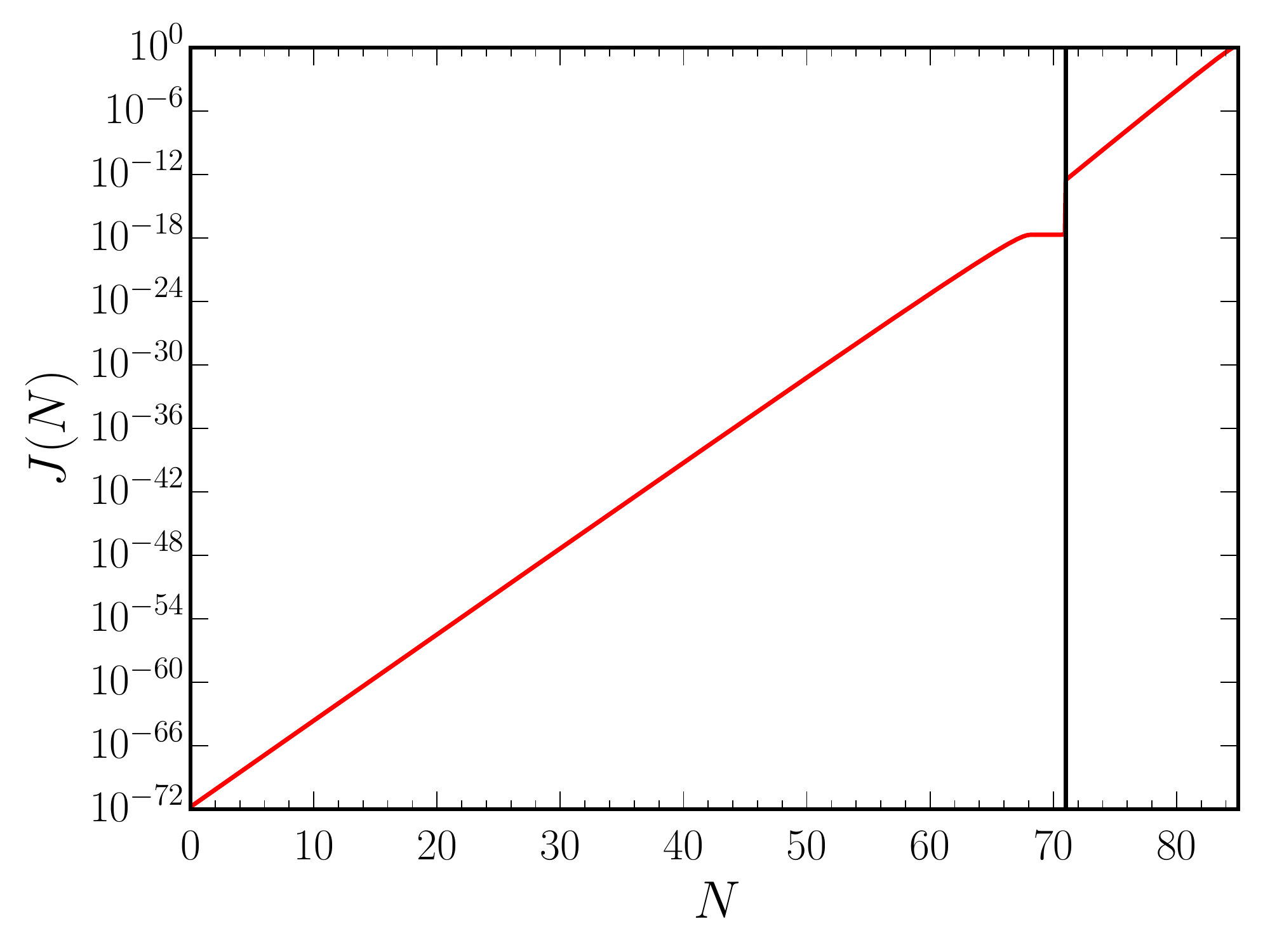}
\includegraphics[width=0.32\linewidth]{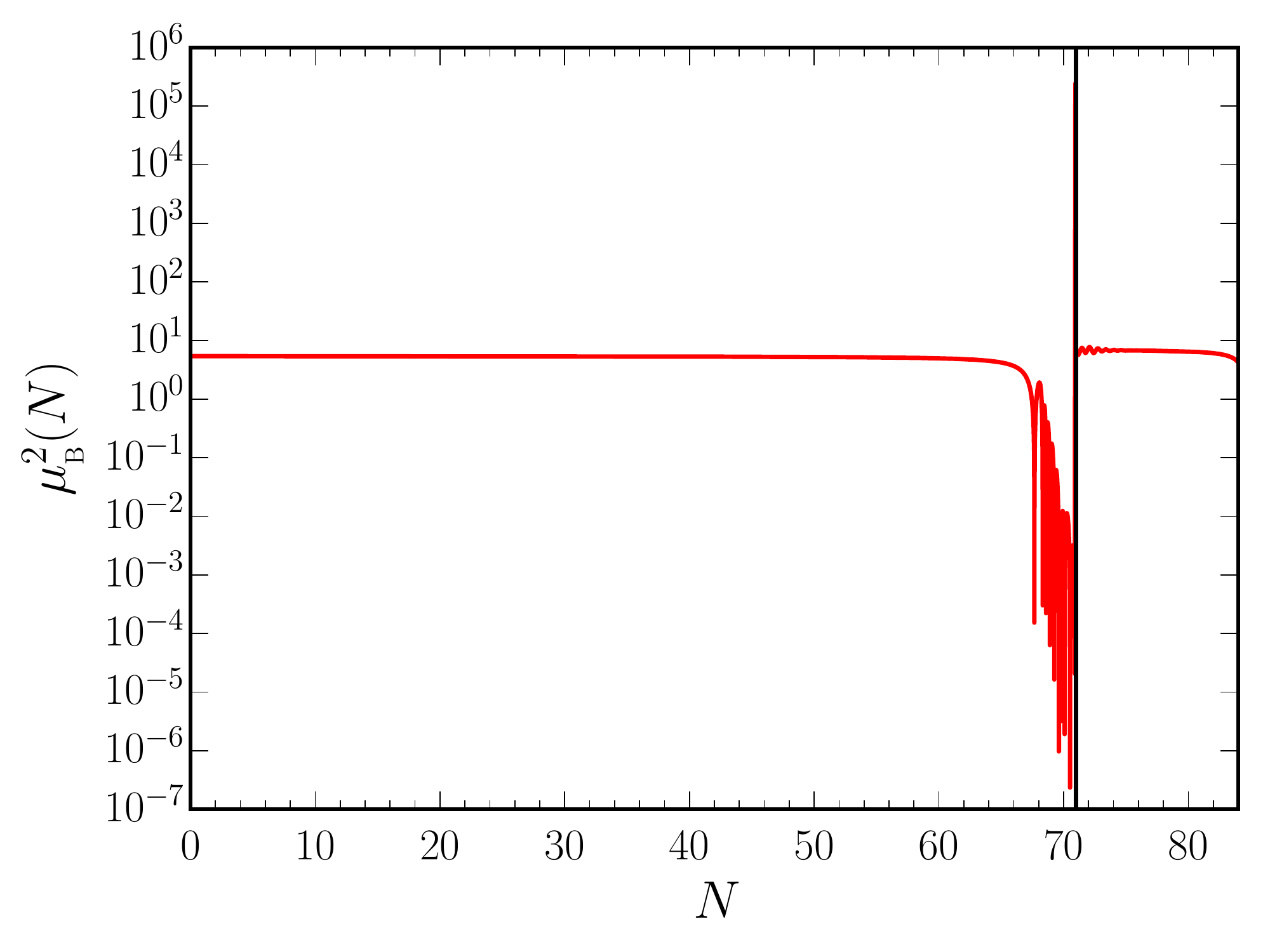}
\includegraphics[width=0.32\linewidth]{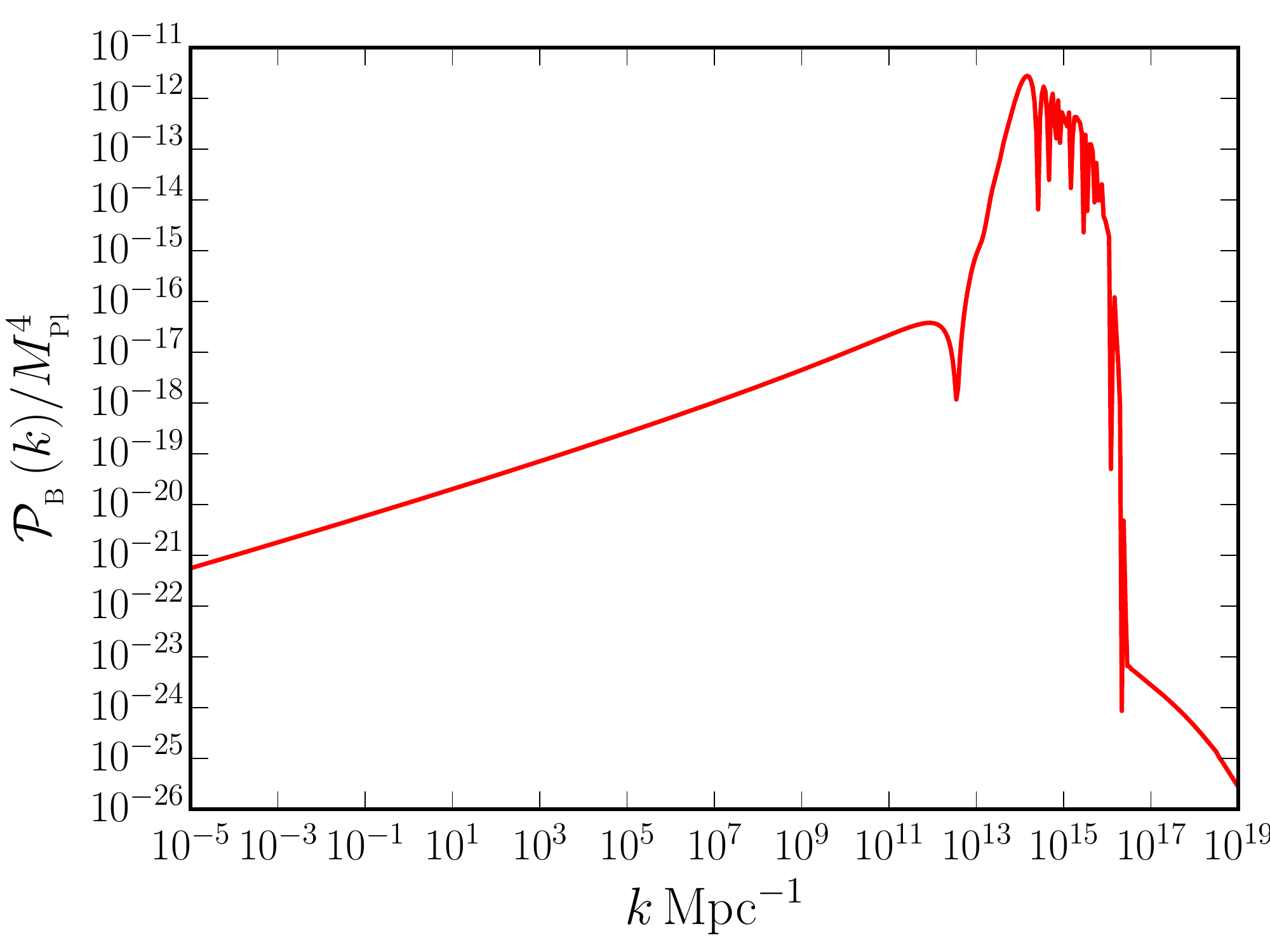}
\caption{The coupling functions $J(N)$ constructed analytically using 
Eq.~\eqref{eq:aJ} (on the left), along with the corresponding $\mub^2(N)$
(in the middle), and the resulting spectra of the magnetic field (on the
right) have been plotted for the models described by the 
potentials~\eqref{eq:sup-ls} (on top) and~\eqref{eq:ss-peak} (at the bottom).
We have indicated the point of transition in the plots of $J(N)$ and $\mub^2(N)$
(as vertical black lines).
Note that the above spectra of the magnetic field differ from the spectra we
had arrived at earlier in Fig.~\ref{fig:pb-pe}.
The differences can be attributed to the inability of the analytical solutions
to capture the dynamics of the fields around the point of transition from the 
first stage to the second stage of slow roll inflation.}\label{fig:J-JppJ-ps}
\end{figure*}

Earlier, in Fig.~\ref{fig:ebg}, we had compared the analytical solutions for
the background scalar fields we have obtained above with the exact numerical 
results.
Clearly, while the analytical solutions are a good approximation to the exact
numerical results in the two domains involving slow roll, they perform poorly 
around the transition. 
In Fig.~\ref{fig:J-JppJ-ps}, we have plotted the non-conformal coupling function
$J$ we have arrived at analytically using the expression~\eqref{eq:aJ} for the 
two models of our interest.
In the figure, we have also plotted the quantity $\mub^2=J''/(J\,a^2\,H^2)$ 
and the resulting power spectra of magnetic fields $\pb(k)$ for the two models. 
As should be evident, though the strengths of magnetic field roughly match the 
numerical results we had obtained earlier (plotted in Fig.~\ref{fig:pb-pe}), the 
shapes of the power spectra are fairly different.
This can be attributed to the discontinuous behavior of the fields around the
point of transition in the analytical case.   


\section{Impact of the choice of the parameters in the non-conformal 
coupling function}\label{app:e-of-p}

Recall that, the non-conformal coupling function $J(\phi, \chi)$ in 
Eq.~\eqref{eq:J-phi-chi} was constructed so that its evolution was 
determined by the field driving the background expansion at any given 
time.
Such a construction had ensured that the function largely behaves in the 
manner that we desire, i.e. as $J(\phi,\chi)\propto a^2$ (see Fig.~\ref{fig:J-mub2}). 
The point at which $J(\phi, \chi)$ switches its dependence on the evolution 
of~$\phi$ to that of~$\chi$ is determined by~$\chi_1$. 
Also, the range over which this switch happens is determined by~$\Delta \chi$. 
Earlier, while arriving at the power spectra of the electromagnetic fields due 
to such a coupling function, we had worked with specific values of these two 
parameters.
In this appendix, we shall discuss the impact of the choice of these parameters 
on the power spectrum of the magnetic field.

It seems natural to choose the value of $\chi_1$ to be the point at which the turn 
in the trajectory in the field space occurs (as marked in Fig.~\ref{fig:ebg}). 
The value of $\Delta \chi$ can be chosen to be that it roughly corresponds to the
duration of the transition. 
However, we ought to consider the effects that may occur due to variation of these 
parameters and quantify the dependence of the features in the spectrum of the magnetic
field on such variations. 
Evidently, we should be cautious so that, even as we try to capture the features 
that arise from the intrinsic dynamics of the fields, we do not end up introducing
features from the very construction and parametrization of the coupling function.

\begin{figure*}
\centering
\includegraphics[width=7.5cm]{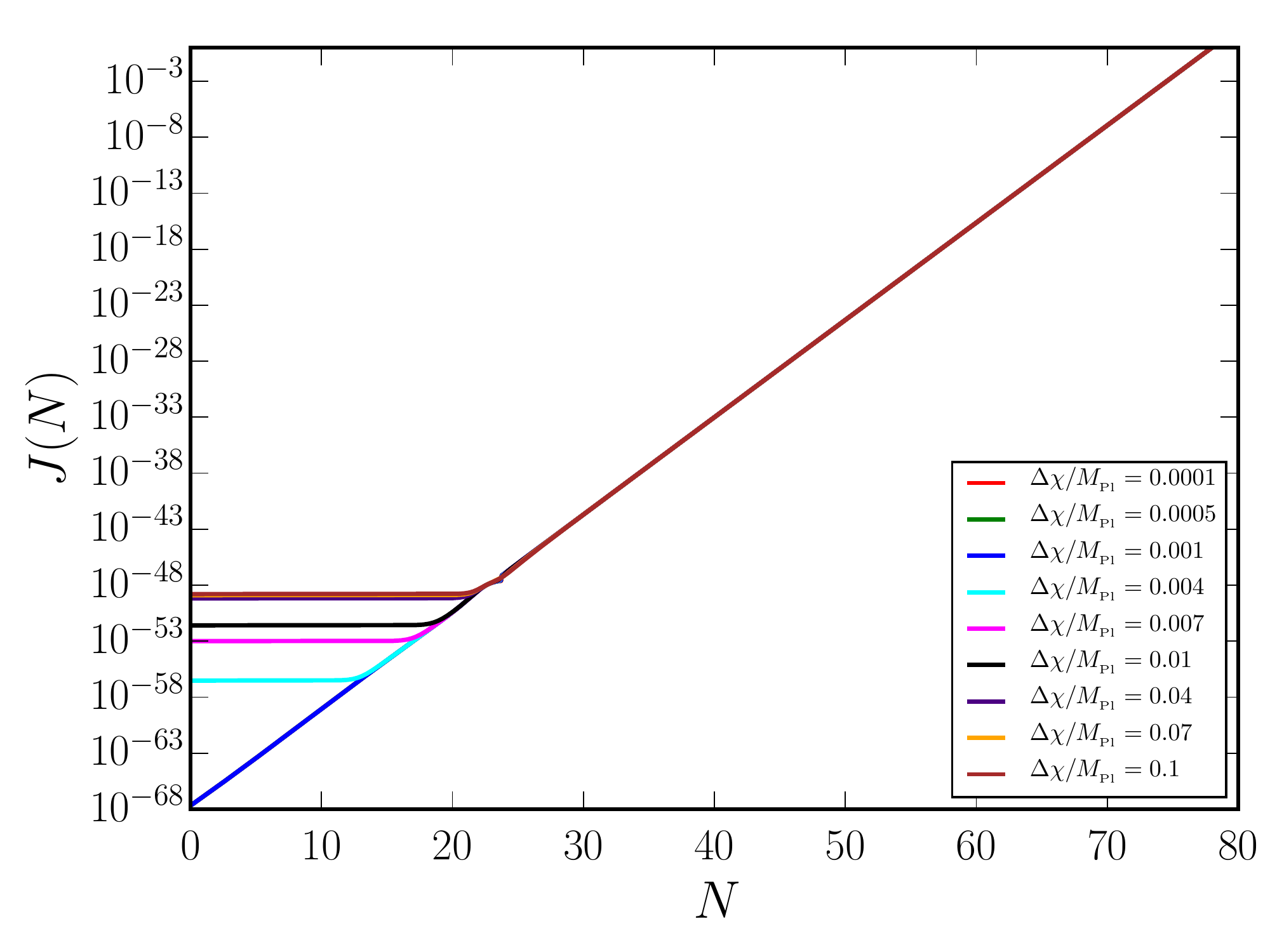}
\includegraphics[width=7.5cm]{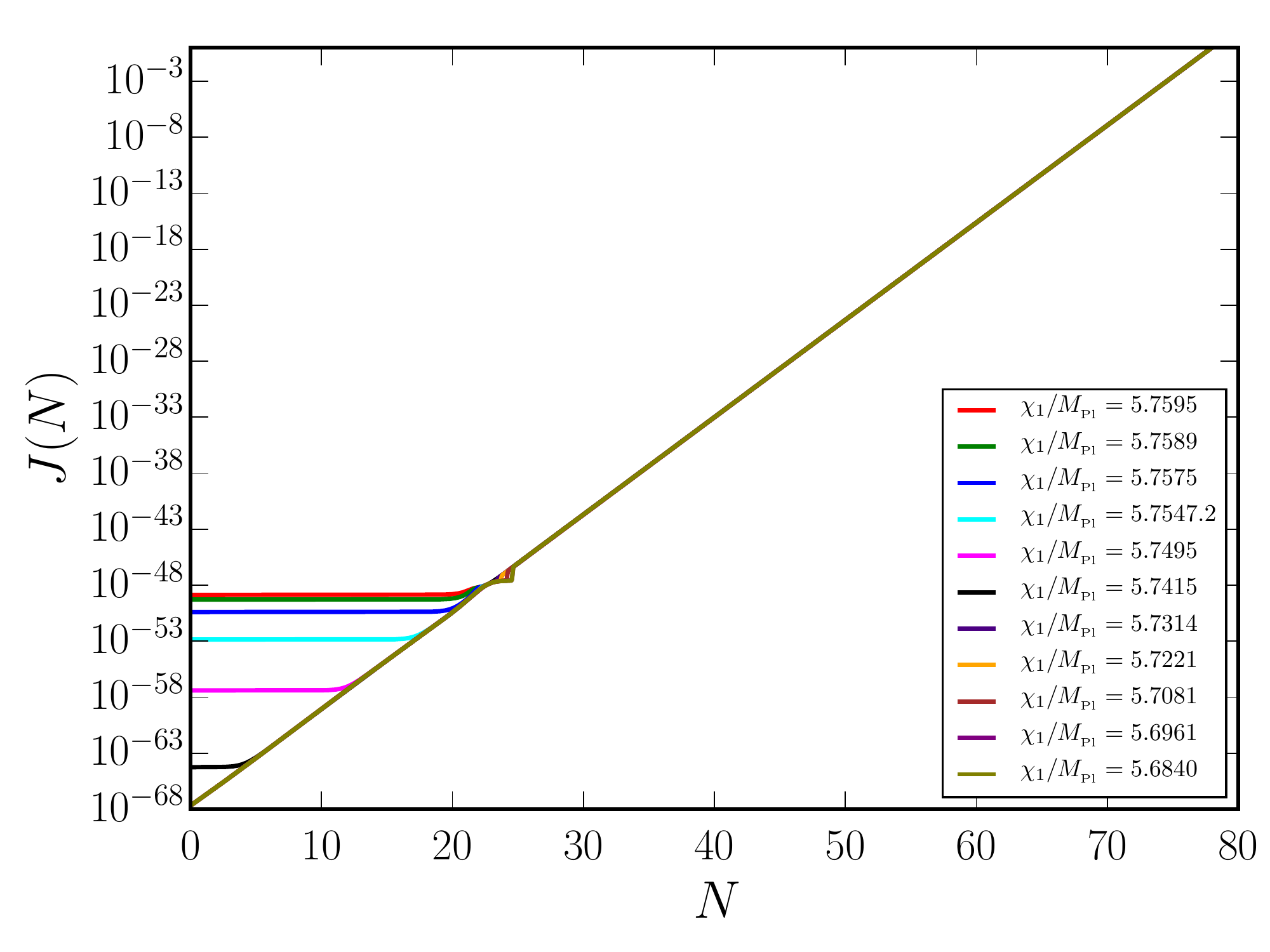}
\includegraphics[width=7.5cm]{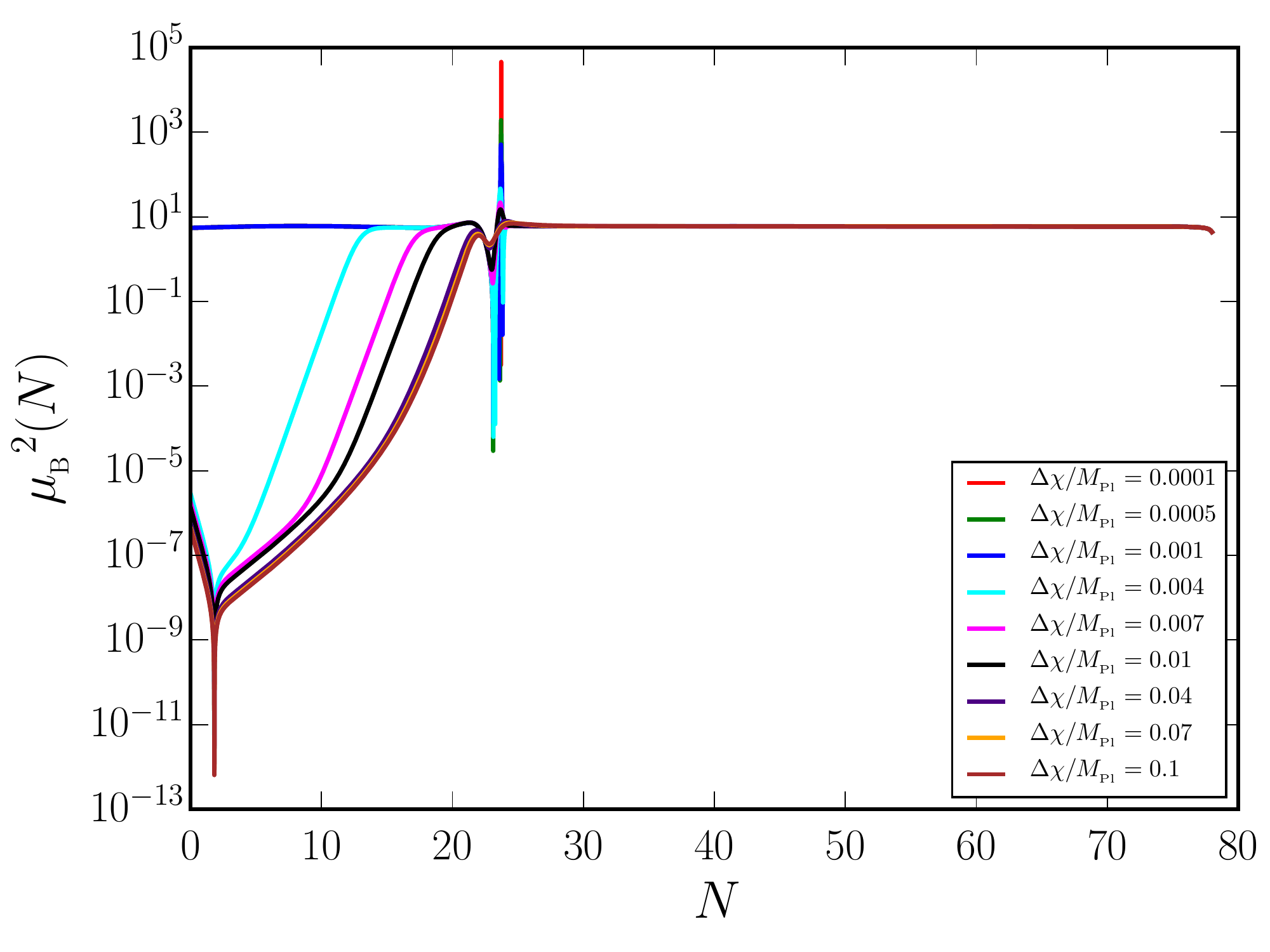}
\includegraphics[width=7.5cm]{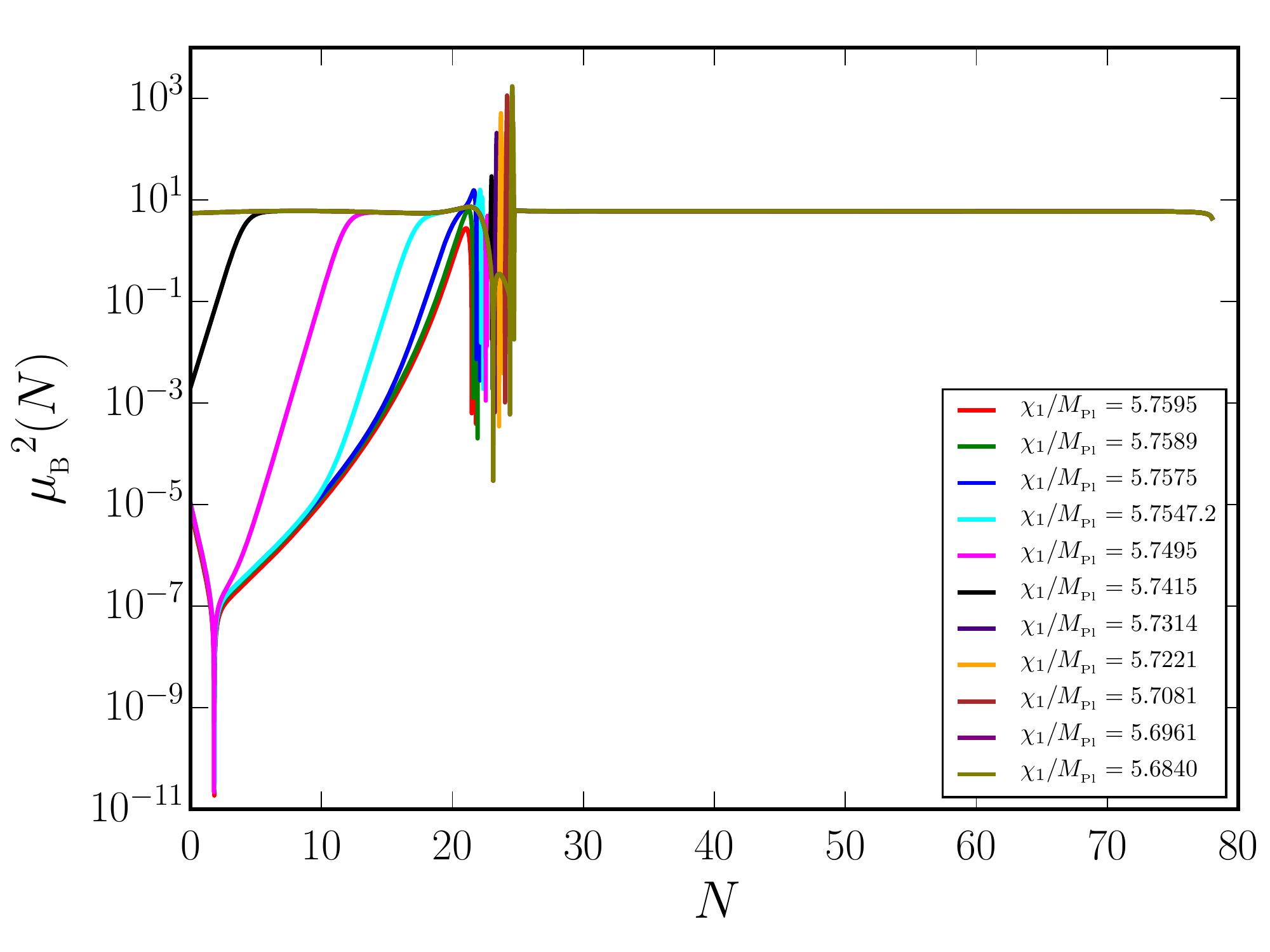}
\includegraphics[width=7.5cm]{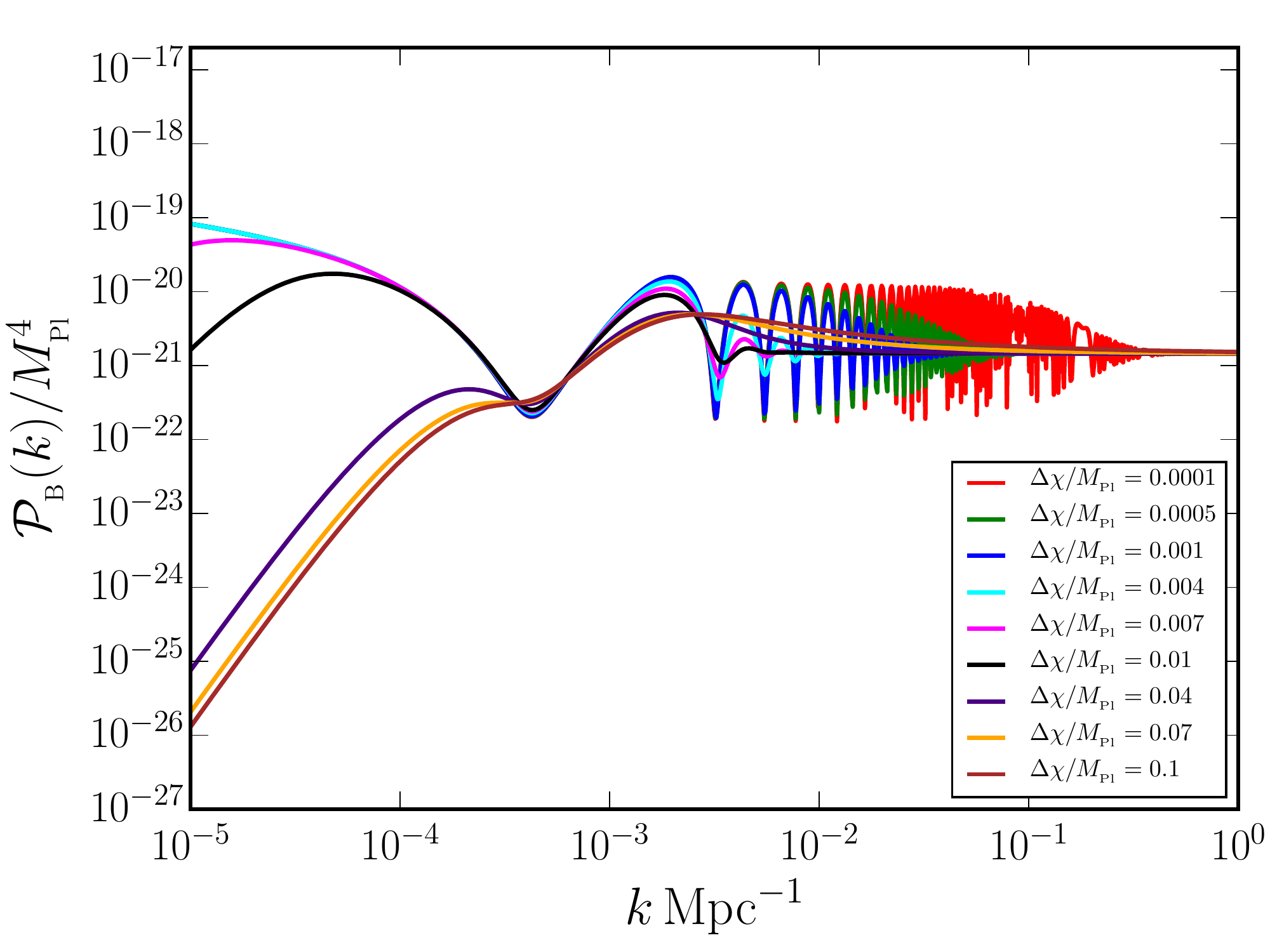}
\includegraphics[width=7.5cm]{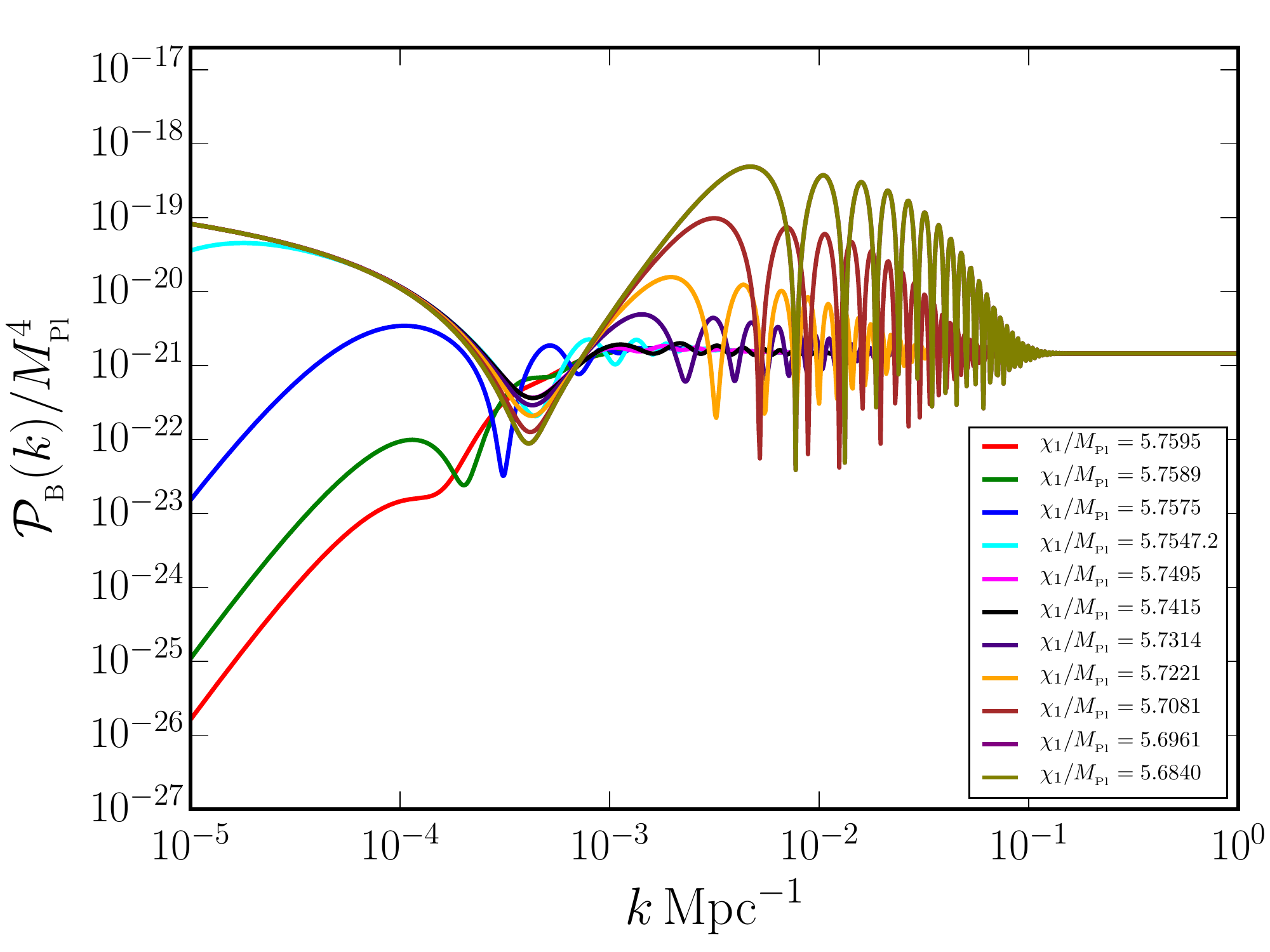}
\caption{We have presented the behavior of $J(\phi, \chi)$, $\mub^2(N)$, and 
the corresponding power spectrum of the magnetic field~$\pb(k)$ (on top, in 
the middle, and bottom rows, respectively) that arise due to the variation of 
the parameters $\Delta \chi$ (on the left column) and $\chi_1$ (on the right column)
in the case of the first model described by the potential~\eqref{eq:sup-ls}.
The parameters have been varied around the values of $\Delta \chi=10^{-3}\,\Mpl$ 
and $\chi_1=5.722\,\Mpl$ that we had considered earlier. 
We should mention that we have retained the original values of the other parameters 
in arriving at these results.
For larger values of $\Delta \chi$, which lead to a smoother transition of 
$J(\phi, \chi)$, we find that $\pb(k) \propto k^4$ over large scales, whereas 
for smaller values of $\Delta \chi$, effecting a sharper transition, we obtain 
a nearly scale invariant spectrum in the asymptotic domains (in wave number) 
with oscillations that extend over a wider range of wave numbers.
Moreover, while larger values of $\chi_1$ lead to the $\pb(k)\propto k^4$ behavior
over large scales, smaller values result in asymptotically (i.e. in wave numbers)
scale invariant spectra with oscillations that are of higher amplitude over the 
intermediate domain.}\label{fig:app2}
\end{figure*}

We have analyzed the effects of the parameters $\chi_1$ and $\Delta\chi$ 
on the spectrum of the magnetic field in the case of the first model described 
by the potential~\eqref{eq:sup-ls}.
We have presented the results of the exercise in Fig.~\ref{fig:app2}.
Note that, in our analysis, while we vary $\chi_1$ and $\Delta\chi$, we have 
retained the original values for parameters of the model and of the fitting 
functions~$N(\phi)$ and $N(\chi)$ [cf. Eqs.~\eqref{eq:N-phi}
and~\eqref{eq:N-chi}].
To begin with, we shall discuss the effects due to variation of $\Delta \chi$.
The value of $\Delta \chi=10^{-3}\,\Mpl$ which we have used earlier, seems to 
be an appropriate choice since, for such a value, we are able to achieve the 
desired behavior of $J(\phi,\chi)\propto a^2$ without considerable deviations 
during the transition.
However, for larger values of the parameter, say that lie in the range $10^{-3}\,
\Mpl \leq \Delta \chi < 10^{-1}\,\Mpl$, we observe that, prior to the transition,
the coupling function $J(\phi,\chi)$ turns to be a constant.
This essentially arises due to the smoothing of the hyperbolic tangent function 
that we had introduced to effect the transition between the two parts of~$J(\phi,\chi)$. 
A smoother hyperbolic tangent function suppresses the contribution due to the 
evolution of $\phi$ before transition and that of $\chi$ after the transition. 
Because of this reason, $J(\phi,\chi)$ settles to a constant over the smoothed 
regime. 
Such a behavior of $J(\phi,\chi)$ leads to extremely small values of $J''/J$,
which invariably results in the spectrum of the magnetic field $\pb(k)$ behaving 
as $k^4$ over large scales (as can seen in the plots in left column of Fig.~\ref{fig:app2}).
Moreover, if we make the transition sharper, i.e. if we choose $\Delta \chi < 
10^{-3}\,\Mpl$, though the spectrum of the magnetic field largely retains its 
shape, there arise oscillations over a wider window of wave numbers between the 
two domains of scale invariance. 
This is expected since a faster transition leads to sharp peak in $J''/J$ between 
the two regimes. 
Hence, we can conclude that, to avoid any artificial features such as either a
suppressed power over large scales or a prolonged burst of oscillations in the 
spectrum of the magnetic field, the choice of $\Delta \chi=10^{-3}\,\Mpl$
seems optimal.

Let us now turn to understanding the effects due to the variation in $\chi_1$. 
Upon choosing the value of $\chi_1$ to be greater than $5.722\,\Mpl$, we observe 
that the non-conformal coupling function $J(\phi,\chi)$ again turns constant during 
the initial epoch, and hence the spectrum of the magnetic field $\pb(k)$ behaves as 
$k^4$ over large scales. 
This is due to the coupling function switching its dependence from $\phi$
to $\chi$ at an earlier time, before the turn in the trajectory in the field space
occurs.
Such a choice suppresses the dependence of $J(\phi,\chi)$ on $\phi$ during the 
initial regime and makes it follow the behavior of $\chi$ which is frozen during 
this epoch. 
As a result, $J''/J$ drops to very small values and, as we have already discussed, 
it leads to the $k^4$ behavior of the spectrum of the magnetic field over large
scales. 
For $\chi_1 \leq 5.722\,\Mpl$, we find that the spectrum regains its near scale 
invariance in the two asymptotic domains, but the amplitude of oscillations over 
the intermediate domain in wave numbers prove to be larger. 
Therefore, the ideal value of $\chi_1$ proves to be around $5.722\,\Mpl$, where 
the turn occurs in the trajectory in the field space. 
Otherwise, one may introduce either a suppression or oscillations with large 
amplitudes, which are clearly artifacts induced by a non-optimal value of~$\chi_1$.


\section{Power spectrum of fluctuations in the energy density of the  
electromagnetic field}\label{app:pem}

Let $\hat{\rho}_{_{\mathrm{EM}}}^{\bm k}(\eta)$ denote the operator associated 
with the energy density corresponding to a given wave vector ${\bm k}$ of the
electromagnetic field.
The power spectrum of fluctuations in the energy density of the electromagnetic 
field for a given mode, say, $P_{_\mathrm{EM}}(k)$, is defined through 
the relation~\cite{Bonvin:2013tba,Bonvin:2011dt}
\begin{eqnarray}
\langle \hat{\rho}_{_{\mathrm{EM}}}^{{\bm k}\dag}(\ee)\,
\hat{\rho}_{_{\mathrm{EM}}}^{{\bm k}'}(\ee)\rangle\,
& & -\,\langle \hat{\rho}_{_{\mathrm{EM}}}^{{\bm k}\dag}(\ee)\rangle\,
\langle\hat{\rho}_{_{\mathrm{EM}}}^{{\bm k}'}(\ee)\rangle\nn\\
& &=(2\,\pi)^3\,P_{_\mathrm{EM}}(k)\,
\delta^{(3)}({\bm k}-{\bm k}'),\qquad\label{eq:P-EM}
\end{eqnarray}
where, as mentioned earlier, $\ee$ denotes the conformal time coordinate close 
to the end of inflation.
Note that the expectation values in the above expression are to be evaluated 
in the Bunch-Davies vacuum.

Recall that, in the non-helical case, for $J \propto a^2$, the energy density
of the electric field is negligible at late times.
Therefore, the total energy density of the electromagnetic field for a given
mode can be expressed in terms of the Fourier modes of the magnetic field, say,
$B_{i{\bm k}}$, as follows:
\begin{equation}
\rem^{\bm k}(\eta)
=\f{J^2(\eta)}{8\,\pi} \int \f{{\rm{d}}^3{\bm q}}{(2\,\pi)^{3/2}}\, 
B_{i\,{\bm q}}(\eta)\,B^i_{({\bm k} - {\bm q})}(\eta),\label{eq:rho-EM}
\end{equation}
where $B_i=\epsilon_{ijl}\,(\pa^{j}A^l)/a$, $B^i=g^{ij}\,B_j$, and 
$B_{i\,{\bm k}}$ denotes the Fourier modes associated with the magnetic field.
For the case wherein the spectrum of the magnetic field is scale invariant 
[i.e. $\pb(k)=9\,\HI^4/(4\,\pi^2)$, see Eq.~\eqref{eq:pb-ne2h}], upon 
substituting Eq.~\eqref{eq:rho-EM} in Eq.~\eqref{eq:P-EM} and using Wick's 
theorem, we find that the power spectrum $P_{_\mathrm{EM}}(k)$ can be 
expressed as
\begin{eqnarray}
P_{_\mathrm{EM}}(k) 
&=& \f{1}{2\,\pi^2}\,\l(\f{3\,\HI^2}{4\,\pi}\r)^4\,
\biggl[\int\f{{\d}^3{\bm q}}{q^3\,\vert\bm{k}-\bm{q}\vert^3}\nn\\
& &+\,\int\f{{\d}^3{\bm q}}{q^5\,\vert\bm{k}-\bm{q}\vert^5}\,
\l[{\bm q}\cdot\l({\bm k}-{\bm q}\r)\r]^2\biggr].
\end{eqnarray}
Upon carrying out the integrals over~${\bm q}$, we obtain that
\begin{equation}
k^3\,P_{_\mathrm{EM}}(k) 
= \f{16}{3\,\pi}\,\l(\f{3\,\HI^2}{4\,\pi}\r)^4\,
{\ln}\l(\f{k}{k_{\mathrm{min}}}\r),\label{eq:pem-fv}
\end{equation}
where we have introduced the infrared cut-off $k_{\mathrm{min}}$ to 
regulate the integral.
It is this result for $P_{_\mathrm{EM}}(k)$ that we have utilized to arrive
at the power spectrum for the curvature perturbations induced by the 
magnetic field, viz. $\mathcal{P}_{\mathcal{R}}^{\mathrm{mag}}(k)$, in
Eq.~\eqref{eq:ps-inf-mag}. 

\bibliographystyle{apsrev4-2}
\bibliography{references}

\end{document}